\documentclass[twocolumn,tighten]{aastex63}
\usepackage{amsmath}
\usepackage{comment}
\usepackage{tabularx}
\usepackage{algorithm2e}
\usepackage{float}
\usepackage{fullwidth}
\usepackage{nccmath}
\usepackage{mathtools}
\usepackage{multirow}
\usepackage{scrextend}
\definecolor{yscol}{rgb}{0.8, 0.6, 1}

\DeclareMathOperator*{\argmax}{arg\,max}

\newcommand{\msun}{\,{\rm M}_\odot}

\newcommand{\kpc}{\,{\rm kpc}}

\newcommand{\s}{\,{\rm s}}
\newcommand{\g}{\,{\rm g}}

\newcommand{\om}{\Omega_{\rm m}}
\newcommand{\sig}{\sigma_{8}}
\newcommand{\asnone}{A_{\rm SN1}}
\newcommand{\asntwo}{A_{\rm SN2}}
\newcommand{\aagnone}{A_{\rm AGN1}}
\newcommand{\aagntwo}{A_{\rm AGN2}}

\newcommand{\btheta}{\boldsymbol{\theta}}
\newcommand{\bx}{\boldsymbol{x}}

\newcommand{\dd}{\mathrm{d}}
\newcommand{\boldf}{\boldsymbol{f}}
\newcommand{\bphi}{\boldsymbol{\phi}}
\definecolor{yscol}{rgb}{0.8, 0.6, 1}

\def\bs#1{{\boldsymbol{#1}}}

\newlength{\Oldarrayrulewidth}

\newcommand{\rv}[1]{{#1}}

\makeatletter
\newcommand{\thickhline}{%
    \noalign {\ifnum 0=`}\fi \hrule height 1pt
    \futurelet \reserved@a \@xhline
}
\newcommand{\thichline}{%
    \noalign {\ifnum 0=`}\fi \hrule height 0.8pt
    \futurelet \reserved@a \@xhline
}
\newcolumntype{"}{@{\hskip\tabcolsep\vrule width 0.8pt\hskip\tabcolsep}} 
\makeatother

\received{June 1, 2019}
\revised{January 10, 2019}
\accepted{\today}
\submitjournal{ApJ}

\graphicspath{{./}{figures/}}

\begin{document}




\title{Calibrating cosmological simulations with implicit likelihood inference using galaxy growth observables}

\correspondingauthor{Yongseok Jo}
\email{kerex@snu.ac.kr}
\author[0000-0003-3977-1761]{Yongseok Jo}
\affiliation{Center for Theoretical Physics, Department of Physics and Astronomy, Seoul National University, Seoul 08826, Korea}

\author{Shy Genel}
\affiliation{Center for Computational Astrophysics, Flatiron Institute, 162 5th Avenue, New York, NY, 10010, USA}
\affiliation{Columbia Astrophysics Laboratory, Columbia University, 550 West 120th Street, New York, NY, 10027, USA}

\author{Benjamin Wandelt}
\affiliation{Sorbonne Universite, CNRS, UMR 7095, Institut d’Astrophysique de Paris, 98 bis boulevard Arago, 75014 Paris, France}
\affiliation{Center for Computational Astrophysics, Flatiron Institute, 162 5th Avenue, New York, NY, 10010, USA}

\author{Rachel S. Somerville}
\affiliation{Center for Computational Astrophysics, Flatiron Institute, 162 5th Avenue, New York, NY, 10010, USA}
\affiliation{Department of Physics and Astronomy, Rutgers University, 136 Frelinghuysen Road, Piscataway, NJ 08854, USA}

\author{Francisco Villaescusa-Navarro}
\affiliation{Center for Computational Astrophysics, Flatiron Institute, 162 5th Avenue, New York, NY, 10010, USA}
\affiliation{Department of Astrophysical Sciences, Princeton University, 4 Ivy Lane, Princeton, NJ 08544 USA}

\author{Greg L. Bryan}
\affiliation{Center for Computational Astrophysics, Flatiron Institute, 162 5th Avenue, New York, NY, 10010, USA}
\affiliation{Department of Astronomy, Columbia University, 550 West 120th Street, New York, NY 10027, USA}

\author{Daniel Anglés-Alcázar}
\affiliation{Department of Physics, University of Connecticut, 196 Auditorium Road, U-3046, Storrs, CT 06269-3046, USA}
\affiliation{Center for Computational Astrophysics, Flatiron Institute, 162 5th Avenue, New York, NY, 10010, USA}

\author{Daniel Foreman-Mackey}
\affiliation{Center for Computational Astrophysics, Flatiron Institute, 162 5th Avenue, New York, NY, 10010, USA}

\author{Dylan Nelson}
\affiliation{Universität Heidelberg, Zentrum für Astronomie, Institut für theoretische Astrophysik, Albert-Ueberle-Str. 2, 69120 Heidelberg, Germany}

\author{Ji-hoon Kim}
\affiliation{Center for Theoretical Physics, Department of Physics and Astronomy, Seoul National University, Seoul 08826, Korea}



\begin{abstract}
In a novel approach employing implicit likelihood inference (ILI), also known as likelihood-free inference, we calibrate the
parameters of cosmological hydrodynamic simulations against observations, which has previously been unfeasible due to the high computational cost of these simulations. 
For computational efficiency, we train neural networks as emulators on $\sim$ 1000 cosmological simulations from the CAMELS project to estimate simulated observables, taking as input the cosmological and astrophysical parameters, and use these emulators as surrogates to the cosmological simulations.
Using the cosmic star formation rate density (SFRD) and, separately, stellar mass functions (SMFs) at different redshifts, we perform ILI on selected cosmological and astrophysical parameters ($\om$, $\sig$, stellar wind feedback, and kinetic black hole feedback) and obtain full 6-dimensional posterior distributions.
In the performance test, the ILI from the emulated SFRD (SMFs) can recover the target observables with a relative error of 0.17\% (0.4\%).
We find that degeneracies exist between the parameters inferred from the emulated SFRD, confirmed with new full cosmological simulations.
We also find that the SMFs can break the degeneracy in the SFRD, which indicates that the SMFs provide complementary constraints for the parameters.
Further, we find that the parameter combination inferred from an observationally-inferred SFRD reproduces the target observed SFRD very well, whereas, in the case of the SMFs, the inferred and observed SMFs show significant discrepancies that indicate potential limitations of the current galaxy formation modeling and calibration framework, and/or systematic differences and inconsistencies between observations of the stellar mass function.
\end{abstract}


\section{Introduction} 
\label{sec:intro}
The significant progress of cosmological simulations and observations has greatly improved our understanding of a wide variety of phenomena, such as the formation and evolution of large-scale structure.
N-body simulations have successfully simulated formation and evolution of large-scale structure  of $\Lambda$CDM universe \citep[]{springel2005Natur.435..629S,boylan-kolchin2009MNRAS.398.1150B,klypin2011ApJ...740..102K,klypin2016MNRAS.457.4340K}.
Furthermore, (magneto-)hydrodynamic simulations that include comprehensive subgrid models such as star formation, stellar winds and active galactic nuclei (AGN) feedback have been performed in a cosmological context and have made significant strides towards reproducing a realistic galaxy population across a range of cosmic epochs \citep[for a review]{volgelsberger2020NatRP...2...42V}. 
These include {\sc Illustris} \citep{illustris2014MNRAS.444.1518V,vogelsberger2014Natur.509..177V,genel2014MNRAS.445..175G,nelson2015A&C....13...12N}, cosmo-OWLS \citep{cosmoOWLS2014MNRAS.441.1270L}, {\sc Magneticum} \citep{magneticum2014MNRAS.442.2304H,magneticum2017A&C....20...52R},  {\sc Horizon-AGN} \citep{dubois2014MNRAS.444.1453D}, {\sc MassiveBlack-II} \citep{massiveblac2015MNRAS.450.1349K}, {\sc Eagle} \citep{schaye2015MNRAS.446..521S}, {\sc BlueTides} \citep{bluetides2016MNRAS.455.2778F}, {\sc Mufasa} \citep{dave2016MNRAS.462.3265D}, {\sc Romulus} \citep{romulus2017MNRAS.470.1121T}, {\sc BAHAMAS} \citep{bahamas2017MNRAS.465.2936M}, {\sc Simba} \citep{dave2019MNRAS.486.2827D}, {\sc IllustrisTNG} \citep{nelson2018MNRAS.475..624N,springel2018MNRAS.475..676S, pillepichconvergence2018MNRAS.473.4077P,naiman2018MNRAS.477.1206N,marinacci2018MNRAS.480.5113M}, {\sc Horizon Run 5} \citep{horizon52021ApJ...908...11L}, and {\sc ASTRID} \citep[]{astrid2022MNRAS.512.3703B,astrid2022MNRAS.513..670N}.

Meanwhile, wide-field and deep surveys have identified samples of many thousands of nearby and distant galaxies, respectively 
\citep{fornax2018A&A...620A.165V,aihara2018PASJ...70S...4A,kuijken2019A&A...625A...2K, erosita2021arXiv210614517B}. 
In addition, high-resolution imaging and spectroscopy have enabled investigations into the structure and kinematics of galaxies \citep{desi2019AJ....157..168D,desi2021SCPMA..6489811W}.
These observational breakthroughs have not only enabled access to a plethora of galaxies from which to construct global distributions of galaxy properties, such as cosmic star formation history \citep{madau2014ARA&A..52..415M} and galaxy stellar-mass functions at different redshifts \citep{smf2012MNRAS.421..621B,lejasmf2020ApJ...893..111L, smfmcleod2021MNRAS.503.4413M},
but also the exploration of many dimensions of galaxy properties, leading to scaling relationships such as 
the (baryonic) Tully-Fisher relation \citep{tullyfisher1977A&A....54..661T,tullyMcGaugh2000ApJ...533L..99M,mcgaugh2021AJ....162..202M}, 
mass-metallicity relation \citep{mzr1979A&A....80..155L,mzrgallazi2005MNRAS.362...41G,mzrFontanot2021MNRAS.504.4481F},
star-forming sequence \citep{sfmainNoeske2007ApJ...660L..43N,msSpeagle2014ApJS..214...15S, msLeja2021arXiv211004314L}, 
size-mass relation \citep{szShen2003MNRAS.343..978S,szvanderwel2014ApJ...788...28V,szmowla2019ApJ...880...57M},
and relations between galaxy properties and the mass of the central massive black hole \citep[]{kormendy1993AJ....105.1793K,kormendy2001AIPC..586..363K,merritt2001ApJ...547..140M}.

The remarkable progress of simulations and observations has provided considerable insights into physical processes for galaxy formation and evolution and has played a crucial role in constraining theoretical models.
However, simulations and observations have not been fully reconciled. 
Contributing factors are observational uncertainties, modeling uncertainties in the simulations, and ad hoc comparisons. 
For instance, cosmological simulations---specifically, the subgrid models such as stellar winds and black hole feedback---have generally been calibrated against only a handful of observables through by-eye comparisons between simulations and observations, along with educated guesses or simple parameter-space search algorithms \citep[]{schaye2015MNRAS.446..521S, pillepichconvergence2018MNRAS.473.4077P,oh2020MNRAS.497.5203O}.
The limits of this conventional calibration approach are as follows:
(1) The dimensions of the subgrid parameter space that one can cover is significantly limited when using by-eye comparisons.
In addition, as the number of parameters of interest increases, it becomes harder to provide educated guesses due to complex and intertwined relations between physical models and observables.
(2) It is challenging to calibrate against numerous observables simultaneously.
(3) The accuracy of the calibrated parameters is hard to determine due to the objective nature of the comparison process. 
(4) The simulation uncertainty, due to such sources as cosmic variance, is generally not taken into consideration.
The cosmological simulations suffer from uncertainty that comes from various sources of randomness such as initial conditions \citep{genel2014MNRAS.445..175G,genel2019ApJ...871...21G,keller2019MNRAS.482.2244K}.
This can lead to appreciable bias in the calibration.

Similarly, semi-analytical models (SAMs), which estimate properties of the galaxy population using parameterized physical models that include a number of free parameters, have rigorously tuned those parameters to reproduce certain observational properties of galaxy population using Bayesian inference together with the Markov chain Monte Carlo method (MCMC) \citep{lu2012MNRAS.421.1779L,lu2014MNRAS.443.1252L,benson2014MNRAS.444.2599B}.
Bayesian inference is a widely-used method of statistical inference that updates one's knowledge---or belief in the Bayesian sense---of the parameters by making new observations.
The majority of the problems of the conventional calibration method for cosmological simulations can be alleviated by the use of Bayesian inference.
For instance,  Bayesian inference is conducted through the likelihood function that can mathematically guarantee the precision of calibration and enables inference from numerous observations simultaneously. 
Also, the probabilistic nature of the Bayesian inference captures the uncertainty of parameters through Bayes' theorem.
However, Bayesian inference usually entails MCMC, which is computationally expensive, for determining the posterior distribution. Moreover, MCMC sometimes fails to retrieve a posterior distribution, usually when the target posterior is complex and high dimensional.
In contrast to SAMs, hydrodynamical cosmological simulations with subgrid models are too computationally costly to perform hundreds of thousands of sequential simulations for MCMC, rendering such an approach impossible in practice.
This is the primary reason, despite all the merits, why calibration of cosmological simulations in the Bayesian framework has not been conducted thus far.
In addition, conventional Bayesian inference has its limitations in terms of its need for an explicit (analytic) likelihood.
Since the likelihood should be explicitly formulated, commonly-used analytic likelihoods such as Gaussians are only an approximation to the true unknown one.

The implicit likelihood inference approach (ILI)---also known as likelihood-free inference or simulation-based inference---provides a framework for performing rigorous Bayesian inference in a computationally efficient way, especially for inferences on computationally expensive simulations \citep[]{nde2016arXiv161003483M,papamakarios2018arXiv180507226P,Alsing_2018,Cranmer2020,durkan2020arXiv200203712D}.
In contrast to the conventional Bayesian inference that requires an explicit (analytic) formulation for the likelihood, ILI learns the likelihood---the conditional distribution of observable given the parameters---directly from simulated parameter-observable pairs using a neural density estimator (NDE).
NDEs are a flexible representation of the likelihood placing only mild assumptions on the likelihood form.

On the computational cost side, the likelihood can be evaluated through the trained NDE without performing further simulations.
The number of emulations required for the inference is equivalent to the number of emulations for training the NDEs, which is generally thousands of simulations \citep[]{alsing2019MNRAS.488.4440A}.
In conventional inference, although it generally depends on the complexity of the problem, the convergent MCMC typically requires at least $10^5$ samples in cosmological applications \citep[]{feroz2008MNRAS.384..449F,trotta2011ApJ...729..106T}, which is significantly more simulations than the NDE requires.
ILI has already started being vigorously exploited for inference and estimation of physical quantities in astrophysics, for example:
inference of the Hubble constant from binary neutron star mergers \citep{Gerardi2021PhRvD.104h3531G}; 
constraints on the cosmological parameters from weak lensing \citep[]{tam2022ApJ...925..145T};
mass estimations of the Milky Way and M31 \citep[]{Lemos2021PhRvD.103b3009L,villaneuva-domingo2021arXiv211114874V};
inference of strong gravitational lensing parameters \citep[]{legin2021arXiv211205278L};
dynamical mass estimation of galaxy clusters \citep[]{kodi2021MNRAS.501.4080K};
inference of reionization parameters from 21 cm power spectrum and light cones \citep[]{zhao2022ApJ...926..151Z,zhao2022arXiv220315734Z}.

In this work, we adopt the {\tt sbi} package \citep[]{sbi2020JOSS....5.2505T}, the successor of the {\tt delfi} package, which is equipped with various NDEs for ILI, to calibrate cosmological simulations against observations.
We also exploit the suite of cosmological simulations of the Cosmology and Astrophysics with MachinE Learning  Simulations (CAMELS) project \citep[]{camels2021ApJ...915...71V} that includes the largest data set designed to train machine learning models and provides more than a thousand simulations for exploring the cosmological and astrophysical parameter space.
Despite the large number of simulations in the CAMELS project, it is still not enough to directly employ our ILI technique, so we build an emulator that is trained on the CAMELS simulations to estimate the target observable taking the cosmological and astrophysical parameters as input.
This provides much flexibility and reduces computational cost during inference, which is in line with \citet[]{elliott2021MNRAS.506.4011E} that uses an emulator to calibrate a semi-analytic model.
Using the emulator as a surrogate to the actual cosmological simulations, we perform ILI  using the observed cosmic star formation history \citep[]{msLeja2021arXiv211004314L} and the observed stellar mass functions \citep[]{lejasmf2020ApJ...893..111L} to infer the parameters that are varied in the CAMELS suite -- two cosmological parameters ($\om$ and $\sig$) and four astrophysical parameters (related to stellar wind feedback and kinetic black holes feedback).

The structure of this paper is as follows. 
In Section \ref{sec:method_camels}, we describe the CAMELS simulations that we use to train the emulators and the details of the target observables.
In Section \ref{sec:method_ILI}, we give a brief review
of the ILI method including the NDE and our emulator design. 
In Section \ref{sec:emulated_sfrd}, we investigate the performance and convergence of the posterior distributions of the cosmological and astrophysical parameters inferred from an emulated star formation rate density (SFRD) as the target observable. 
In Section \ref{sec:observation_sfrd}, we perform the inference from the observed SFRD and study how the inferred SFRD matches the observed one.
In Section \ref{sec:performance_smf}, we investigate the performance and convergence of the posterior distributions from an emulated stellar mass function (SMF) as the target observable and in Section \ref{sec:observation_smf}, we perform the inference from observed SMFs and study the discrepancies between those and the inferred ones.
In Section \ref{sec:discussion}, we discuss the properties of the inferred posteriors and mismatch between inferences and observations with respect to the correlation between the parameter-observable pairs and physical analysis of cosmological simulations.
In Section \ref{sec:summary}, we present a summary of the results and findings.

\section{Cosmological Simulations: the CAMELS project}
\label{sec:method_camels}
\subsection{Overview}
Cosmology and Astrophysics with MachinE Learning Simulations (CAMELS)\footnote{\url{https://www.camel-simulations.org}} is a suite of 4,233 cosmological simulations: 2,184 (magneto-)hydrodynamic simulations with the {\sc AREPO} and {\sc GIZMO} codes, and 2,049 N-body simulations \citep{camels2021ApJ...915...71V}. 
Each simulation contains $256^3$ dark matter particles of mass $6.49 \times 10^7 (\Omega_m - \Omega_b)/0.251 h^{-1} \msun$ and $256^3$ gas cells with an initial mass of $1.27 \times 10^7 h^{-1}\msun$ in a periodic box of comoving volume of $(25 h^{-1}\mathrm{Mpc})^3$, which results in a resolution comparable to but slightly lower than that of {\sc TNG300} simulation of the {\sc IllustrisTNG} project \citep[]{pillepichconvergence2018MNRAS.473.4077P, marinacci2018MNRAS.480.5113M, naiman2018MNRAS.477.1206N, nelson2018MNRAS.475..624N, springel2018MNRAS.475..676S}.
The CAMELS project has been exploring a wide cosmological and astrophysical parameter space for the applications of machine learning in astrophysics.
The cosmological and astrophysical parameters of interest are $\Omega_\mathrm{m}$, $\sigma_8$, $A_\mathrm{SN1}$, $A_\mathrm{SN2}$, $A_\mathrm{AGN1}$, and $A_\mathrm{AGN2}$ (refer to Equations \ref{eq:camels1} to \ref{eq:camels6} for details).
The suite of (magneto-)hydrodynamic CAMELS simulations comprises four different sets for each of the {\sc AREPO} and {\sc GIZMO} codes, as follows: 
1) the LH set consists of 1000 simulations with different initial conditions varying all parameters sampled from a latin hypercube;
2) the 1P set consists of 61 simulations with the same initial condition varying only one parameter at a time;
3) the CV set consists of 27 simulations with fixed cosmology and astrophysics that sample cosmic variance using different initial conditions;


The simulations run with the {\sc AREPO} and {\sc GIZMO} codes use the {\sc TNG} and {\sc SIMBA} models, respectively, and we refer to these suites as the TNG and SIMBA runs. 
Throughout this work, we adopt the {\sc TNG} suites of the CAMELS simulations unless specified otherwise.
The LH set of the TNG suites is exploited to train the emulator (Section \ref{sec:emulator_as_surrogate}).
The CV set of the TNG suites is used to model simulation uncertainty (Appendix \ref{apx:cosmicvariance_butterflyeffect}). 
The 1000 simulations of the LH set are run with $\Omega_\mathrm{m} \in [0.1, 0.5]$, $\sigma_8 \in [0.6,1.0]$, $A_\mathrm{SN1}\in [0.25, 4.0]$, $A_\mathrm{SN2}\in [0.5,2.0]$, $A_\mathrm{AGN1}\in [0.25,4.0]$, and $A_\mathrm{AGN2}\in [0.5,2.0]$ arranged in a latin hypercube.
The 27 simulations of the CV set are run with $\Omega_\mathrm{m} = 0.3$, $\sigma_8 = 0.8$, $A_\mathrm{SN1}=A_\mathrm{SN2}=A_\mathrm{AGN1}=A_\mathrm{AGN2}=1$ but with different initial conditions.
In the meantime, the following cosmological parameters are fixed across all simulations: $\Omega_\mathrm{b} = 0.049,\, h=0.6711,\, n_\mathrm{s}=0.9624,\, M_\nu =0.0eV,\, w = -1$ and $\Omega_\mathrm{K}=0$. 
The {\sc TNG} suite of the CAMELS simulations implements the subgrid physics models of {\sc IllustrisTNG} \citep{weinberger2017MNRAS.465.3291W,pillepichconvergence2018MNRAS.473.4077P}. 
These simulations employ the {\sc AREPO} code\footnote{\url{https://arepo-code.org}} \citep{springel2010MNRAS.401..791S, weinberger2020ApJS..248...32W} to solve gravity (TreePM) and magneto-hydrodynamics using a Voronoi moving-mesh approach. 
The {\sc IllustrisTNG} physics includes various subgrid models: radiative cooling and heating, star-formation, stellar evolution, feedback from galactic winds, the formation and growth of the supermassive black holes (SMBH), and feedback from AGN.

\subsection{Physics of Astrophysical Parameters}
We have four astrophysical parameters that control the strengths of star formation-driven galactic winds and SMBH feedback. 
The star formation-driven galactic wind is isotropically injected in a kinetic form \citep[for details]{pillepichrescaling2018MNRAS.475..648P}.
The winds are characterized by a mass loading factor $\eta_w$ which is defined by
\begin{equation}
\label{eq:camels1}
    \eta_w \equiv \frac{\dot M_w}{\dot M_{\rm SFR}} = \frac{2}{v_w^2}e_w(1-\tau_w),
\end{equation}
where $\dot M_w$ and $\dot{M}_\mathrm{SFR}$ are the rate of gas mass to be converted into wind particles and the instantaneous, local star formation rate, respectively. 
With a fixed thermal energy fraction $\tau_w$, the mass loading factor is determined by the total energy injection rate per unit star-formation $e_w$ and the wind speed $v_w$ that involve $A_\mathrm{SN1}$ and $A_\mathrm{SN2}$, respectively as follows:
\begin{equation}
\begin{split}
    e_w = A_\mathrm{SN1}\times \bar{e_w} \left[f_{w,Z} + \frac{1-f_{w,Z}}{1+(Z/Z_{w,ref})^{\gamma_w,Z}} \right]\\
    \times N_\mathrm{SNII} E_\mathrm{SNII,51} 10^{51} \mathrm{erg} \msun^{-1}
\end{split}
\end{equation}
and
\begin{equation}
   v_w = A_\mathrm{SN2} \times \max\left[\kappa_w \sigma_\mathrm{DM}\left(\frac{H_0}{H(z)}\right)^{1/3}, v_{w,\mathrm{min}}\right],
\end{equation}
where details on the parameters $\bar{e_w}$, $f_{w,Z}$, $Z_{w,ref}$, $Z$, $\gamma_{w,Z}$, $N_\mathrm{SNII}$, $E_\mathrm{SNII,51}$, $\kappa_{w}$, $\sigma_8$, and $v_{w,\mathrm{min}}$ can be found in \citep[Table 1]{pillepichconvergence2018MNRAS.473.4077P}.

A SMBH particle with mass $M_{\rm seed} = 8 \times 10^5 h^{-1} \msun$ is seeded on-the-fly at the center of any halo with mass $M_{\rm FoF} > 5 \times 10^{10} h^{-1} \msun$ that does not yet contain a SMBH.  
To prevent the SMBH particles from artificially wandering around the galaxy, the SMBH particles are kept close to the potential minimum of their host dark matter haloes using an ad hoc prescription. 

{\sc IllustrisTNG} adopts the Bondi-Hoyle-Lyttleton accretion \citep{hoylelytttleton1939PCPS...35..405H, bondihoyle1944MNRAS.104..273B, bondi1952MNRAS.112..195B} with the Eddington cap for the growth of SMBHs. 
The states of SMBHs are distinguished into high accretion (a classical thin disc) and low accretion (hot accretion flow) based on a threshold of 
\begin{equation}
\label{equation:accretion_threshold}
\chi = \min\left[0.002\left(\frac{M_{\rm BH}}{10^8\msun}\right)^\bs{2},0.1\right]
\end{equation}
in units of the Eddington accretion limit.
According to the state of the accretion, the feedback mode is determined. For the high-accretion state, the feedback energy is injected as pure thermal energy into the vicinity of the SMBH (thermal mode). 
For the low-accretion state, feedback energy is released kinetically in a random direction (kinetic mode) as
\begin{equation}
    \dot{E}_{\rm low} = A_\mathrm{AGN1} \epsilon_{\rm f,kin} \dot{M}_\mathrm{BH}c^2,
\end{equation}
where 
\begin{equation}
    \epsilon_{\rm f,kin} = \min \left[\frac{\rho}{0.05\rho_{\rm SFthresh}}, 0.2\right].
\end{equation}
Here, $\rho$ and $\rho_{\rm SFthresh}$ are the gas density around the SMBH and the density threshold for star-formation.
The injection of the kinetic feedback occurs every time the accumulated energy has reached the energy threshold since the last feedback.
The energy threshold for the kinetic feedback is parameterized as
\begin{equation}
\label{eq:camels6}
    E_\mathrm{inj, min} = A_\mathrm{AGN2} \times f_\mathrm{re}\frac{1}{2}\sigma^2_\mathrm{DM} m_\mathrm{enc},
\end{equation}
where $\sigma^2_{\rm DM}$ is  the  one-dimensional  dark  matter  velocity  dispersion around the SMBH, $m_\mathrm{enc}$ is the enclosed gas mass within the feedback sphere, and $f_{\rm re}$ is a free parameter which is set to 20 for the fiducial TNG model.
$A_\mathrm{AGN2}$ controls the frequency and speed of the SMBH feedback.
The details of the prescription for the SMBH physics in {\sc IllustrisTNG} are described in \citep{weinberger2017MNRAS.465.3291W}.


\subsection{Target Observables: \\Cosmic Star Formation Rate Density \\ \& Stellar Mass Function}
\label{sec:method_sfrd_smf}
In this work, the observables from which the cosmological and astrophysical parameters are inferred are the cosmic star formation rate density (SFRD) and the stellar mass functions (SMF).
For a fair comparison between observations and simulations, we take into consideration the consistency between SFRD and SMFs.
For instance, the cosmological simulations, by nature, can guarantee consistency between the evolution of star formation rate and stellar mass. 
That is, the SFRD $f(z) = \dd M_{\star}(z)/\dd z + \dot{M}_{\rm return}$, where $M_{\star}(z)=\int_M \phi(M,z)$, $\phi(M,z)$ is the galaxy stellar mass function, and $\dot{M}_{\rm return}$ is the rate of mass return from evolving stellar populations.
On the observational side, the consistency depends on e.g. the modeling for each observable, and it is not guaranteed that the SFRD and the SMFs at different redshifts are consistent with one another (for more details, refer to Section 5 of \citet{leja2019ApJ...877..140L}).

To circumvent this, we adopt the SFRD of \citet[]{msLeja2021arXiv211004314L} ranging from $z=3$ to 0.5 and the five SMFs of \citet{lejasmf2020ApJ...893..111L} at $z =$ 0.5, 1, 1.5, 2.0, and 2.5, both of which are inferred with {\tt Prospector-$\alpha$} \citep[]{leja2019ApJ...877..140L} using galaxies in the 3D-HST \citep[]{skelton2014ApJS..214...24S} and COSMOS-2015 \citep[]{laigle2016ApJS..224...24L} catalogs  in a way that SFRD and SMFs are consistent with each other \citep[]{leja2019ApJ...877..140L}.
The galaxies are selected above the stellar mass-completeness limit taken from \citet[]{tal2014ApJ...789..164T} for the 3D-HST and \citet[]{laigle2016ApJS..224...24L} for COSMOS-2015.
In the case of the SFR estimation, the mass-completeness limits are adjusted upwards by $\mathcal{O}(0.1)$ dex since red galaxies, which can have an impact on SFR, is more likely to be excluded.
However, \citet[]{msLeja2021arXiv211004314L} finds that the resultant SFRD is not sensitive to these adjustments.
In addition, neither SFRD nor SMF have error bars since the statistical uncertainties are negligible, and the true uncertainties are systematic in nature.

On the simulation side, the SFRD is constructed from the global star formation rate per unit co-moving volume for 21 snapshots (21 redshifts), matching the redshifts between simulations and observations. 
The SMFs are obtained from the stellar mass of the galaxy catalog, with each binned into 13 bins in the range $[10^{8.9}, 10^{11.4}]\msun$ which \citet[]{lejasmf2020ApJ...893..111L} aims at.

\subsection{Resolution Effects: Convergence \& Rescaling}
\label{sec:method_rescaling}
The resolution convergence and effects in the TNG simulations have been extensively studied in \citet[Appendix B]{weinberger2017MNRAS.465.3291W}, \citet[Appendix A]{pillepichconvergence2018MNRAS.473.4077P}, and \citet[Section 3.3]{pillepichrescaling2018MNRAS.475..648P}. 
In general, observables such as the SMF at different resolutions are not converged. 
Figure \ref{fig:rescaling} shows that there is a shift of order tens of percent in both SMFs and SFRDs between TNG100-1 and TNG100-2. 
Note that the fiducial parameters are calibrated against observations for the TNG100 simulation of the {\sc IllustrisTNG} project, and the CAMELS parameter variations of the LH set are chosen around the fiducial values of the TNG100 simulation.
Hence, based on the approach of \citep[Appendix A]{pillepichrescaling2018MNRAS.475..648P}, we re-scale the SFRD and SMFs with a mass-modulated re-scaling factor. 
Since the resolution of CAMELS simulations are comparable to TNG100-2, we construct a re-scaling factor using the star formation rate-halo mass relation and the stellar mass-halo mass relation from TNG100-1 and TNG100-2 (details in Appendix \ref{appendix:rescaling}).
The re-scaled CAMELS simulations in Figure \ref{fig:rescaling} demonstrates that this procedure reduces the resolution effects at some level.
One remark is that we find that the re-scaling depends on the cosmological and astrophysical parameters.
However, in this work, we ignore this effect.
Further will be discussed in Section \ref{sec:discussion_resolution_effect}.

\subsection{Uncertainties in Simulation: \\Cosmic Variance \& Butterfly Effects}
\label{sec:cosmic_variance_buttefly_effect}
In this section, we focus on {\it simulation uncertainty} that is modeled as mock uncertainty and added to emulators in Section \ref{sec:emulator_as_surrogate}.
The {\it simulation uncertainty} is the intrinsic uncertainty of cosmological simulations.
In cosmological simulations, randomness in the initial conditions that correspond to the density fluctuations of the early universe leads to {\it cosmic variance}.
The minute position differences of the initial conditions owing to random seeds manifest as differences in the large-scale structure that directs to the galaxy populations.
On an observational side, {\it cosmic variance} can be attributed to the limited volume of the surveys.
Meanwhile, the {\it butterfly effect} stems from the chaotic behaviors of cosmological simulations. 
In dynamical systems, we can quantify chaotic or stochastic behaviors using a quantity called Lyapunov exponent.
If the positive Lyapunov exponent, the minute perturbations evolve exponentially with Lyapunov timescale and manifest as macroscopic differences. 
The chaotic behavior of the galactic dynamical systems amplifies minute fluctuations, seeded by randomness in the numerical computations such as stochasticity in subgrid models and floating errors, into appreciable differences in later times \citep{genel2019ApJ...871...21G}.
Also, \citet[]{keller2019MNRAS.482.2244K} studied stochasiticity of galaxy propertices in terms of using the particle-base code {\sc Gasoline} and the grid-base code {\sc Ramses}.
For details of quantification of simulation uncertainty, refer to Section Appendix \ref{apx:cosmicvariance_butterflyeffect}.

\begin{figure*}[ht!]
    \centering
    \includegraphics[width=\textwidth]{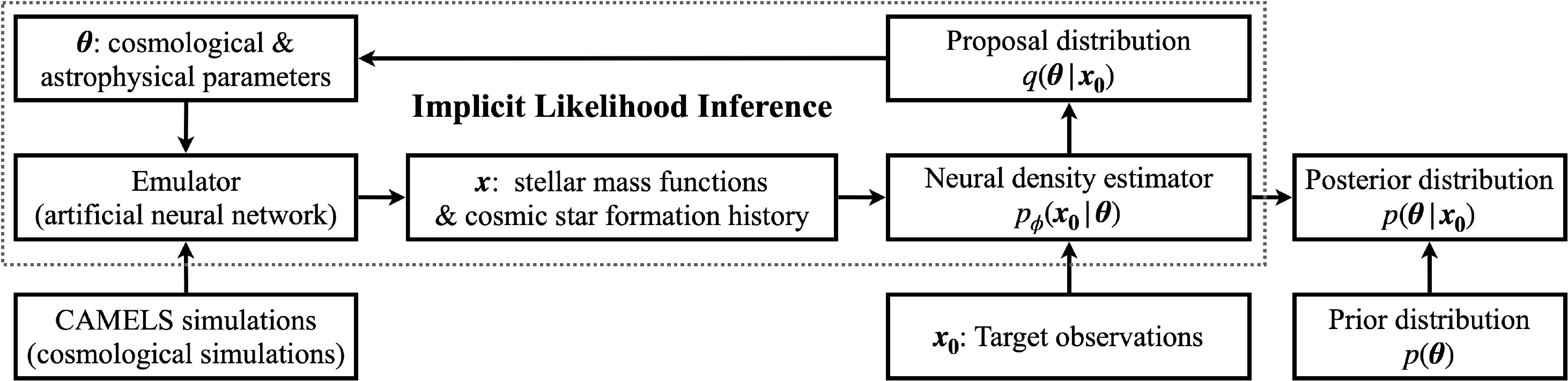}
    \caption{
    Diagram for the pipeline of this work that aims at calibrating cosmological simulations against observations.
    We use the CAMELS simulations (Section \ref{sec:method_camels}) to train emulators (Section \ref{sec:emulator_as_surrogate}) that estimate stellar mass functions and cosmic star formation history, $\bx$, taking cosmological and astrophysical parameters, $\btheta$, as input.
    Using the emulators as a surrogate to cosmological simulations, we perform ``implicit likelihood inference'' (Section \ref{sec:method_ILI}) from observations, $\boldsymbol{x_0}$, (Section \ref{sec:method_sfrd_smf}) to retrieve posterior distributions, $p(\btheta|\boldsymbol{x_0})$, of the cosmological and astrophysical parameters given the observations.}
    \label{fig:diagram}
\end{figure*}

\section{Implicit Likelihood Inference With Emulator}
\label{sec:method_ILI}
Implicit likelihood inference (ILI) aims at identifying the regions of parameter space of forward models to match observations \citep[]{sbi2020JOSS....5.2505T,Cranmer2020}.
In Bayesian terminology, ILI retrieves the posterior distribution over the parameters given an observation. 
In contrast to the conventional Bayesian inference, ILI requires no assumption or ansatz for likelihood $p(\bx|\btheta)$ so that it can also be applicable when analytical formulation for the likelihood is not accessible. 
Here, $\bx$ and $\btheta$ generally stand for observable and relevant parameters, respectively.

In this work, we adopt {\tt sbi}\footnote{https://github.com/mackelab/sbi}, a package designed to retrieve the posterior distribution $p(\btheta|\bx)$ \citep{sbi2020JOSS....5.2505T}. 
Instead of analytical probability distributions such as the Gaussian distribution, {\tt sbi} employs a neural network to output a probability distribution, called a neural density estimator (NDE). 
The NDE is a neural network that takes data points $\bx$ as input and returns a conditional probability distribution $\tilde{p}_{\bphi}(\btheta|\bx)$ over the simulation parameters such that $\int \tilde{p}_{\bphi}(\btheta|\bx) d\bx = 1$ where $\bphi$ represents neural network parameters. 

The {\tt sbi} package also aims at training the NDE with the least number of evaluations of forward models such as cosmological simulations, which are in general computationally expensive, by concentrating on the relatively small parameter space around parameter values of interest.
To this end, we must choose the proposal distribution $q(\btheta)$ from which we draw the sets of parameters for the new simulations in the next iteration.
In general, the choice of the optimal proposal for a specific problem is an open question.
Since we are interested in the high probable regions of parameter space, it might be natural to adopt the current approximate posterior density as a proposal density \citep{papamakarios2018arXiv180507226P, durkan2020arXiv200203712D}.
On the other hand, \citet[]{alsing2018proposalarXiv180806040A} adopts the geometric mean of prior and the current approximate posterior density in the context of sequential Approximate Bayesian Computation (ABC).
It might increase the probability of exploring parameter space more broadly beyond the posterior density.

\rv{Lastly, the {\tt sbi} package provides three different ways of computing an NDE: 
1. Sequential Neural Posterior Estimation (SNPE) trains an NDE to directly compute the posterior distribution;
2. Sequential Neural Likelihood Estimation (SNLE) trains an NDE to estimate the likelihood, which allows to evaluate the posterior distribution with the prior; 
3. Sequential Ratio Estimation (SRE) trains a binary classifier as an NDE to estimate density ratios, which can be used to estimate the posterior distribution (details will be discussed in Section \ref{sec:SRE}).
Of the three, we adopt the ratio estimation method because it generally requires a smaller amount of computational resources for training an NDE than others do, without losing much accuracy \citep[]{durkan2020arXiv200203712D}. 
We have compared the performances of three different methods in our setup based on the posteriors inferred from one of our observables using all three methods, but any noticeable differences cannot be found. 
}

\subsection{Neural Density Estimator: Learning the Density Ratio}
\label{sec:nde}
The neural density estimator (NDE) estimates the posterior density $p(\bs{\theta}|\bs{x})$ by computing the density ratio $r(\bs{\theta}, \bs{x}) =\frac{p(\bs{x}|\bs{\theta})}{p(\bs{x})} = \frac{p(\bs{\theta}|\bs{x})}{p(\bs{\theta})}$, where a simulator defines a valid probability density function $p(\bs{x}|\bs{\theta})$ over observations $\bs{x}$ \citep[]{sugiyama2012density,nde2016arXiv161003483M,hermans2019arXiv190304057H}.


Consider a binary random variable $Y:\Omega\rightarrow\mathbb{R}$ where $\Omega \in \{y=0, y=1\}$ and each outcome is equally likely a priori (i.e., $p(y=0)=p(y=1)$).
Then,
\begin{equation}
\begin{split}
    p(y=1|\boldsymbol{x}, \btheta) &= \frac{p(\boldsymbol{x},\btheta|y=1)}{p(\boldsymbol{x},\btheta|y=1)+p(\boldsymbol{x},\btheta|y=0)} \\
    &=\frac{ r(\bs{x},\btheta)}{1+r(\bs{x},\btheta)}
\end{split}
\end{equation}
where the density ratio $r(\boldsymbol{x},\btheta)$ is defined by $r(\boldsymbol{x},\btheta)\equiv\frac{p(\boldsymbol{x},\btheta|y=1)}{p(\boldsymbol{x},\btheta|y=0)}$ 
(refer to the derivation \footnote{
By definition of conditional probability, $p(y=1|\bx)=p(\bx|y=1)p(y=1)/p(\bx)$.
Using the law of total probability, we can rewrite
$p(y=1|\bx) = p(\bx|y=1)p(y=1)/(p(\bx|y=0)p(y=0)+p(\bx|y=1)p(y=1))$. 
Since each class is equally likely a priori, 
$p(y=1|\bx) = p(\bx|y=1)/(p(\bx|y=0)+p(\bx|y=1))$ . 
}).
That is, the binary classifier that computes $p(y=1|\boldsymbol{x},\btheta)$ or $p(y=0|\boldsymbol{x},\btheta)$ can be exploited to compute the ratio density $r(\bs{x},\btheta)$.

In case of the conditional probability density, $y=1$ class represents $(\boldsymbol{x}, \bs{\theta}) \sim p(\boldsymbol{x}, \bs{\theta})$ that $\bs{\theta}$ is drawn from the given prior $p(\bs{\theta})$ and $\boldsymbol{x}$ is obtained subsequently by the simulations with the sampled parameters.
The $y=0$ class consist of pairs $(\boldsymbol{x}, \bs{\theta}) \sim p(\boldsymbol{x})p(\bs{\theta})$ with parameters and observations sampled independently.
By training the neural classifier $\tilde{p}_{\phi}(\bx, \btheta)$ that takes $(\bs{x}, \bs{\theta})$ as input and outputs the probabilities $p(y=1|\boldsymbol{x}, \bs{\theta})$ or $p(y=0|\boldsymbol{x}, \bs{\theta})$ where $\phi$ stands for hyper-parameters (e.g., weights and biases) of the neural network, we can compute the density ratio 
\begin{equation}
    r(\boldsymbol{x},\bs{\theta}) = \frac{p(\bs{x},\bs{\theta})}{p(\bs{x})p(\bs{\theta})} = \frac{p(\bs{x}|\bs{\theta})}{p(\bs{x})} = \frac{p(\bs{\theta}|\bs{x})}{p(\bs{\theta})}.
\end{equation}
Therefore, if the prior $p(\bs{\theta})$ is known and can be evaluated, the posterior density $p(\bs{\theta}|\bs{x})$ can be obtained as $p(\bs{\theta}|\bs{x}) = r(\bs{x},\bs{\theta})p(\bs{\theta})$. 



\subsection{Sequential Ratio Estimation}
\label{sec:SRE}
We employ and modify the sequential ratio likelihood estimation (SRE) \citep[]{durkan2020arXiv200203712D} implemented in {\tt sbi}.
The SRE retrieves a posterior density using the neural density estimator described in Section \ref{sec:nde}.
The workflow of the SRE is as follows (see also Algorithm \ref{alg:SRE}):
(1) The SRE samples $M$ sets of parameters $\btheta_m$ from a prior $p(\btheta)$ or a proposal density $q^{(n)}(\btheta|\bx_0)$ and generates $M$ observables $\bx_m$ with the parameters $\btheta_m$ where $m = 1, \ldots, M$;
(2) The SRE constructs a training set $\mathcal{T}$ that consists of two classes of pairs: $(\bx_m, \btheta_m)$ for $y=1$ and $(\bx_k, \btheta_m)$ for $y=0$ by permutations where $k \neq m$, where $y$ simply represents a binary class (see Section \ref{sec:nde}).
(3) The NDE $\tilde{p}_{\phi}(\bx, \btheta)$ is trained on $\mathcal{T}$ until it converges;
(4) The structure of trained NDE is saved as an external file for future use.
(5) The posterior and the proposal densities are updated as $p^{(n)}(\btheta|\bx_0) \propto \tilde{p}_{\phi}(\bx_0|\btheta) p(\btheta)$ and $q^{(n)}(\btheta|\bx_0) \propto p^{(n)}(\btheta|\bx_0)p(\btheta)$, respectively.
(6) The SRE repeats from (1) with the newly updated proposal.
During the SRE, we use {\tt emcee}\footnote{https://github.com/dfm/emcee}, an MCMC package, for sampling parameters from the proposal \citep[]{emcee2013PASP..125..306F}. 
However, for the final production plots in this paper, parameters are drawn from the externally saved NDE using {\tt zeus} \citep[]{karamanis2020ensemble, karamanis2021zeus}.
The choice of an optimal proposal density is a crucial element for the SRE. 
However, this still remains as an open question. 
The posterior inferred in the previous round is the most common for proposal density for the next round \citep[]{papamakarios2018arXiv180507226P, durkan2020arXiv200203712D}.
In this work, we adopt the geometric mean of the prior and posterior density of the current epoch, inspired by \citet{alsing2018proposalarXiv180806040A}.

The SRE has two hyper-parameters subject to optimization: the number of simulations per iteration $M$ and the number of iterations $N$.
Since $M$ determines the amount of information that the training set can carry, it can strongly affect the accuracy of the posterior density. 
For instance, if $M$ is biased, the resultant posterior density can also be biased.
$M$ should be set to a sufficiently large value that the sampled training set can carry enough information about the trained posterior.
We set $M$ empirically via many trials as follows: new 100 (200) emulated SFRD that are generated by the emulator, with parameters sampled from the proposal density, are added to the training dataset.
$N$ is directly related to the convergence and stability of the inference. 
We perform the SRE without imposing definite $N$ and stop it whenever there is no improvement in training for 20 epochs---e.g., converged.

On the other hand, the structure of the NDE can be more decisive and has critical hyper-parameters in terms of accuracy of the inferred posterior density. 
The residual network (ResNet) is adopted for the baseline neural network for the NDE. 
We determine the complexity of the NDE depending on correlations between parameters and observables. 
Since the SMF-parameters pair has more correlated behaviour than the SFRD-parameters pair (Section \ref{sec:correlation_between_obs_and_params}), we adopt a deeper network for the SFRD than for the SMFs as follows:
the NDE, $\tilde{p}_{\phi}(\bx, \btheta)$, that we adopt for the SFRD (SMF) is a feed-forward residual network that consists of two residual blocks of 250 (100) hidden units (for detailed structures of the NDE, refer to Appendix $\ref{appendix:arch}$).

\RestyleAlgo{ruled}
\SetKwInput{KwInput}{Input}                
\SetKwInput{KwOutput}{Output}              
\SetKwInput{KwUpdate}{Update}              
\SetKwInput{KwInit}{Initialize}              
\SetKwInput{KwSave}{Save}              
\begin{algorithm}[!t]
\label{alg:SRE}
\caption{Sequential Ratio Estimation}
\KwInput{Simulator $\tilde{p}(\bx|\btheta)$, prior $p(\btheta)$, target observation $\bx_0$, neural density estimator $\tilde{p}_{\phi}(\bx, \btheta)$, iterations N, simulations per iteration M}
\KwInit{Proposal $q^{(1)}(\btheta|\bx_0)=p(\btheta)$, training set $\mathcal{T} =\{\}$}
\For{$n = 1$ to $N$}{
Draw $\btheta_{m} \sim$ $q^{(n)}(\btheta|\bx_0)$, $m=1, \,\ldots,\, M$;\\
Simulate $\bx_{m} \sim \tilde{p}(\bx|\btheta_m)$, $m=1, \,\ldots,\, M;$\\
Construct $\mathcal{T} = \mathcal{T} \cup \{(\bx_m, \theta_m)|m=1, \ldots, M\}$;\\
\While{$\tilde{p}_{\phi}$ not converged}{
    Sample mini-batch $\{(\bx_b, \btheta_b)\} \sim \mathcal{T}$;\\
    Optimize a neural density estimator $\tilde{p}_{\phi}(\bx, \btheta)$ using stochastic gradient descent;
  }
  Save the neural density estimator $\tilde{p}_{\phi}(\bx, \btheta)$\\
  Update posterior $p^{(n)}(\btheta|\bx_0) \propto \tilde{p}_{\phi}(\bx_0|\btheta) p(\btheta)$\\
  Update proposal $q^{(n)}(\btheta|\bx_0) \propto 
  \sqrt{p^{(n)}(\btheta|\bx_0)p(\btheta)}$
}

\KwOutput{Posterior $p^{(N)}(\btheta|\bx_0) \propto \tilde{p}_{\phi}(\bx_0|\btheta) p(\btheta)$}
\end{algorithm}

\subsection{Emulator: Surrogate for Cosmological Hydrodynamic Simulations}
\label{sec:emulator_as_surrogate}
The number of simulations required to retrieve the posterior density is highly correlated with dimensions and complexity of a problem. 
Nevertheless, ILI generally requires more than $\mathcal{O}(1000)$ simulations \citep[and Figures \ref{fig:emulated_sfrd_stability} and \ref{fig:emulated_smf_stability} in this work]{hermans2019arXiv190304057H,durkan2020arXiv200203712D,Huppenkothen2021MNRAS.tmp.3294H,dalmasso2021arXiv210703920D}, which exceeds the total number of the CAMELS simulations.
We, thus, circumvent this issue by adopting an emulator as a surrogate simulation.
The emulator is constructed upon a fully-connected neural network that is faster than hydrodynamic simulations by several of orders of magnitude.
We split the LH set of the CAMELS simulations into {\it training} (750), {\it test} (150), and {\it validation} (150) sets. 
Six independent neural {\it emulators} are trained on the training sets to estimate the SFRD and the five SMFs at five different redshift $z=0.5,1.0,1.5,2.0,2.5$, respectively, as a function of six cosmological and astrophysical parameters: $(\Omega_{\rm m},\, \sigma_8, \,A_{\rm SN1}, \,A_{\rm SN2}, \,A_{\rm AGN1}, \,A_{\rm AGN2})$ (further details of the six emulators are discussed in Appendix \ref{appendix:arch}).
We use {\tt Optuna} \citep{optuna2019arXiv190710902A}, an automatic hyper-parameter optimization tool, to train and optimize the emulators.
The hyper-parameters subject to optimization include learning rate, weight decay, the number of layers and the number of neurons.
During training, both inputs (cosmological and astrophysical parameters) and outputs (the SFRD and the SMFs) are normalized, using linear scaling\footnote{$x'=(x-x_{\rm min})/(x_{\rm max}-x_{\rm min})$ where $x'$ is the normalized input, and $x_{\rm min}$ and $x_{\rm max}$ are the minimum and maximum of the inputs, namely the edge values of the parameter ranges (refer to Section \ref{sec:method_camels}).} and z-score\footnote{$x'=(x-\mu)/\sigma$ where $x'$ is the normalized output and $\mu$ and $\sigma$ are the mean and the standard deviation of the outputs, respectively.}, respectively.
We measure the accuracy of the emulators with the mean square error (MSE) and the Pearson correlation coefficient.
The MSE for emulators are 0.0007 dex (SFRD) and 0.0011 dex (SMF). 
The Pearson coefficient correlations are 0.98 (SFRD) and 0.94 (SMF).

\subsubsection{Connection between Emulator and Cosmological Hydrodynamic Simulation}
\label{sec:connection}
The emulator can be the best option for a surrogate simulation with reasonable accuracy in terms of computational cost.
However, uncertainty of cosmological simulations such as cosmic variance and butterfly effects is missing in an emulator since our emulators are based on a simple fully-connected neural network that output predictions in a deterministic way without having any randomness (refer to Section \ref{sec:cosmic_variance_buttefly_effect} and Appendix \ref{apx:cosmicvariance_butterflyeffect} for details of {\it simulation uncertainty}).
The {\it simulation uncertainty} can play a significant role in a probabilistic inference such as ILI, especially in quantifying uncertainty in the inferred parameters, which depends on uncertainty of observable.
In this section, we focus on (1) how emulators marginalize the {\it simulation uncertainty} and (2) the mock uncertainty that is implemented in emulators to mimic the {\it simulation uncertainty}.

First, to investigate how much the emulators marginalize the {\it simulation uncertainty}, 
we use a deviation of the emulator $f(\btheta)$ from the uncertainty-marginalized ideal simulation $\bar{g}(\btheta)$ which denotes the ideal, uncertainty-free, infinite-volume simulation, defining the deviation $\Delta(\btheta)\equiv$ $f(\btheta)-\bar{g}(\btheta)$.
We approximate the deviation by taking an average over the ensembles of the thousand simulations in the LH set (note that the uncertainty-marginalized ideal simulation---e.g., the infinite-volume simulation---is unobtainable).
Here, two assumptions are made: 
(1) The mean of the infinite number of simulation ensembles converges to the uncertainty-marginalized ideal simulation.
(2) The {\it simulation uncertainty} is constant across the cosmological and astrophysical parameter space.
The details of the assumptions are further described in Appendix \ref{apx:marginalization}.

Using the deviation $\Delta(\btheta)$, we estimate the mean of the deviations over the LH set that the emulators are trained on (i.e., the bias $b_{\rm LH}=\left<\Delta(\btheta)\right>_{\rm LH}$) and the standard deviation over the LH set ($\hat{\sigma}_{\rm LH}=\sqrt{\left<\Delta(\btheta)^2\right>_{\rm LH}}$) where $\left<\cdot\right>_{\rm LH}$ averages over all the parameters $\btheta$ in the LH set.
If $b_{\rm LH}=0$ and $\hat{\sigma}_{\rm LH}=0$, the emulator perfectly marginalizes the {\it simulation uncertainty} and emulated prediction follows the uncertainty-marginalized ideal simulation.
In this work, the emulators have the bias of $b_{\rm LH}=0.003$ dex that can indicate that the emulators predict the observables with a relatively high accuracy {\it on average}.
However, provided that the standard deviations from emulators ($\hat{\sigma}_{\rm LH}=0.066$ dex) and simulations ($\sigma_{\rm sim, sfr}=0.057$ dex, refer to Appendix \ref{apx:marginalization} for details) are comparable, it is less likely for the emulators to marginalize the {\it simulation uncertainty}.
In other words, the emulator predictions are in agreement with the uncertainty-marginalized ideal simulations {\it on average but} each emulator prediction of a point in parameter space has a similar variance to {\it simulation uncertainty} with respect to the uncertainty-marginalized ideal simulations as the actual cosmological simulations does.

Without proper marginalization, the implementation of uncertainty in the emulator leads to a greater uncertainty in the parameters as well as the observables. 
In addition, this weakens the connection between simulations and emulators to a large extent in terms of physical interpretation of the parameters. 
Hence, in this work, we treat an emulator as ground truth or the mean of {simulation uncertainty}
and implement a mock {\it simulation uncertainty} on top of the emulators.
We model the {\it simulation uncertainty} using the multivariate Gaussian noise with minor modifications.
Then, the {\it mock simulation uncertainty}---hereafter {\it mock uncertainty}---is added to the emulators manually (refer to Appendix \ref{apx:cosmicvariance_butterflyeffect} for details of implementation).

\section{Inference from The History of the Cosmic Star Formation Rate Density}

\subsection{Inference from Emulated Histories of the Cosmic Star Formation Rate Density}
\label{sec:emulated_sfrd}
In this section, we study the performance and properties of ILI on the cosmic star formation rate density (SFRD) using an {\it emulator-based} SFRD rather than an observed SFRD. 
To this end, we use the emulator to predict SFRDs as functions of the cosmological and astrophysical parameters and adopt an {\it emulated} SFRD as a target observation from which the cosmological and astrophysical parameters can be inferred.
As discussed in Section \ref{sec:emulator_as_surrogate}, the emulators are adopted as surrogates for the hydrodynamic simulations and considered to be the ground truth throughout this work unless it is specified otherwise.

In contrast to deterministic approaches, probabilistic inference such as Bayesian inference and ILI have a significant flexibility in that the inferred posterior distributions can take on versatile structures of probability distributions depending on the nature of the problems.  
The probabilistic distributions have three main beneficial aspects in our application: 
(i) the variance of the posteriors can be interpreted as the error bars of inferred parameters (Section \ref{sec:performance_sfrd});
(ii) the posterior can have an appreciable volume of parameter space that reproduces the same target observable within some accuracy (i.e. degeneracy, refer to Section \ref{sec:bimodality_sfrd});
(iii) we can measure confidence intervals of each parameter under the presence of uncertainty in the observations (Section \ref{sec:uncertainty_sfrd}).
Lastly, in Section \ref{sec:observation_sfrd}, we apply our ILI machinery to a measurement of an observationally-derived estimate of SFRD and study how well the inferred posterior distribution can match the observations.

\begin{figure*}[t!]
    \centering
    \includegraphics[width=\textwidth]{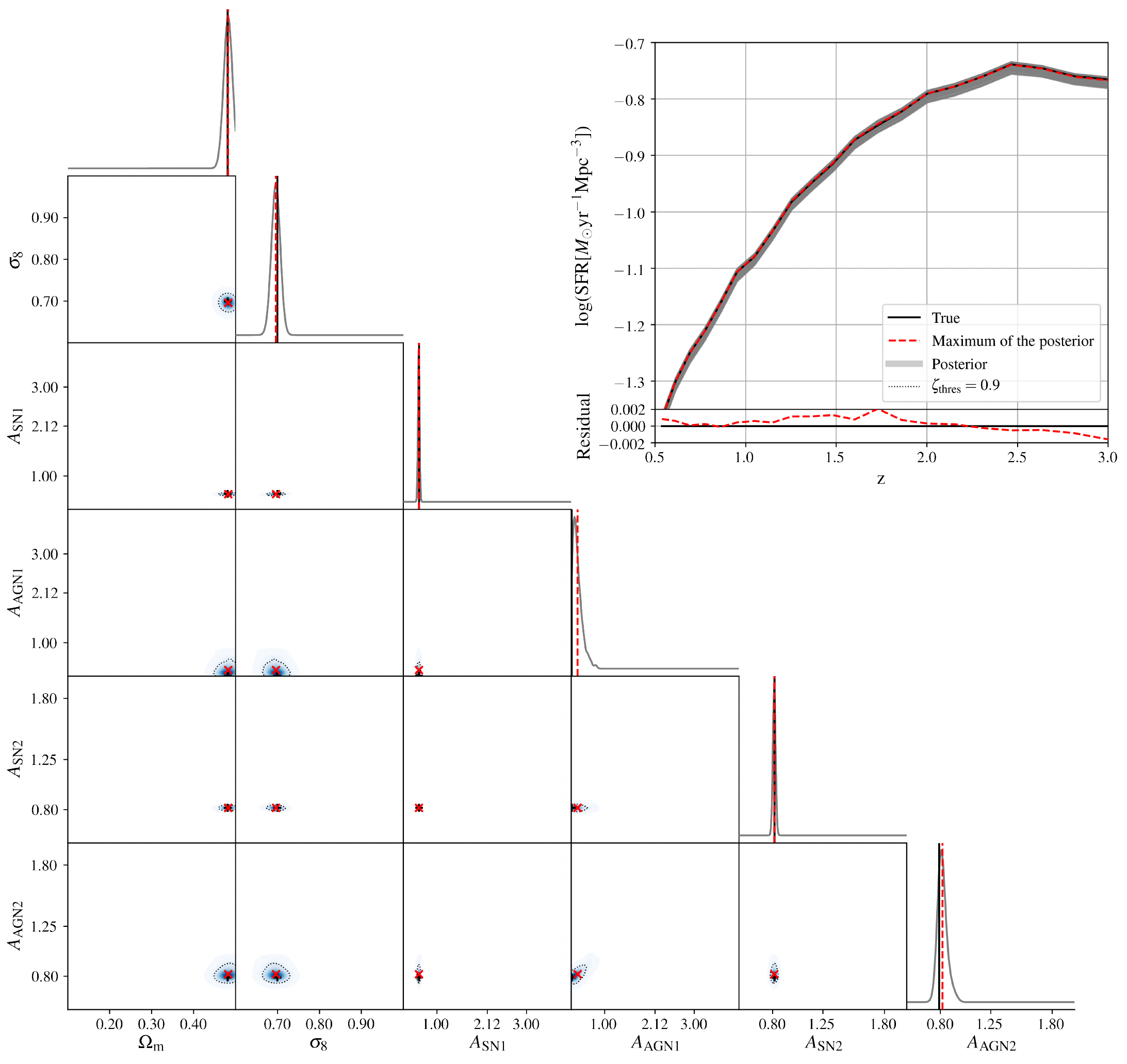}
    \caption{
    {\it Top right}: {\it Emulated} cosmic star formation rate densities (SFRDs) from the inference without the mock uncertainty. Shown are the inferred posterior ({\it grey}), the maximum of the posterior ({\it red dashed}), and the emulator-based target SFRD ({\it black solid}). 
    The emulator-based target SFRD is generated by the emulator taking as input one of the parameter combinations from the LH set (details in Section \ref{sec:performance_sfrd}).
    {\it Bottom left}: Two-dimensional distribution of the inferred posterior. 
    The {\it black} and {\it red} cross-hairs represent the values of target and maximum of the posterior, respectively. 
    The marginal distributions are obtained by kernel density estimation.
    The {\it black solid} and {\it red dashed} vertical lines indicate the true values and the maximum values of the inferred posterior, respectively.
    The inferred parameters ({\it red dashed}) and their true values ({\it black solid}) are nearly on top of each other with small errors and variances (see Table \ref{table:table_sfrd}.A for details).
    The inferred SFRD and the target SFRD are also in a good agreement with a relative error of 0.17\% (see Section \ref{sec:performance_sfrd}).
    }
    \label{fig:emulated_sfrd_no_uncertainty}
\end{figure*}

\begin{figure}[t!]
    \centering
    \includegraphics[width=0.45\textwidth]{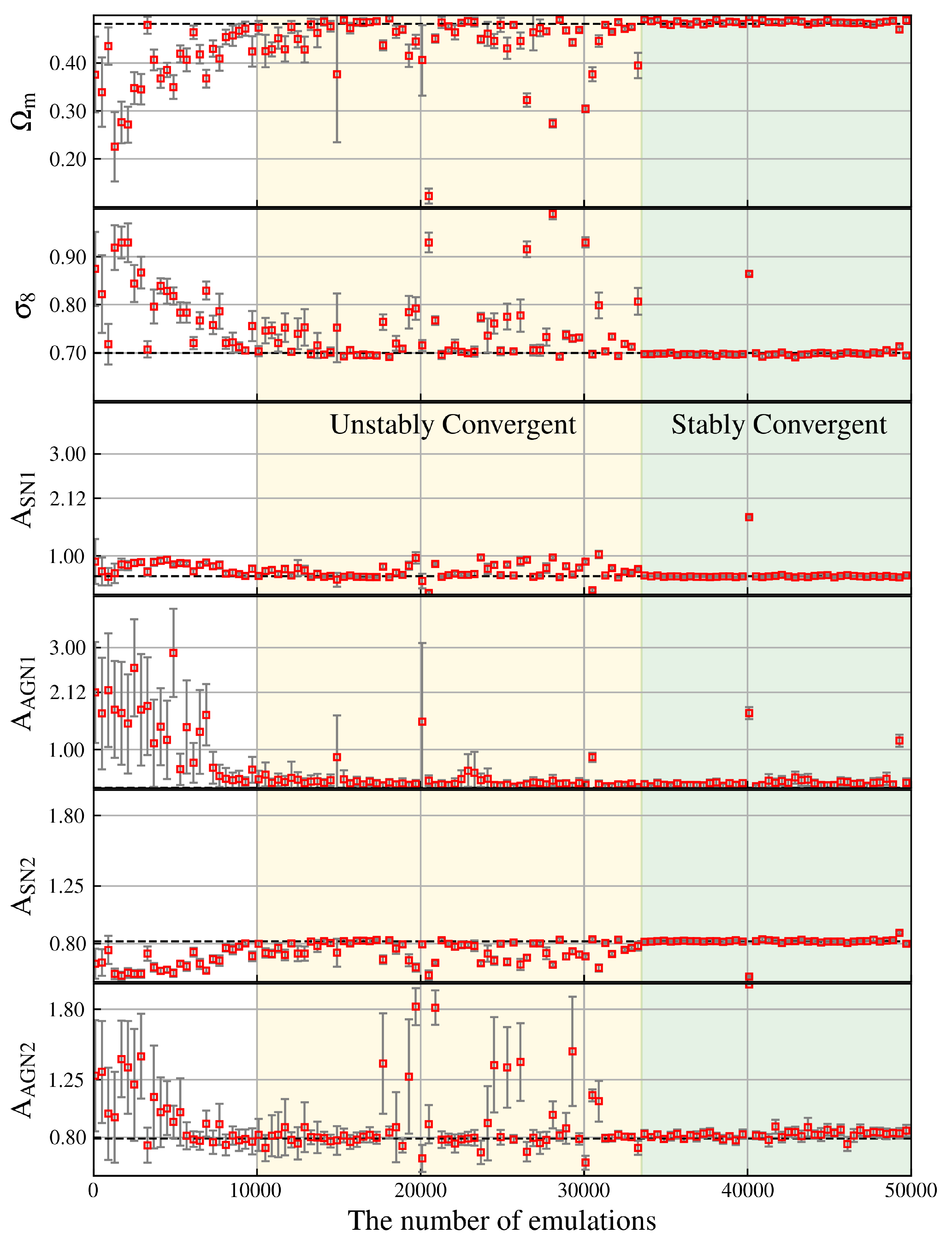}
    \caption{
    The convergence of each parameter as a function of the number of emulations used in the ILI on the emulator-based target SFRD used in Figure \ref{fig:emulated_sfrd_no_uncertainty}. 
    The {\it red squares} and {\it grey error bars} present the maximum and the standard deviation of the posterior density. 
    The {\it black dashed} lines show the true values.
    The convergence of ILI from the SFRD is divided into the unstably convergent stage ({\it yellow region}) and the stably convergent stage ({\it green region}) (see Section \ref{sec:performance_sfrd}).
    In the unstably convergent stage, the maxima of the inferred posteriors jump around from the true values occasionally while the deviation of the corresponding SFRDs from the target is relatively small, which leads to degeneracy in Section \ref{sec:bimodality_sfrd}.
    As the number of emulations for ILI exceeds $\sim 30000$, the inferred posteriors reach the stably convergent stage.
    }
    \label{fig:emulated_sfrd_stability}
\end{figure}

\subsubsection{Performance of Inference}
\label{sec:performance_sfrd}
We first investigate the accuracy of our ILI and how stably the inferred posterior density converges without the mock uncertainty.
The performance of ILI with the mock uncertainty will be discussed in Section \ref{sec:uncertainty_sfrd}.
Using the suite of CAMELS simulations, we train an emulator that takes as input six cosmological and astrophysical parameters $\btheta$ and predicts the SFRD $\bx$ (for details, refer to Section \ref{sec:emulator_as_surrogate}). 
Then, we perform ILI to retrieve the posterior density over six cosmological and astrophysical parameters $\btheta$ given an {\it emulator-based} target SFRD $\bx_0$.
Here, the {\it emulator-based} target SFRD $\bx_0$ is generated by the emulator with $\btheta_0$ that is one of the data points in the LH set.

Shown in Figure \ref{fig:emulated_sfrd_no_uncertainty} are two-dimensional projections of the inferred posterior ({\it bottom left}) and the SFRD plot ({\it top right}) that includes the corresponding SFRDs drawn from the posterior as well as the emulator-based target SFRD. 
In this example, a total of 40000 emulations is used to retrieve the posterior density $p(\btheta|\bx_0)$ ({\it grey}) given the SFRD $\bx_0$ ({\it black solid}).
The location of the maximum of the posterior density\footnote{The set of parameters, giving the maximum value of the posterior density, is drawn from the MCMC samples. i.e. $\btheta_{\rm max} = \argmax_{\btheta}p(\btheta|\bx_0)$.\label{footnote:maximum_of_posterior}} ({\it red dashed} or {\it red cross-hair}) matches the true values ({\it black solid} or {\it black cross-hair}) almost perfectly for all six parameters with small relative errors (see Table \ref{table:table_sfrd} for the values of the relative errors, standard deviations, and coefficients of variation\footnote{The coefficient of variation, also known as relative standard deviation, is a standardized measure of dispersion of a probability distribution or frequency distribution.
It is usually defined as $c_{\rm v} = \frac{\sigma}{\mu}\times100$, but in this work we adopt a definition using the maximum of posterior $\theta_{\rm max}$ instead of the mean $\mu$, namely $c_{\rm v} = \frac{\sigma}{\theta_{\rm max}}\times100$.
\label{footnote:coefficient_of_variantion}}).
The values of the standard deviations imply that the inferred posterior density has very small variances, in that compared to the volume of parameter space covered by the prior of $\mathcal{O}(10)$, the volume of the posterior density approximates to $\mathcal{O}(10^{-12})$.
In the {\it top right} panel, the SFRD from the maximum of the posterior ({\it red dashed}) coincides with the target SFRD ({\it black line}) with a relative error of 0.17\%.
The SFRDs from the full posterior ({\it grey region}) have an exceedingly narrow distribution with the standard deviation of 0.003 dex.

Figure \ref{fig:emulated_sfrd_stability} illustrates the convergence of the ILI on the SFRD and its stability.
Each panel shows the maximum of the posterior ({\it red squares}) and standard deviations ({\it grey error bars}) of each parameter as a function of the (cumulative) number of emulations used for training so far. 
The panel for $\asnone$ ({\it third} row) shows that the $\asnone$ parameter converges to the truth almost right after the beginning. 
The rest of the parameters ($\om, \sig, \aagnone, \asntwo, \aagntwo$) can seemingly come to convergence after 10000 emulations. 
However, the convergence can be divided into two different stages: the ``unstably convergent stage'' in $[10000, 33500]$ emulations ({\it yellow region}) and the ``stably convergent stage'' in $[33500, 50000]$ emulations ({\it green region}).

In the unstably convergent stage ({\it yellow region}), the inferred parameters jump around occasionally without a particular period (unstably) but rapidly return to the truth in the next iteration (convergent).
The average relative errors\footnote{
The average relative error of a parameter $\theta$ is defined as $\bar{\delta}_{\theta} \equiv \frac{1}{N}\sum_{n} \left(\theta_{\rm truth}-\argmax_{\theta}{p^{(n)}(\theta|\bx_0)}\right)/\theta_{\rm truth}\times 100$ where $n$ enumerates over a range of emulations in the ILI process (e.g.~in Figure \ref{fig:emulated_sfrd_stability}) and $N = \sum_{n}1$. $\bx_0$ is the target observation.
\label{footnote:averaged_relative_error}
}, standard deviations, and the coefficients of variation over $[10000, 33500]$ emulations can be found in Table \ref{table:table_sfrd} ({\it first}, {\it third}, and {\it fifth} rows).
All three measures above are an order of magnitude greater than that of the converged posterior in Figure \ref{fig:emulated_sfrd_no_uncertainty}.
However, the mean of the relative error of all the SFRDs including not only the maximum of posteriors but also the posteriors themselves, at the unstably convergent stage, approximates to 0.95\%, which is clearly greater than that of one single convergent SFRD (0.17\%) but the value itself seems acceptable. 
The most interesting feature of the stage is that although the inferred parameters are jumping around, the corresponding SFRDs are relatively well converging to the truth.
This is attributed to the weak correlation between the parameters and the SFRD (see Section \ref{sec:correlation_between_obs_and_params} for a discussion).
More importantly, this implicitly indicates the possibility of a multi-modal distribution that can reproduce the same observable from different sets of the parameters, as further discussed in Section \ref{sec:bimodality_sfrd}.

\newcolumntype{?}{!{\vrule width 1pt}}
\begin{table*}[ht!]
\centering
\caption{
Table for the average relative errors $\bar{\bs{\delta}}$, the standard deviations $\bs{\sigma}$, and the coefficients of variation $\bs{c}_{\rm v}$\footref{footnote:coefficient_of_variantion} for each inference in Section \ref{sec:emulated_sfrd}.
This table includes the values of the well-converged posterior ({Cnvg.}, see Figure \ref{fig:emulated_sfrd_no_uncertainty}), tests of convergence and stability of the ILI showing unstably convergent and stably convergent stages ({U.C.} and {S.C.}, see Figure \ref{fig:emulated_sfrd_stability}), the bimodal posterior ({Bmd.}, see Figure \ref{fig:emulated_sfrd_bimodality}), and the posterior inferred with the mock uncertainty ({Uncrt.}, see Figure \ref{fig:emulated_sfrd_uncertainty}). 
Since the bimodal posterior has two peaks, the corresponding columns (Bmd.) include two values.
}
\begin{tabular}{?c?c|c|c|c|c?c|c|c|c|c?c|c|c|c|c?}
\thickhline
&\multicolumn{5}{c?}{$\bar{\bs{\delta}}$ (\%)}&\multicolumn{5}{c?}{$\bs{\sigma}$}&\multicolumn{5}{c?}{$\bs{c}_{\rm v}$\footref{footnote:coefficient_of_variantion}(\%)}\\ 
\thickhline
&Fig.\ref{fig:emulated_sfrd_no_uncertainty}&\multicolumn{2}{c|}{Fig.\ref{fig:emulated_sfrd_stability}}&Fig.\ref{fig:emulated_sfrd_bimodality}&Fig.\ref{fig:emulated_sfrd_uncertainty}&Fig.\ref{fig:emulated_sfrd_no_uncertainty}&\multicolumn{2}{c|}{Fig.\ref{fig:emulated_sfrd_stability}}&Fig.\ref{fig:emulated_sfrd_bimodality}&Fig.\ref{fig:emulated_sfrd_uncertainty}&Fig.\ref{fig:emulated_sfrd_no_uncertainty}&\multicolumn{2}{c|}{Fig.\ref{fig:emulated_sfrd_stability}}&Fig.\ref{fig:emulated_sfrd_bimodality}&Fig.\ref{fig:emulated_sfrd_uncertainty}\\
\cline{2-16}
           &Cnvg.&U.C.&S.C.&Bmd.&Uncrt.&Cnvg.&U.C.&S.C.&Bmd.&Uncrt.&Cnvg.&U.C.&S.C.&Bmd.&Uncrt.\\
\thickhline
\multirow{2}{*}{$\om$}&\multirow{2}{*}{0.02}&\multirow{2}{*}{7.9} &\multirow{2}{*}{0.9} &0.45&\multirow{2}{*}{-}&\multirow{2}{*}{0.004}&\multirow{2}{*}{0.016}&\multirow{2}{*}{0.004}&0.015&\multirow{2}{*}{0.073}&\multirow{2}{*}{0.79}&\multirow{2}{*}{3.2} &\multirow{2}{*}{0.8} &3.1&\multirow{2}{*}{16.0}\\
\cline{5-5}\cline{10-10}\cline{15-15}
&&&&-&&&&&0.018&&&&&4.8&\\
\hline

\multirow{2}{*}{$\sig$     }&\multirow{2}{*}{0.49} &\multirow{2}{*}{6.1 }&\multirow{2}{*}{1.0 }&0.36&\multirow{2}{*}{-}&\multirow{2}{*}{0.003}&\multirow{2}{*}{0.014}&\multirow{2}{*}{0.003}&0.014 &\multirow{2}{*}{0.100}&\multirow{2}{*}{0.42}&\multirow{2}{*}{2.0 }&\multirow{2}{*}{0.4 }&2.0&\multirow{2}{*}{14.7}\\
\cline{5-5}\cline{10-10}\cline{15-15}
&&&&-&&&&&0.025&&&&&3.0&\\
\hline

\multirow{2}{*}{$\asnone$  }&\multirow{2}{*}{1.36} &\multirow{2}{*}{18.8}&\multirow{2}{*}{6.3 }&0.55&\multirow{2}{*}{-}&\multirow{2}{*}{0.014}&\multirow{2}{*}{0.043}&\multirow{2}{*}{0.015}&0.042 &\multirow{2}{*}{0.353} &\multirow{2}{*}{2.33}&\multirow{2}{*}{7.1 }&\multirow{2}{*}{2.5 }&7.0  &\multirow{2}{*}{60.7}\\
\cline{5-5}\cline{10-10}\cline{15-15}
&&&&-&&&&&0.046&&&&&4.9&\\
\hline

\multirow{2}{*}{$\aagnone$ } &\multirow{2}{*}{49.8} &\multirow{2}{*}{52.1}&\multirow{2}{*}{52.1}&30.8&\multirow{2}{*}{-}&\multirow{2}{*}{0.105}&\multirow{2}{*}{0.138}&\multirow{2}{*}{0.072}&0.090 &\multirow{2}{*}{1.158}&\multirow{2}{*}{40.5}&\multirow{2}{*}{53.0}&\multirow{2}{*}{27.5}&26.5 &\multirow{2}{*}{450.6}\\
\cline{5-5}\cline{10-10}\cline{15-15}
&&&&-&&&&&0.039&&&&&13.1&\\
\hline

\multirow{2}{*}{$\asntwo$ }&\multirow{2}{*}{0.25} &\multirow{2}{*}{8.5 }&\multirow{2}{*}{1.6 }&0.41&\multirow{2}{*}{-}&\multirow{2}{*}{0.007}&\multirow{2}{*}{0.026}&\multirow{2}{*}{0.009}&0.032&\multirow{2}{*}{0.145}&\multirow{2}{*}{0.76}&\multirow{2}{*}{3.2 }&\multirow{2}{*}{1.1 }&3.9   &\multirow{2}{*}{17.8}\\
\cline{5-5}\cline{10-10}\cline{15-15}
&&&&-&&&&&0.020&&&&&3.4&\\
\hline

\multirow{2}{*}{$\aagntwo$ }&\multirow{2}{*}{3.65} &\multirow{2}{*}{17.8}&\multirow{2}{*}{8.2 }&5.7&\multirow{2}{*}{-}&\multirow{2}{*}{0.040}&\multirow{2}{*}{0.117}&\multirow{2}{*}{0.036}&0.087&\multirow{2}{*}{0.358}&\multirow{2}{*}{5.01}&\multirow{2}{*}{14.8}&\multirow{2}{*}{4.6 }&10.5   &\multirow{2}{*}{47.0}\\
\cline{5-5}\cline{10-10}\cline{15-15}
&&&&-&&&&&0.148&&&&&8.1&\\
\thickhline
\end{tabular}
\label{table:table_sfrd}
\end{table*}

On the other hand, the posterior distributions are stably convergent with relatively small variances after $\sim 30000$ emulations ({\it green region}). 
All six maxima ({\it red squares}) for the six parameters stably converge to the truth ({\it black dotted line}) with the average relative errors\footref{footnote:averaged_relative_error}
and their standard deviations
over $[33500, 50000]$ emulations.
The variances of the posterior density ({\it grey error bars}) are convergent as well (refer to the {\it S.C.} columns of Table \ref{table:table_sfrd}).
These values are an order of magnitude lower than the values at the unstably convergent stage, which can be a clear sign of transition.
Also, in comparison to the values for the posterior in Figure \ref{fig:emulated_sfrd_no_uncertainty}, the average relative errors of stably convergent posteriors are somewhat greater but the standard deviations and coefficients of variation are comparable.
That is, the peaks or the maxima of the posteriors have appreciable scatter compared to the truth at the stably convergent stage, whereas the widths of the posteriors are consistent.
As for the variances, the convergence properties depend not only on the number of emulations but also differ by parameter. 
In the convergence of $\aagnone$, there is no evident transition from unstably convergent to stably convergent in terms of the relative errors, standard deviations, or the coefficients of variation. 
On the other hand, the stellar feedback parameters ($\asnone$ and $\asntwo$) converge rapidly as soon as the training begins, leading to seamlessly smooth transition to the unstably convergent stage.
However, the transition to the stably convergent stage takes place drastically both visually and quantitatively.
In the meantime, the AGN feedback parameters ($\aagnone$ and $\aagntwo$) have notable scatter even in the stably convergent stage ({\it green region}) compared to the other four parameters.
Hence, we can conclude that the strengths of the correlations between each parameter and the SFRD are in the following order: $\asnone \simeq \asntwo > \om \simeq \sig > \aagnone \simeq \aagntwo$.
An indication for such correlations can be also found in \citet[Figure 11 and Figure 12]{camels2021ApJ...915...71V}.

Thus far, we have not implemented any uncertainties in the inferences. 
Therefore, assuming that the emulators are injective (one-to-one functions), a negligible variance is expected for all six parameters;
i.e. $p(\bx, \btheta) = \delta(\bx - f(\btheta))$ and $\btheta$ is unique.
However, a tiny amount of variance exists in both the posterior and the inferred SFRDs ({\it grey} in Figure \ref{fig:emulated_sfrd_no_uncertainty}).
This can be attributed to (1) physical degeneracy, which is discussed in Section \ref{sec:bimodality_sfrd}, and (2) inaccuracy of the NDE. 
In both Figures \ref{fig:emulated_sfrd_no_uncertainty} and \ref{fig:emulated_sfrd_stability}, $\Omega_{\rm m}$, $\sigma_8$, $A_{\rm SN1}$, and $A_{\rm SN2}$ show high convergence and precision for both maxima and variances whereas $\aagnone$ and $\aagntwo$ have larger variances on average.
The magnitude of the variance indicates how strongly the observable can constrain each parameters,
or how intimately each parameter and observable correlate with each other. 
The greater the variance, the weaker the correlation is.
Due to the weak correlations, $\aagnone$ and $\aagntwo$ require more simulations to converge stably and tend to have a larger variance than other parameters.
Nevertheless, the relative errors of both parameters and SFRDs are less than 1\% on average with a total of 34000 emulations.

\begin{figure*}[t!]
    \centering
    \includegraphics[width=\textwidth]{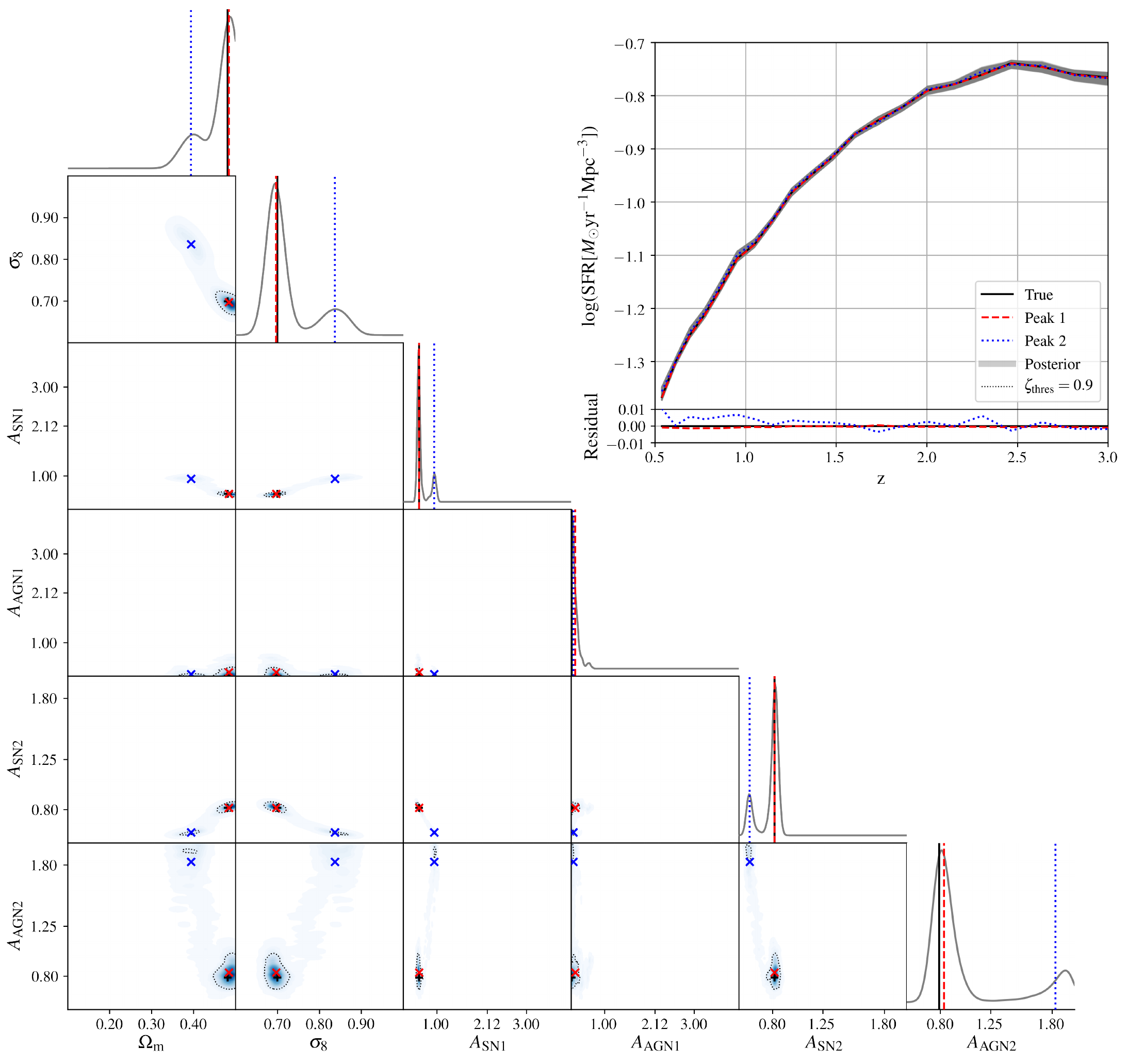}
    \caption{
    {\it Top right}: {\it Emulated} star formation rate densities (SFRD) from the inferred posterior ({\it grey}), two peaks of the posterior ({\it red dashed} and {\it blue dotted}), and the emulator-based target SFRD used in Figure \ref{fig:emulated_sfrd_no_uncertainty} ({\it black solid}). 
    {\it Bottom left}: 2D contour projections of the inferred posterior. 
    The inferred posterior contains two strong peaks, both of which reproduce the SFRD well, with relative errors of 0.35\% ({\it red dashed}) and 0.98\% ({\it blue dotted}).
    This indicates a degeneracy in the SFRD (see Section \ref{sec:bimodality_sfrd}).
    }
    \label{fig:emulated_sfrd_bimodality}
\end{figure*}

\begin{figure*}[t!]
    \centering
    \includegraphics[width=\textwidth]{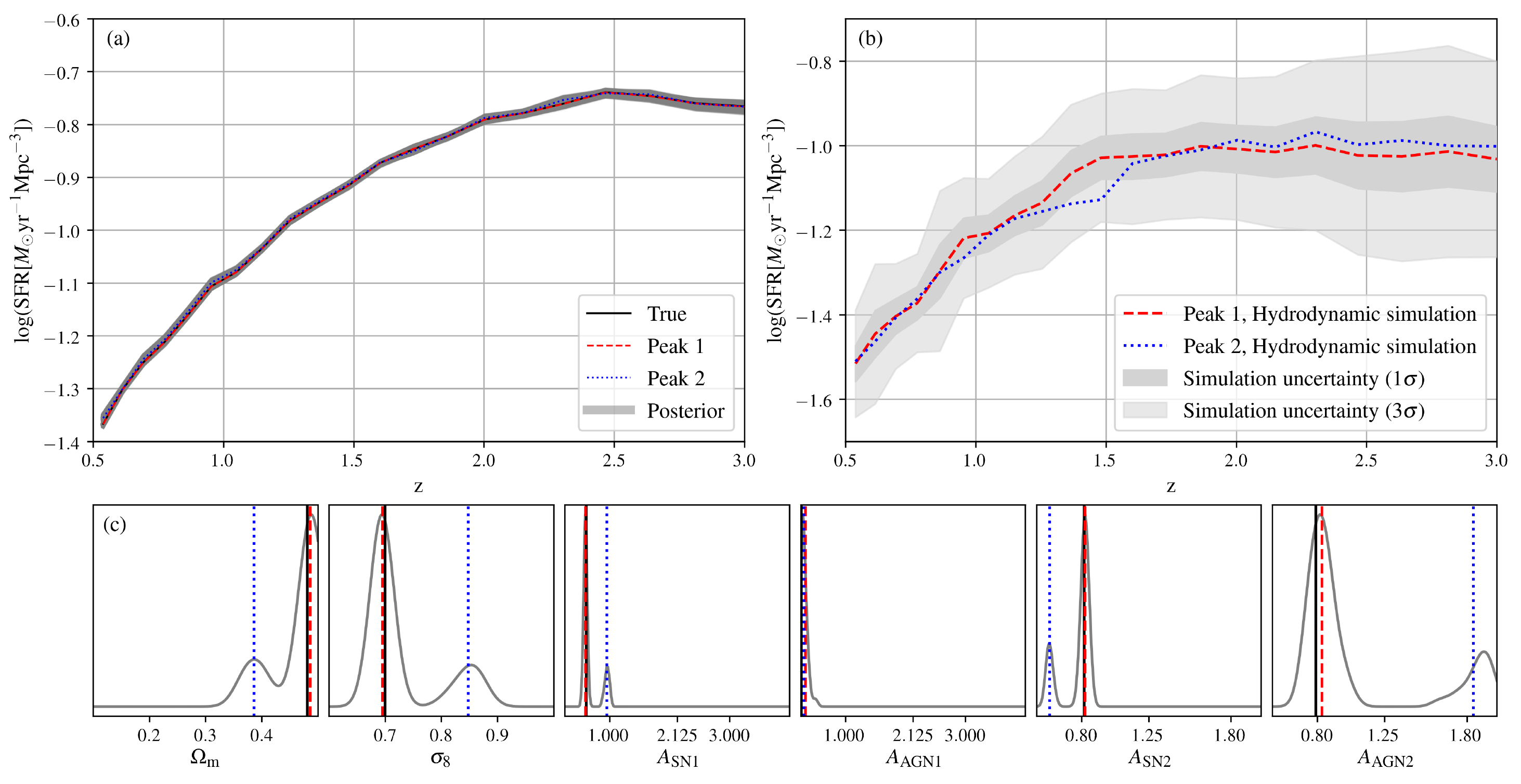}
    \caption{
    (a) Cosmic star formation rate density from the emulator, representing the posterior and its two peaks; 
    (b) Cosmic star formation rate density from cosmological simulations using the points in parameter space that are the peaks of the posterior;
    (c) One-dimensional projections of the inferred posterior ({\it grey}) based on Figure \ref{fig:emulated_sfrd_bimodality}.
    The two peaks are drawn with the {\it red dashed} and {\it blue dotted} lines consistently across Figure \ref{fig:emulated_sfrd_bimodality} and panels (c).
    However, the marginals of the {\it degenerate} posterior distribution in panel (c) are reconstructed with a modified probability density, namely that which is limited to only the region within $\zeta_{\rm degen}=0.9$ in Figure \ref{fig:emulated_sfrd_bimodality}.
    The {\it grey} and {\it light grey} regions in panel (b) indicate the $\sigma$ and $3\sigma$ confidence regions of the simulation uncertainty.
    Given that both simulated SFRDs from the two peaks lie within the $3\sigma$ region of the simulation uncertainty, we conclude that these cosmological simulations themselves are degenerate in terms of the SFRD (see Section \ref{sec:bimodality_sfrd}).
    }
    \label{fig:sfrd_degeneracy}
\end{figure*}

\subsubsection{Bimodality in the Posterior Distribution}
\label{sec:bimodality_sfrd}
In this section, we present one of the bimodal posterior distributions that can be found in the unstably convergent stage ({\it yellow region}) in Figure \ref{fig:emulated_sfrd_stability}.
Figure \ref{fig:emulated_sfrd_bimodality} illustrates two-dimensional projections of the inferred posterior ({\it bottom left}) as well as the corresponding SFRDs from the posterior and the ({\it emulator-based}) target SFRD ({\it top right}).
The most intriguing feature in Figure \ref{fig:emulated_sfrd_bimodality} is the {\it bimodal peaks} ({\it red} and {\it blue}) in the posterior density.
Not only do the two peaks exist in the posterior density, but they also reproduce the target observations within a barely appreciable margin of error.

We use $k$-means clustering to divide the posterior density into two unimodal distributions.
The $k$-means clustering partitions samples into $k$ clusters in which each sample belongs to the cluster with the nearest mean. 
Each peak is defined by the center of each cluster.
The relative errors of the two SFRDs with respect to the true SFRD are 0.35\% for peak 1 and 0.98\%  for peak 2.
Peak 2 has slightly larger errors than peak 1 but still less than 1\%.
Furthermore, peak 1 ({\it red dashed}) accurately coincides with the truth ({\it black solid}) with a small relative errors (see Table \ref{table:table_sfrd}).
We measure the standard deviations and coefficients of variation with respect to each peak by using the result of the $k$-means clustering (see Table \ref{table:table_sfrd}).
There is no marked difference between the variances of peak 1 and peak 2.
The level of deviation is slightly higher than that of the stably convergent posterior in Figure \ref{fig:emulated_sfrd_no_uncertainty} but similar to the unstably convergent ones shown in Figure \ref{fig:emulated_sfrd_stability}.

Having two strong peaks in the posterior distribution is the result of a physical {\it degeneracy}, namely a situation where more than a single set of parameters reproduces the same observable. 
To study this, we propose a definition of degeneracy in a mathematically consistent way using a given posterior distribution.
The set of degenerate points in parameter space, $\Theta_{\rm degen}$, is defined such that it satisfies $\int_{\btheta \in \Theta_{\rm degen}}p(\btheta)\mathrm{d}\btheta=\zeta_{\rm thres}$ where $p(\btheta \in \Theta_{\rm degen}) > p(\btheta \notin \Theta_{\rm degen})$, assuming that the inferred posterior distribution $p(\btheta)$ is normalized.
Here, $\zeta_{\rm thres}$ is a free parameter, and the degenerate-parameter set $\Theta_{\rm degen}$ collects parameters according to their probability density in a descending order until the integration of the probability over the degenerate set becomes equal to $\zeta_{\rm degen}$ (refer to Appendix \ref{apx:definition_degeneracy} for a precise and detailed definition). 
In this work, we set $\zeta_{\rm degen}$ to $0.9$.
Also, every two-dimensional projection of posterior distributions in this paper includes a contour line for $\zeta_{\rm degen}=0.9$.

Shown in panel (c) of Figure \ref{fig:sfrd_degeneracy} are the marginals of the {\it degenerate} posterior distribution that is reconstructed with the probability density only enclosed within $\zeta_{\rm degen}=0.9$ through the Gaussian kernel density estimation. 
On the other hand, two peaks ({\it red dashed} and {\it blue dotted}) are identical to the ones in Figure \ref{fig:emulated_sfrd_bimodality} that illustrates details of the posterior distribution with $\zeta_{\rm thres}=0.9$ ({\it black dotted contour}) in the two-dimensional projection plot ({\it bottom left}).
The two emulated SFRDs that correspond to the two peaks of the posterior distribution ({\it red dashed} and {\it blue dotted}) in panel (a) are approximately on top of each other. 
The distribution of SFRDs ({\it grey}) is also sufficiently concentrated with a variance of 0.003 dex that is similar to that of the SFRDs from the stably convergent posterior in Figure \ref{fig:emulated_sfrd_no_uncertainty}.
Hence, we can conclude that the two peaks are degenerate in terms of the emulated SFRDs.

However, the presence of degeneracy in the emulated SFRD does not necessarily demand that a degeneracy exists in the actual cosmological simulations as well.
Therefore, we test this with new simualtions.
Panel (b) in Figure \ref{fig:sfrd_degeneracy} shows the {\it simulated} SFRDs performed with each set of parameters from the two peaks ({\it red dashed} and {\it blue dotted}).
Here, we investigate whether the two sets of parameters are also degenerate in the simulations given the simulation uncertainty (see Section \ref{sec:cosmic_variance_buttefly_effect}).
We use $\pm 1\sigma$ ({\it grey}) and $\pm 3\sigma$ ({\it light grey}) regions that correspond to 68.1\% and 99.7\% confidence levels assuming that the simulation uncertainty follows the Gaussian distribution, respectively.
The standard deviations of the Gaussian distribution are directly calculated from the suite of simulations in the CV set along redshift (refer to Appendix \ref{apx:cosmicvariance_butterflyeffect}).
In panel (b), the simulated SFRDs from the parameter combinations corresponding to the degenerate peaks are consistent within $1\sigma$.
Thus, the two simulated SFRDs are highly likely to be degenerate, sharing the same parameters, while having small discrepancies originating from the simulation uncertainty.

Above all, the emulator result demonstrates the clear signs of degeneracy in the SFRD having two approximately identical SFRDs from two different sets of parameters.
On the simulation side, cosmological simulations with the two sets of parameters also produce two SFRDs that are close to each other, though simulation uncertainty leads to small but notable differences in the simulated SFRDs, which are nevertheless not statistically significant. 
The SFRD lies approximately inside the 64.8\% ($1\sigma$) confidence regions and is completely enclosed within the 99.7\% ($3\sigma$) confidence regions of the simulation uncertainty.
This is strong evidence that both emulator and cosmological simulations have degeneracies in the SFRD, in spite of the emulator not being a perfect representation of the simulations.

\begin{figure*}[th!]
    \centering
    \includegraphics[width=\textwidth]{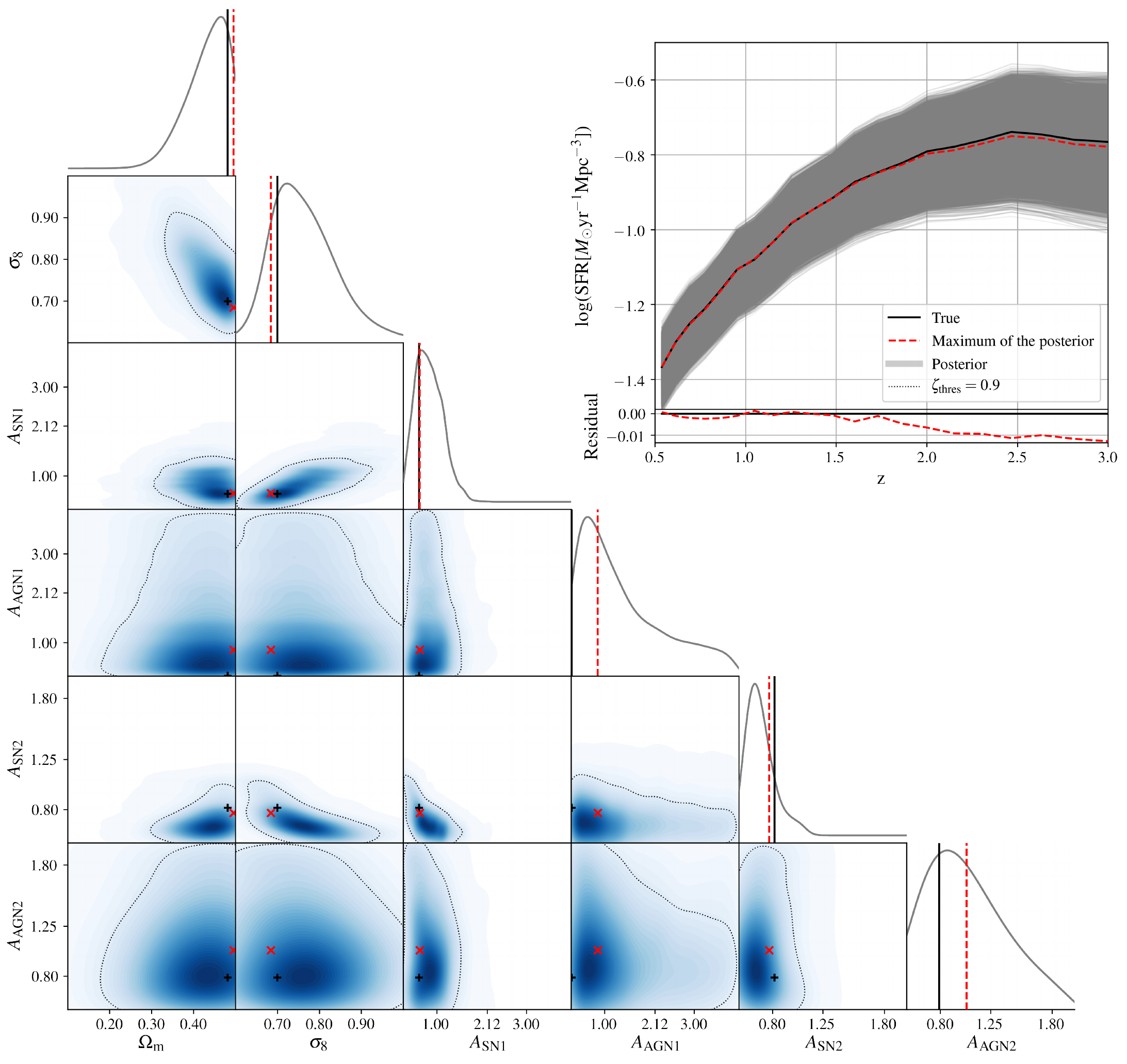}
    \caption{{\it Bottom left}: 2D projections of the posteriors ({\it blue}) inferred from the emulator-based target SFRD used in \ref{fig:emulated_sfrd_no_uncertainty} with the mock uncertainty. {\it Top right}: Cosmic star formation rate density from the inferred parameters ({\it grey}) and the truth ({\it black solid}). The {\it red} vertical lines indicate the maximum values of the inferred posterior. 
    The mock uncertainty in ILI leads to large variances in the inferred marginal distributions compared to the sharp marginal distributions of Figure \ref{fig:emulated_sfrd_no_uncertainty} inferred without the mock uncertainty (see Section \ref{sec:uncertainty_sfrd} and for a quantitative comparison see Table \ref{table:table_sfrd}).
    }
    \label{fig:emulated_sfrd_uncertainty}
\end{figure*}

\subsubsection{Response to Mock Uncertainty}
\label{sec:uncertainty_sfrd}
Thus far, we have performed ILI without including any uncertainty in the forward model, namely the emulator, but such uncertainties do exist in full cosmological simulations, as discussed in Section \ref{sec:cosmic_variance_buttefly_effect}. 
In this section, we include the {\it simulation uncertainty} that originates from various sources of randomness in the cosmological simulations using the mock uncertainty on top of the emulators (refer to Section \ref{sec:connection} and Appendix \ref{apx:cosmicvariance_butterflyeffect} for a technical description of the mock uncertainty).

Shown in Figure \ref{fig:emulated_sfrd_uncertainty} are two-dimensional projections of the posterior density inferred from the emulated SFRD ({\it bottom left}) with the mock uncertainty.
The mock uncertainty that we impose is modelled to have the same standard deviation as the simulation uncertainty (see Appendix \ref{apx:cosmicvariance_butterflyeffect}).
In comparison to the inferences without the mock uncertainty, the posterior densities inferred with the uncertainty cover much larger regions of parameter space (see Table \ref{table:table_sfrd}).
In comparison to Figure \ref{fig:emulated_sfrd_bimodality}, the bimodal peaks are merged to a single oval or banana-shape distribution ({\it bottom left}).
Although the deviations of the inferred $\aagnone$ and $\aagntwo$ become notable, the relative error of SFRD is \rv{1.0\%, and the deviation is only 0.0044 dex}.
The variance of the inferred SFRDs (0.057 dex) is comparable to the variance of the mock uncertainty (0.061 dex).
This demonstrates that the inferred posterior densities and the corresponding observables successfully reproduce the mock uncertainty in terms of the variances.
Thus, concerning the uncertainty propagation from observation to parameters, the variance of the inferred parameters can be reliable.
On the other hand, the inclusion of the mock uncertainty has led to an increase in the standard deviation of the posterior as well as the relative errors of the inferred observables to the truth.
However, given the size of standard deviations, the inferred parameters are still accurate.
We can relate it to the stochasticity of sampling of {\it mock uncertainty}.
In every iteration of the ILI, additional training data $(\bx, \btheta)$ is generated from the proposal density.
Here, the observable $\bx$ is emulated as a function of the sampled parameters together with the mock uncertainty $Z(\eta)$. 
Due to the finite size of sampling, the mean of observables $\left<\bx+Z(\eta)\right>_{\rm samples}$ cannot be the same as the ideal (theoretical) mean $\left<\bx+Z(\eta)\right>=\left<\bx\right>$, leading to a bias in the sampled data.
Notice that for an infinite number of samples (ideal case), $\left<Z(\eta)\right>=0$ since $Z(\eta)$ is the Gaussian noise that has a mean of zero.
The bias in the newly generated training data is highly likely to result in the inaccuracy of the inferred posterior density.

\begin{figure*}[t!]
    \centering
    \includegraphics[width=\textwidth]{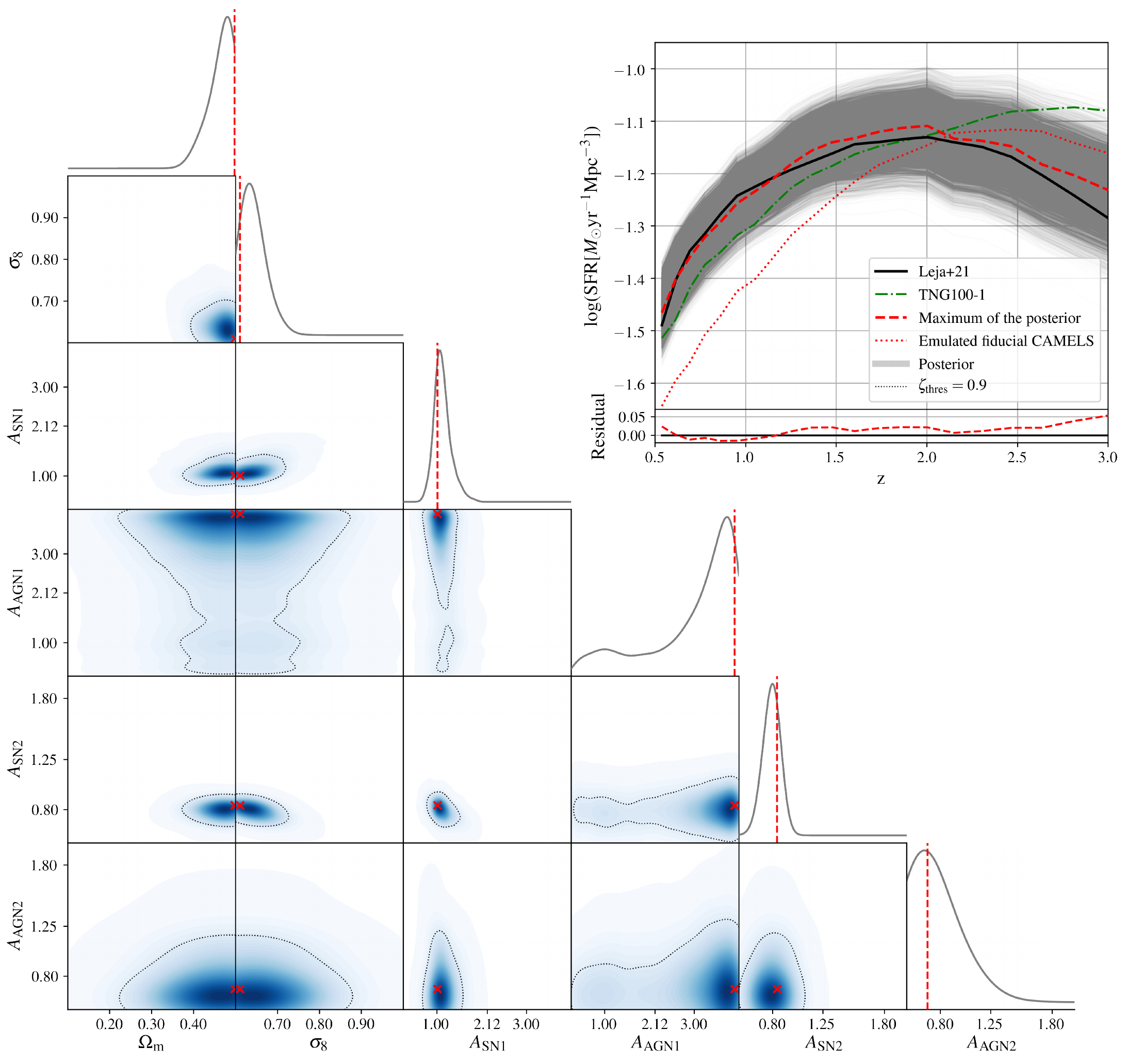}
    \caption{
    {\it Top right}: The inferred cosmic star formation rate densities (SFRDs) ({\it red dashed}) and the observationally-driven target SFRD \citep[]{msLeja2021arXiv211004314L} ({\it black solid}). 
    {\it Bottom left}: 2D projections of the inferred posteriors ({\it grey}) from the observationally-driven target SFRD. 
    The {\it red dashed} vertical lines indicate the maximum values of the inferred posterior in the six dimensional parameter space. 
    For comparison, the SFRD of the {\sc TNG100-1} simulation and the emulated fiducial SFRD of the CAMELS simulations are drawn ({\it green dotdashed} and {\it red dotted}, respectively).
    The inferred SFRD matches the observed SFRD with a relative error of 4.1\% and a deviation of 0.04 dex, whereas the deviation of the fiducial CAMELS from the observed SFRD is $\sim 0.15$ dex (see Section \ref{sec:observation_sfrd}).
    The inferred cosmological parameters are quite off from the common sense due to e.g., resolution effect and degeneracies with the astrophysical parameters  (see Section \ref{sec:discussion_physical_limits})
    }
    \label{fig:observed_sfrd}
\end{figure*}
\subsection{Inference from the Observed Star Formation Rate Density}
\label{sec:observation_sfrd}
We now apply our framework to actual observational data. 
Here, we perform ILI from the observationally-driven SFRD of \citet{msLeja2021arXiv211004314L} with an identical setup to the previous section that includes the mock uncertainty.
Figure \ref{fig:observed_sfrd} illustrates the two dimensional projections of the posterior density inferred from the observed SFRD. 
Note that we have not included the uncertainty of the observation data (see Section \ref{sec:method_sfrd_smf}); rather, only the mock uncertainty is adopted.
The observed SFRD ({\it black solid}) lies completely within the region of the inferred SFRDs ({\it grey}), and also the SFRD from the maximum of the posterior ({\it red dashed}) matches it with a relative error of 4.1\%.
The standard deviations and the coefficients of variation are (0.040, 0.039, 0.224, 1.475, 0.078, 0.181) and (8.3, 5.6, 37.1, 566.9, 9.7, 23.0)\%, respectively (for comparison, see Table \ref{table:table_sfrd}).

This can be thought of as a successful inference given that the emulated fiducial SFRD in Figure \ref{fig:observed_sfrd} ({\it red dotted}) shows entirely different trends from the target observation over all redshifts, with an average deviation of $\sim 0.1$ dex \rv{and a relative error of 22.8\%}.
Here, the emulated fiducial SFRD ({\it green dotdashed}) is an SFRD generated by the emulator from the fiducial parameters. 
Despite the huge discrepancy between the fiducial SFRD and the target observation, the inferred SFRD follows the observation relatively well with a mean deviation less than 0.02 dex \rv{(relative error of 4.1\%)} and even matching the observed peak precisely.
\rv{That being said, not only the inferred parameters ({\it red cross-hairs}) have significant discrepancy with the standard values (e.g., standard cosmology), but also the posterior distribution itself hardly includes the standard values}. 
\rv{This inconsistency mostly stems from} several reasons such as resolution effects and degeneracy between cosmological and astrophysical parameters, \rv{but not from inference procedure} (see Section \ref{sec:discussion_physical_limits} \rv{for details}).
\rv{In short, the fiducial astrophysical parameters of the CAMELS simulations were obtained by calibrating the TNG100-1 simulation to the observations. 
However, the fiducial CAMELS simulation and TNG100-1 simulation show non-negligible discrepancy due to the resolution effect and simulation box size (Section \ref{sec:discussion_resolution_effect}).
This would inevitably lead to a discrepancy between the inferred parameters and the standard values even in the hypothetical case of inference from the same observations against which the TNG100-1 was calibrated. In addition, our target SFRD and SMFs \citep{lejasmf2020ApJ...893..111L,msLeja2021arXiv211004314L} are different from those used for calibrating TNG100-1.
Lastly, inconsistency of cosmological parameters can be attributed to the compensatory reaction to lack of some astrophysics (e.g., Section \ref{sec:observation_individual_smf} and \ref{sec:mismatch_SMFs}).
}

In the case of inference from observations, the intrinsic limit of the emulators should also be accounted for.
The most significant difference between the inferences from the emulated SFRD and observed SFRD is that any emulated SFRD can be predicted precisely by the emulator for sure, whereas we cannot ensure whether there exists a data point in parameter space that can reproduce the observation perfectly.
We can define the image of the emulator as the set of all SFRDs and SMFs it can produce (and similarly for simulation predictions).
The image of an emulator might not coincide with the set of all of the physically possible SFRDs, including observed SFRDs: the codomain of the SFRDs.
In other words, the emulators and simulations might not be able to reproduce any possible universe.
The necessary condition for the successful precision inference is that the observed SFRD be a member of the image of the emulator.
Concerning the limit of the emulator and the simulations that the emulator is trained on, there are fundamental issues:
e.g., (1) the limited dimension of parameter space (domain);
(2) the gap between the emulators and the simulations;
(3) the limits of the physical models in the simulations.
The above will be discussed in Section \ref{sec:discussion_physical_limits}.

\begin{figure*}[t!]
    \centering
    \includegraphics[width=\textwidth]{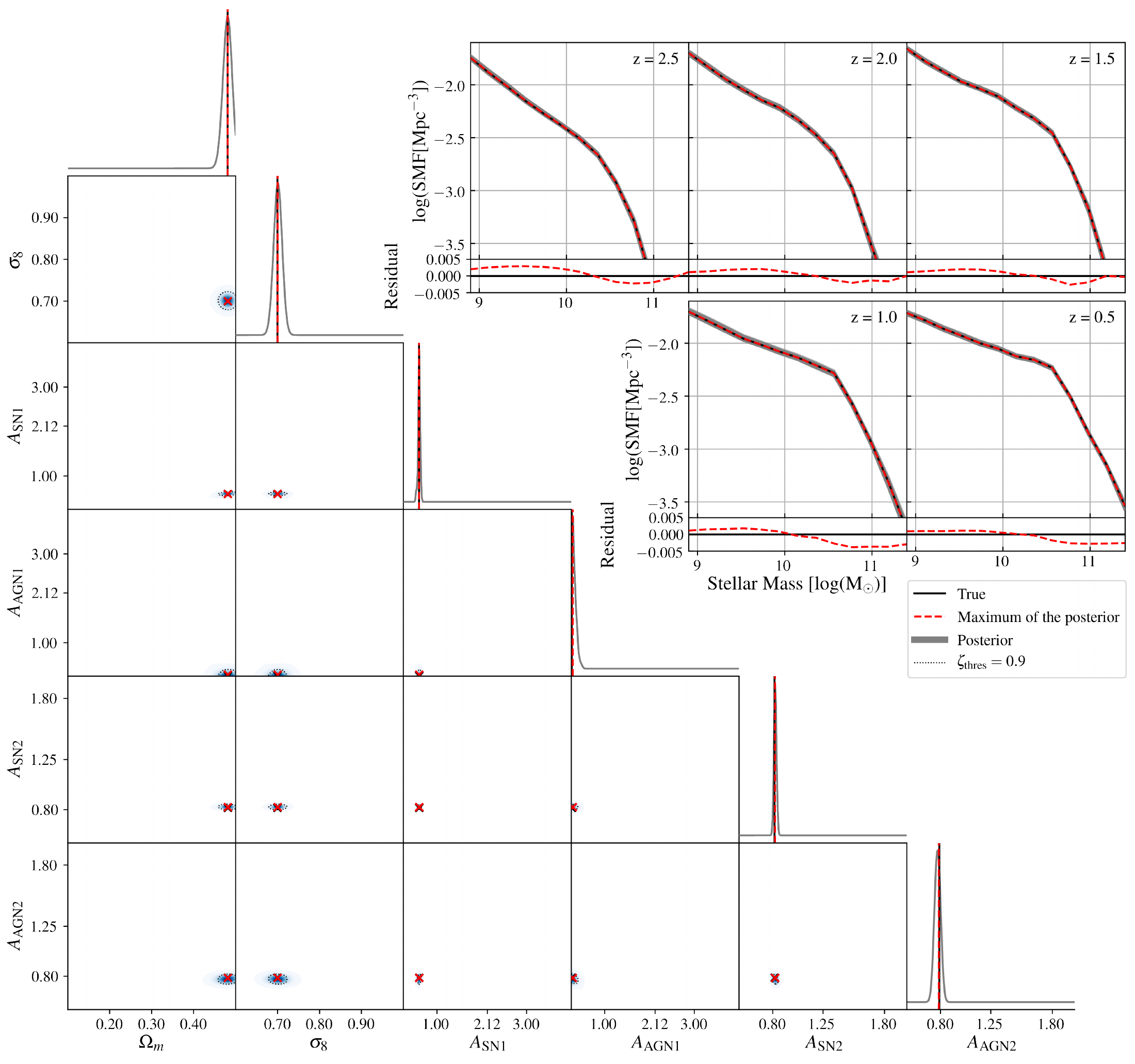}
    \caption{
    {\it Top right}: {\it Emulated} stellar mass functions (SMFs) from five different redshifts from the inferred posterior ({\it grey}), the maximum of the posterior ({\it red dashed}), and the emulator-based target SMFs ({\it black solid}). 
    Here, the emulator-based target SMFs is generated by the emulator with the same set of the parameters that is used in Section \ref{sec:emulated_sfrd} and Figure \ref{fig:emulated_sfrd_no_uncertainty}.
    For comparison, the SMFs of the {\sc TNG100-1} simulation are drawn ({\it green dotdashed}).
    {\it Bottom left}: 2D projections of the inferred posterior. 
    The {\it black} and {\it red} cross-hairs represent the values of target and mean of the posterior, respectively. 
    The {\it black solid} and {\it red dashed} vertical lines indicate the true values and the maximum values of the inferred posterior, respectively.
    The inferred parameters ({\it red dashed}) and true values ({\it black solid}) are nearly on top of each other with small errors and variances (see Table \ref{table:table_smf} for details).
    The inferred SMF and the target SMF are also in a good agreement with a relative error of 0.4\% (see Section \ref{sec:performance_smf}). 
    }
    \label{fig:emulated_smf_no_uncertainty}
\end{figure*}

\section{Inference From Stellar Mass Functions}
\label{sec:smf}
We now turn to inference on stellar mass functions (SMFs) as the target observable.
We study the dependence of the properties of the ILI and the inferred posteriors on the choice of observable with a comparison to the cosmic star formation rate density (SFRD). 
The latter covers the evolutionary history of the universe ranging from $z=3$ to $z=0.5$ whereas a single SMF contains information of only one iteration.
To be consistent with the SFRD, we concatenate five SMFs at $z = 0.5$, $1.0$, $1.5$, $2.0$, and $2.5$ and each SMF is binned with 13 bins in the mass range $[10^{8.9}, 10^{11.4}]\msun$ (refer to Section \ref{sec:method_sfrd_smf}).
Hereafter, ``SMFs'' denotes the five concatenated SMFs from the five different redshifts throughout the paper unless specified otherwise.
In Section \ref{sec:performance_smf}, we first investigate the performance and convergence of the inference on the emulator-based target SMFs compared to that of the SFRDs.
Then, we perform ILI from observationally-driven five concatenated SMFs (Section \ref{sec:observation_concatenated_smf}). 
Finally, in Section \ref{sec:observation_individual_smf}, we study the ILI from one {\it individual} observationally-driven SMF (at a single redshift) at a time.

\begin{figure}[t!]
    \centering
    \includegraphics[width=0.45\textwidth]{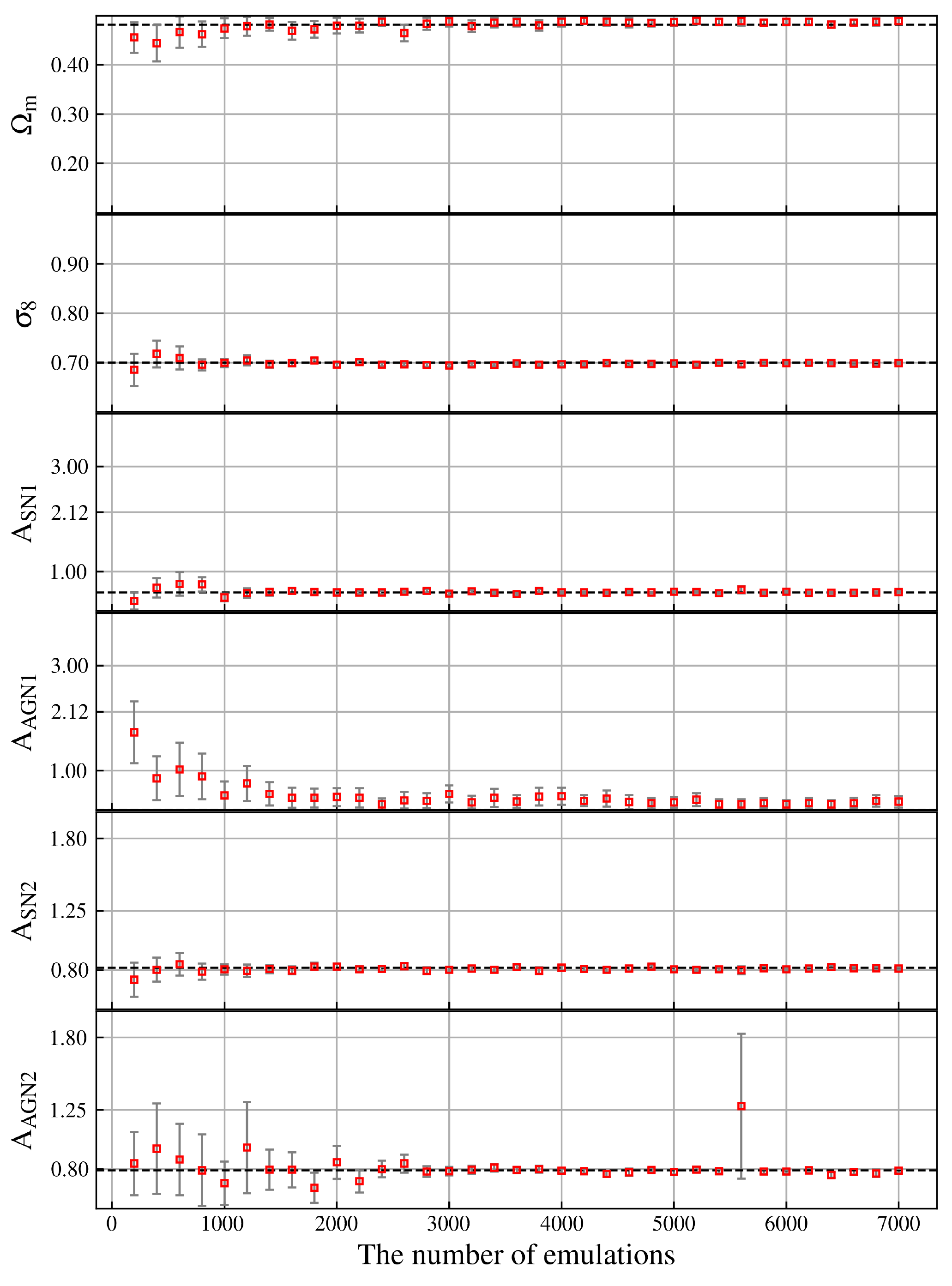}
    \caption{
    The convergence of each parameter as a function of the number of emulations for ILI from emulator-based target stellar mass functions (SMFs). 
    The {\it red square} and {\it grey error bar} present the maximum and the standard deviation of the posterior density. 
    The {\it black dotted} lines show the true values.
    The posteriors inferred from the SMFs rapidly and stably converge (see Section \ref{sec:performance_smf} and Table \ref{table:table_smf}).
    }
    \label{fig:emulated_smf_stability}
\end{figure}

\begin{figure*}[t!]
    \centering
    \includegraphics[width=\textwidth]{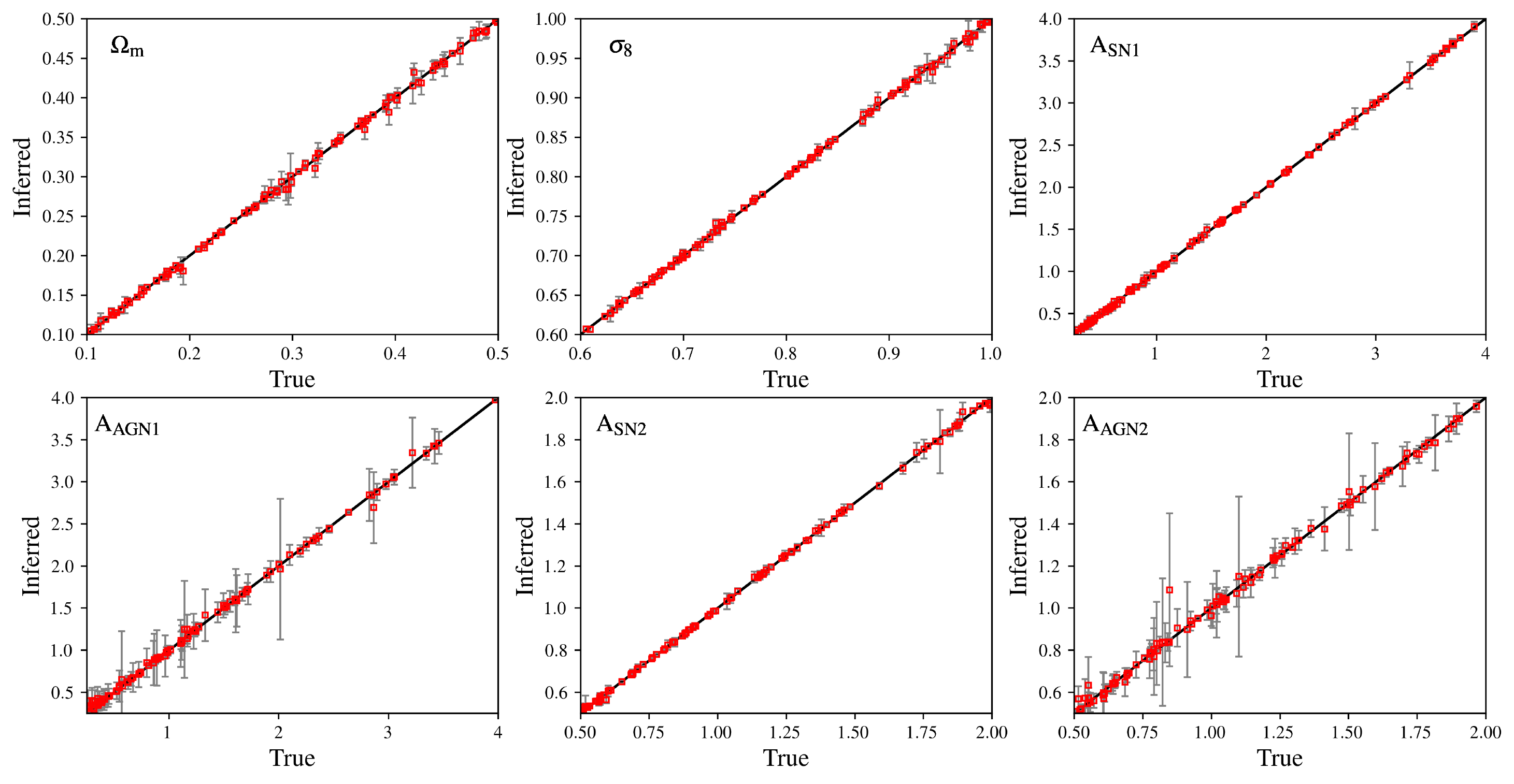}
    \caption{
    Comparison between true values and the corresponding inferred values from posteriors for 100 stellar mass functions generated by the emulator using 100 different combinations of parameters without mock uncertainty.
    Each {\it red square} represents the maximum value of the inferred posterior against its true value. Each {\it grey bar} represents the standard deviation of the inferred posterior.
    The black solid line represents inferred equals true.
    ILI can infer parameters from SMFs accurately with a relative error of $\sim 1\%$ on average across parameter space (see Section \ref{sec:performance_smf}).
    }
    \label{fig:paco_errorbar}
\end{figure*}

\subsection{Inference from Emulated Stellar Mass Functions}
\label{sec:performance_smf}
We first investigate the performance of the ILI and how stably the inferred posterior converges in terms of the SMFs.
As in Section \ref{sec:performance_sfrd}, we train emulators that take as input six cosmological and astrophysical parameters $\btheta$ and predict the SMFs $\bx$ (for details, refer to Section \ref{sec:emulator_as_surrogate}). 
Using the emulator, we perform the ILI to retrieve the posterior density for six cosmological and astrophysical parameters $\btheta$ given the emulator-based target SMFs $\bx_0$.

\begin{table}[t!]
\centering
\caption{
Table for the average relative errors $\bar{\bs{\delta}}$, the standard deviations $\bs{\sigma}$, and the coefficients of variation $\bs{c}_{\rm v}$ for each inference in Section \ref{sec:performance_smf}.
This table includes the values of the well-converged posterior ({Cnvg.}, see Figure \ref{fig:emulated_smf_no_uncertainty}) and Figure \ref{fig:emulated_smf_stability} that tests convergence and stability of ILI (Stb.). 
}
\begin{tabular}{?c?c|c?c|c?c|c?}
\thickhline
&\multicolumn{2}{c?}{$\bar{\bs{\delta}}$ (\%)}&\multicolumn{2}{c?}{$\bs{\sigma}$}&\multicolumn{2}{c?}{$\bs{c}_{\rm v}$(\%)}\\ 
\thickhline
&Fig.\ref{fig:emulated_smf_no_uncertainty}&Fig.\ref{fig:emulated_smf_stability}&Fig.\ref{fig:emulated_smf_no_uncertainty}&Fig.\ref{fig:emulated_smf_stability}&Fig.\ref{fig:emulated_smf_no_uncertainty}&Fig.\ref{fig:emulated_smf_stability}\\
\cline{2-7}
           &Cnvg.&Stb.&Cnvg.&Stb.&Cnvg.&Stb.\\
\thickhline
$\om$  &0.15&1.4&0.006&0.007&0.9&0.2\\
\hline
$\sig$ &0.01&0.3&0.002&0.002&0.2&0.1\\
\hline
$\asnone$&0.08&1.2&0.024&0.022&2.7&2.6\\
\hline
$\aagnone$&4.9&34.1&0.076&0.076&17.7&4.9\\
\hline
$\asntwo$ &0.72&0.6&0.006&0.006&0.6&0.7\\
\hline
$\aagntwo$ &0.12&1.9&0.018&0.028&2.3&20.2\\
\thickhline
\end{tabular}
\label{table:table_smf}
\vspace{2mm}
\end{table}

Shown in Figure \ref{fig:emulated_smf_no_uncertainty} are two-dimensional projections of the inferred posterior ({\it bottom left}) and {\it emulated} SMFs from the posterior along with the emulator-based target SMFs ({\it top right}). 
Here, the emulator-based target SMFs $\bx_0$ is generated by the emulator with the parameters that we used to generate the emulated SFRD in Section \ref{sec:emulated_sfrd}.
A total of 6000 emulations are used to retrieve the posterior density $p(\btheta|\bx_0)$ given the emulator-based target SMFs $\bx_0$.
The maximum of the posterior density\footref{footnote:maximum_of_posterior} ({\it red dashed} or {\it red cross-hair}) match the true values ({\it black solid} or {\it black cross-hair}) nearly perfectly for all six parameters (see Table \ref{table:table_smf} for details).
In the {\it top right} panel, the SMFs from the maximum of the posterior ({\it red dashed}) coincide with the true SMFs ({\it black solid}) with a relative error of 0.4\%.
The SMFs from the full posterior ({\it grey region}) have an exceedingly narrow distribution with the mean standard deviation of 0.007 dex.  
In comparison to the SFRD in Section \ref{sec:performance_sfrd}, the relative errors of SMFs are slightly better except $\asntwo$, whereas the standard deviations of the SMFs are comparable to that of the SFRD.
In general, the error of inference---e.g., relative errors--- and the size of uncertainty region---e.g., standard deviations---can be related to the correlation between the parameters and the observable. 
We discuss why $\asntwo$ can be more precisely predicted from the SFRDs in Section \ref{sec:correlation_between_obs_and_params}, where we show how surprisingly the errors presented here are well explained by the correlations.

Figure \ref{fig:emulated_smf_stability} illustrates the convergence and its stability for the inference from the emulated SMFs.
Each panel shows maxima ({\it red squares}) and variances ({\it grey error bars}) of the posterior for a given parameter as a function of the (cumulative) number of emulations used for training so far. 
All six maxima ({\it red squares}) stably converge to the true values ({\it black dotted lines}) over $[4800, 8000]$ emulations (see Table \ref{table:table_smf} for details).
Hence, the convergence of the inference requires at least $\sim 4000$ emulations.

In contrast to the SFRD, the unstably convergent stage is absent in the SMFs.
To be stably convergent, the SMFs require only 4000 emulations whereas $\sim 40000$ emulations are needed for the SFRD, an order of magnitude difference.
Moreover, the average variances at the stably convergent stage, $\bar{\sigma}_{\rm smf}\sim\mathcal{O}(1)$ for the SMFs  and $\bar{\sigma}_{\rm sfrd}\sim\mathcal{O}(10)$ for the SFRD, demonstrate that the SMFs converge with less fluctuations. 
Hence, we conclude that the SMFs converge far more rapidly and stably to the truth compared to the SFRD.
Furthermore, the NDE of the SFRD has more hidden units ($N_{\rm hid, sfrd} = 250$) than that of the SMFs ($N_{\rm hid,smf}=100$).
Given that the more neurons the easier it converges, we can conclude that the SMFs-parameters pairs are more easily and tightly to be mapped than the SFRD-parameters pairs. 
This indicates that the degree of correlation between the SMFs and the parameters is stronger than between the SFRD and the parameters (as discussed further in Section \ref{sec:correlation_between_obs_and_params}).

We also investigate the accuracy of the inference for various SMFs over the parameter space. 
Figure \ref{fig:paco_errorbar} shows the maxima ({\it red squares}) and variances ({\it grey error bars}) of 100 posterior densities inferred from 100 {\it emulated} SMFs.
Here, 100 SMFs are generated by the emulators with 100 randomly-sampled sets of parameters from the LH set (refer to Section \ref{sec:method_camels}).
We consistently perform the ILI with the total of 4800 emulations over all 100 SMFs without any further convergence tests.  
Due to computational cost, the minimal (necessary but perhaps not sufficient) number of emulations for the convergence is adopted\footnote{For the same reason, we cannot produce this plot for the SFRD since the SFRD requires significantly more simulations than the SMFs and owing to numerous hidden units of the NDE for the SFRD, the computational time is at least 10 times greater than that of the SMFs.}, deduced from the previous convergence test in Figure \ref{fig:emulated_smf_stability}.
Most maxima ({\it red square}) are on top of the ideal prediction line ({\it black solid}) or are indistinguishably close to it.
The average relative errors\footref{footnote:averaged_relative_error} of the maximum are (1.1$\pm$1.2, 0.2$\pm$0.2, 1.3$\pm$2.2, 4.0$\pm$8.5, 0.6$\pm$0.8, 1.8$\pm$3.4)\%.
These values are in line with the relative errors from the convergence test.
Thus, the ILI can be performed on the SMFs stably with relatively small, constant errors regardless of the choice of parameters.

\subsubsection{No Response to Uncertainty}
\label{sec:smf_uncertainty}
We perform ILI on the SMFs with the mock uncertainty (for details of mock uncertainty for the SMFs, refer to Section \ref{sec:emulator_as_surrogate} and Appendix \ref{apx:cosmicvariance_butterflyeffect}).
However, unlike the case of the SFRD in Section \ref{sec:uncertainty_sfrd} which shows appreciable variances in the inferred posterior, the posterior density inferred from the SMFs {\it with} the mock uncertainty is essentially {\it identical} to the posterior density inferred {\it without} the mock uncertainty. 
Since we cannot tell any difference both visually and quantitatively, the corresponding figure for the {\it with-uncertainty} case is not presented.
In Section \ref{sec:various_types_of_uncertainties}, we explain this result by studying  in detail how the inferred posteriors respond to various types of uncertainties and which type of uncertainty is suitable for the ILI.

\begin{figure*}[ht!]
    \centering
    \includegraphics[width=\textwidth]{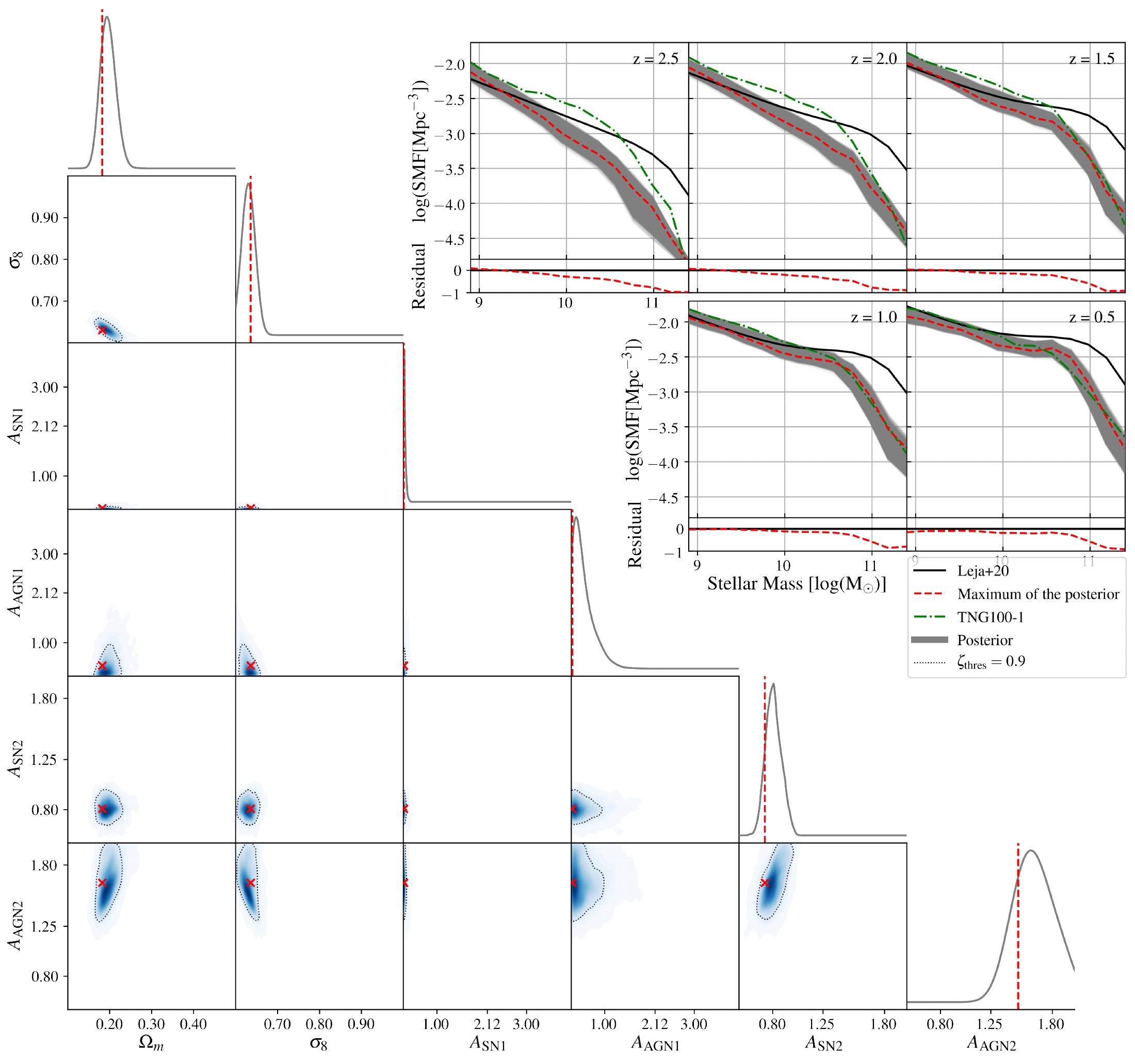}
    \caption{
    {\it Top right}: the stellar mass functions (SMFs) from the inferred posteriors ({\it grey} and {\it red}) and the observationally-driven target SMFs ({\it black}).
    {\it Bottom left}: 2D projections of the inferred posteriors for the observationally-driven target SMFs from five different redshifts ($z=$2.5, 2.0, 1.5, 1.0, and 0.5) \citep{lejasmf2020ApJ...893..111L}. 
    The {\it red} vertical lines indicate the maximum values of the inferred posterior.
    The inferred SMFs and the target SMFs show considerable discrepancy with a relative error of 41.1\% (see Section \ref{sec:observation_concatenated_smf}).
    }
    \label{fig:obs_smf}
\end{figure*}
\subsection{Inference from Observed Stellar Mass Functions}
\label{sec:observation_smf}
\subsubsection{Five Concatenated Stellar Mass Functions}
\label{sec:observation_concatenated_smf}
Figure \ref{fig:obs_smf} shows two-dimensional projections of the posterior density inferred from the observationally-driven target SMFs \citep{lejasmf2020ApJ...893..111L} ({\it bottom left}) and the corresponding SMFs ({\it top right}). 
The discrepancy between the observation ({\it black solid}) and the maximum of posterior ({\it red dashed}) is considerable with a mean relative error of 41.1 \% over all mass ranges. 
The discrepancy of the high mass end of the SMFs is dominant with relative error for stellar masses $\gtrsim 10^{11}\msun$ being $\sim 80.9$\% versus 13.7\% at the lower mass ends.
The high mass ends ($\gtrsim 8\times10^{9}\msun$) of the observed SMFs are located completely outside the region of the SMFs from the full posterior ({\it grey}) across all five redshifts.
The population of massive galaxies in the observed SMF is far greater than that in the inferred SMFs.

At the low mass ends, even though the observed SMFs lie within the ({\it grey}) region, the emulated and observed SMFs have different, distinct characteristics: (1) the slope of SMFs at high redshift and (2) the evolutionary rate of the SMFs.
At $z=2.5$ and $z=2.0$, the slopes of the inferred SMFs are notably steeper than the observed SMFs. 
The observed SMFs do not show the appreciable difference in slope across the redshifts, whereas the slopes of inferred SMFs become less steep as redshift decreases.
Also, there is a significant difference in the evolutionary rate of SMFs.
The evolution of the inferred SMFs is barely notable, but the apparent growth can be seen in the observation.
For example, the differences between SMFs at $M_{\star}=10^{8.9}\msun$ at $z=2.5$ and $z=0.5$ are 0.20 dex for the inference and 0.47 dex for the observation. 

\begin{table}[t!]
\centering
\caption{The maximum of the posteriors, $\btheta_{\rm max}$, ({\it first-fourth rows}) in Figures \ref{fig:obs_smf} and \ref{fig:obs_smf_seperated} and their standard deviations $\sigma$ ({\it fifth-eighth rows}) with respect to the maximum of the posteriors.}
\begin{tabular}{?c|c?c|c|c|c|c|c?}
\thickhline
&& $\om$ & $\sig$ & $\asnone$ & $\aagnone$ & $\asntwo$ & $\aagntwo$\\ 
\thickhline
\multirow{4}{*}{\rotatebox[origin=c]{90}{$\btheta_{\rm max}$}}
&ALL\footnote{The parameters are inferred from all five redshifts simultaneously ({\it red cross-hair} in Figure \ref{fig:obs_smf}).\label{footnote:table_all}}
& 0.18 & 0.64 & 0.26 & 0.27& 0.73& 1.50\\ 
 \cline{2-8}
&$z=0.5$& 0.22 & 0.97 & 0.41 & 1.50 & 0.51 & 0.52\\ 
 \cline{2-8}
&$z=1.5$& 0.10 & 1.00 & 0.47 & 0.25& 0.58 & 1.63\\ 
 \cline{2-8}
&$z=2.5$& 0.11 & 1.00 & 0.72 & 2.77& 0.83 & 0.61\\ 
\thichline
\multirow{4}{*}{\rotatebox[origin=c]{90}{$\sigma$}}
&ALL\footref{footnote:table_all}
&0.02 & 0.01 & 0.03 & 0.31 & 0.10 & 0.22\\
 \cline{2-8}
&$z=0.5$&0.02 & 0.04 & 0.11 & 1.32 & 0.18 & 0.71\\
 \cline{2-8}
&$z=1.5$&0.01 & 0.03 & 0.04 & 1.38 & 0.04 & 0.35\\
 \cline{2-8}
&$z=2.5$&0.05 & 0.03 & 0.11 & 1.02 & 0.06 & 0.82\\
\thickhline
\end{tabular}
\label{table:obs_smf_param}
\end{table}

\begin{figure*}[t!]
    \centering
    \includegraphics[width=\textwidth]{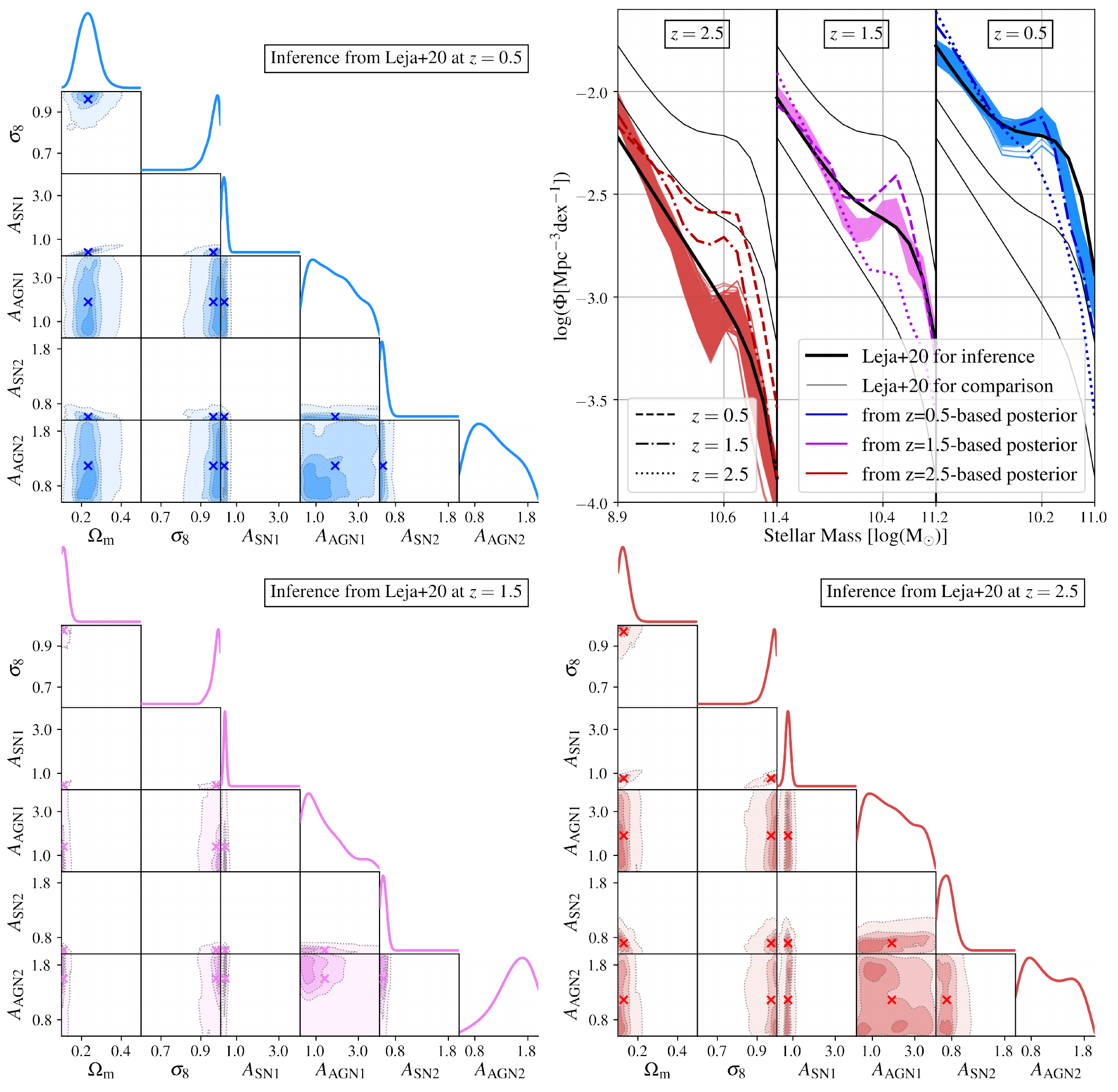}
    \caption{
    Two-dimensional projections of the posteriors inferred from the individual (observationally-driven target) stellar mass functions (SMFs; \citet[]{lejasmf2020ApJ...893..111L}) at the three different redshifts: $z=0.5$ ({\it top left sky blue}), $z=1.5$ ({\it bottom left, violet}), and $z=2.5$ ({\it bottom right, red}).
    The {\it top right} panel shows three inference results from the observationally-driven target SMFs ({\it thick black solid}) at different redshifts $z=2.5$ ({\it left}), $1.5$ ({\it middle}), and $0.5$ ({\it right}).
    Here, {\it thin} black solid lines are the observed SMFs that are not involved in the inferences at a particular redshift.
    The {\it dashed}, {\it dotdashed}, and {\it dotted} lines are the emulated SMFs at $z=0.5$, $1.5$, and $2.5$ from the maximum of the posteriors inferred from the observed SMFs at $z=0.5$ ({\it dark blue}), $1.5$ ({\it dark violet}), and $2.5$ ({\it dark red}), respectively.
    The relative errors of each SMF with respect to each observationally-driven target SMF are 17.7\% for z = 2.5, 10.3\% for z = 1.5, and 10.3\% z = 0.5.
    Compared to the inference from the concatenated SMFs, the accuracy has been increased, but still, the inferred SMFs cannot precisely match the observationally-driven target SMFs compared to the SFRD case (see Section \ref{sec:observation_individual_smf}).
    }
    \label{fig:obs_smf_seperated}
\end{figure*}

\subsubsection{Three Individual Stellar Mass Functions}
\label{sec:observation_individual_smf}
To isolate the problem, we perform inferences from one SMF at a time for each redshift ($z=0.5,$ $1.5,$ and $2.5$) separately.
Figure \ref{fig:obs_smf_seperated} illustrates the three posterior densities inferred from the observed SMFs ({\it thick solid lines}) at $z=0.5$ ({\it sky blue}), $z=1.5$ ({\it violet}), $z = 2.5$ ({\it red}) and the corresponding SMFs ({\it top right}).
The separate inferences lead to significantly higher accuracy than the previous inference, with the relative errors of SMFs of 17.7\% for $z=2.5$, 10.3\% for $z=1.5$, and 10.3\% $z=0.5$ (the relative error of the previous inference is 31.4\%).
In comparison to the previous inference (Figure \ref{fig:obs_smf}), the locus of the posterior densities are notably different, especially in $\sig$ and $\asnone$ (refer to Table \ref{table:obs_smf_param}).
The standard deviations of the AGN parameters in Figure \ref{fig:obs_smf_seperated} are notably larger than those from the inference based on the five concatenated SMFs in Figure \ref{fig:obs_smf}, whereas the standard deviation of the inferred SMFs are similar.
This might indicate that the impact of the AGN parameters on the SMFs is trivial.
In general, the impact of the AGN parameters on the SMFs is negligible in this analysis, given the extensive variances that cover almost the entire parameter space. 
The average variances of $\aagnone$ and $\aagntwo$ over the three redshifts are 1.23 and 0.62, whereas the variances of the other parameters are less than 0.01 on average.

$\sig$ values from the separate inferences are considerably higher than from the combined-SMF inference in Section \ref{sec:observation_concatenated_smf}, whereas $\om$ shows relatively small changes.
In both emulators and simulations, the AGN parameters only have a minor effect on the stellar populations of galaxies (Sections \ref{sec:correlation_between_obs_and_params} and \ref{sec:agn_weak_correlation}).
As a compensatory action, the emulated universes have exploited the cosmological parameters, especially $\sig$, to control the populations of massive galaxies in the context of structure formation. 
Meanwhile, to sustain the density of low mass galaxies, stronger supernova feedback is required, which we can see in Figure \ref{fig:obs_smf_seperated} as increases in $\asnone$.

In the {\it top right} panel of Figure \ref{fig:obs_smf_seperated}, each  of the three subpanels corresponds to the inference based on an observed SMF from a different redshift, $z_{\rm obs}$=(0.5,1.5,2.5) (thick black curves). SMFs generated by the emulators for $z_{\rm emu}=(0.5,1.5,2.5)$ with the maximum of the corresponding posteriors are drawn as dashed, dash-dotted and dotted curves, respectively, and in each panel the presented SMFs for the corresponding redshift sample the full posterior in thin colored curves that appear largely as continuous shaded regions. 
Note that we have three independent emulators that estimate the SMFs at three different redshifts.
For example, the emulated SMF at $z=0.5$ from the maximum of the $z=1.5$-based posterior ({\it bottom left, violet}) is drawn with a {dark violet dashed} line in the middle subpanel.
Each subpanel exhibits the discrepancies of the slope and evolution of SMFs across different redshift from a single posterior.
Notice that although the separate inferences can predict observations with much improved accuracy compared to the inference from the combined SMFs, the discrepancies are not resolved.
Further discussion of the discrepancies between inferred SMFs and observed SMFs can be found in Section \ref{sec:mismatch_SMFs}.

On the simulation side, the significant mismatch between the inference and observation can be attributed to the following:
(1) the limited volume of simulations; (2) low resolution of the simulations and the failure of re-scaling; (3) limitations of the physical models of simulations; (4) inaccuracy of the emulators.
The size of the simulation is directly linked to the level of simulation uncertainty that we fail to realize in the SMFs (refer to Section \ref{sec:smf_uncertainty} and \ref{sec:various_types_of_uncertainties} for details).
Since we cannot account for the simulation uncertainty, the inference results suffer from sampling bias.
Also, the lack of massive galaxies is prevalent in low-resolution simulations (Appendix A. in \citet[]{pillepichrescaling2018MNRAS.475..648P,pillepichconvergence2018MNRAS.473.4077P} and Appendix \ref{appendix:rescaling} in this paper).
Even though we apply re-scaling to alleviate the resolution effects, the rescaled SMFs are still subject to resolution convergence (refer to Section \ref{appendix:rescaling}).
Inside the parameter space that we adopt in this work, there might not exist a set of parameters that can reproduce the observations simultaneously across redshifts.
Lastly, since we are using the emulators as a surrogate for cosmological simulations, the inference results are also limited by the accuracy of the emulators.
Note that observational uncertainties, such as systematic biases and physical limits, can also be crucial factors for the mismatch but will not be discussed in this paper.

\section{discussion}
\label{sec:discussion}
\subsection{Correlations between Observables and Parameters}
\label{sec:correlation_between_obs_and_params}
Correlations that measure how dependent two variables are on each other play a significant role in mapping between domain (input) and codomain (output)
The accuracy of machine learning can be proportional to the degree of correlation between input and output features.
For example, the less correlated they are, the larger the variances and vice versa. 
In this work, we have already encountered several trails of the correlation between parameters and observables in the inference results.
One of the most evident examples is that the SFRD requires $\mathcal{O}(10^4)$ simulations to reach the stable convergence, whereas only $\mathcal{O}(10^3)$ simulations are sufficient for the SMFs (see Sections \ref{sec:performance_sfrd} and \ref{sec:performance_smf} for the SFRD and the SMFs, respectively).
In other words, with the same size of training data---i.e. the same number of emulations, the SMF would have attained a more accurate posterior with a smaller variance.

For a quantitative analysis, we measure the correlation between parameters and observables using the mutual information (MI), which is a fundamental measure for the inter-dependence or relationship between two variables.
In contrast to linear correlation coefficients, such as the Pearson correlation coefficient, MI captures non-linear statistical dependencies \citep[]{kinney2014PNAS..111.3354K}.
MI is defined by 
\begin{equation}
    I(X,Y) = D_{\rm KL}(P_{XY}\Vert P_{X}\otimes P_{Y})
\end{equation}
where $D_{\rm KL}(\cdot\Vert\cdot)$ is the Kullback-Leibler divergence\footnote{
A Kullback-Leibler divergence is defined as 
$D_{\rm KL}(P\Vert Q)$ $\equiv$ $\sum_{x\in \mathcal{X}}P(x)\log(P(x)/Q(x))$ or $\int_{-\infty}^{\infty}p(x)\log(p(x)/q(x))\dd x$. 
This is also referred as to a statistical distance between $Q(x)$ and$ P(x)$.
\label{footnote:KL_divergence}}, and $P_{XY}$ and $P_X$ are joint and marginal distributions, respectively.
Here, $\otimes$ denotes the outer product that spans the probability space from $X$ and $Y$ to $(X, Y)$.
$I(X,Y)$ quantifies a statistical distance between the joint probability and the product of marginals.
The MI is zero if and only if $X,\;Y$ are independent.

\begin{table}[t]
\centering
\caption{The mutual information (MI) between the parameters and obseravbles ({\it first-fourth} rows). 
The relative errors $\bs{\delta}$ of the convergent SFRD and SMFs from Section \ref{sec:performance_sfrd} and \ref{sec:performance_smf}, respectively ({\it fifth} and {\it sixth} rows).
`{\it Sim}' and `{\it Emu}' stands for ``simulated'' and ``emulated''. 
The values of the MI are multiplied by 100 for convenience and the unit for the relative error is percentage.
Note that the higher the MI is, the more the parameters-observable pairs are correlated, whereas the MI is zero when they are independent.
}
\begin{tabular}{?c?c|c|c|c|c|c?}
\thickhline
& $\om$ & $\sig$ & $\asnone$ & $\aagnone$ & $\asntwo$ & $\aagntwo$\\ 
\thickhline
{\it SimSFR}& 1.19 &0.62 & 0.54 & 0.29 & 2.11 & 1.43\\ 
\cline{1-7}
{\it SimSMF}& 12.5 & 9.27 & 36.9& 1.17& 4.69 & 0.41\\ 
\thickhline
{\it EmuSFR}& 11.4 & 6.82 & 30.0 & 0.7& 17.1 & 0.85\\ 
\cline{1-7}
{\it EmuSMF}& 14.6 & 12.8 & 41.4& 1.49& 5.79 & 0.30\\ 
\thickhline
{$\bs{\delta}_{\rm sfr}$ (\%)}& 0.20 & 0.49 & 1.36 & 49.8& 0.25 & 3.65\\ 
\cline{1-7}
{$\bs{\delta}_{\rm smf}$ (\%)}& 0.15& 0.01 & 0.08& 4.9& 0.72 & 0.12\\ 
\thickhline
\end{tabular}
\label{table:mutual_info}
\end{table}

However, the estimation of MI is challenging and only tractable for discrete variables or when probability distributions are known \citep[]{paninski10.1162/089976603321780272}.
Thus, we adopt the mutual information regression function in the {\tt sklearn} package \citep[]{kraskov2004PhRvE..69f6138K,scikit-learn,ross2014PLoSO...987357R}. 
Using the package, we estimate the MI for (1) simulated observables and parameters from the LH set, 
(2) emulated observables and parameters from the LH set.
Here, the observables and the parameters are normalized to reduce the effects of the difference in magnitude of values in the same way as used to train the emulator (Section \ref{sec:emulator_as_surrogate}).

Table \ref{table:mutual_info} shows the estimated MI of each observable-parameter pair.
In the simulations, the MI of the SMFs ({\it SimSMF}) is significantly higher than that of the SFRD ({\it SimSFR}) over all parameters except $\aagntwo$.
Interestingly, the gap has been drastically reduced in the emulators ({\it EmuSMF} and {\it EmuSFR});
The MI goes up for both, but more so for the SFRD, resulting in narrowing of the gap.
The increase of MI in the emulators (especially in the SFRD) can be attributed to the training of the emulators.
During training, the emulator (neural network) naturally suppresses the noise in the data, making the correlations stronger \citep[]{goldfeld2018arXiv181005728G, gabrie2019JSMTE..12.4014G}.
As a result, input and {\it emulated} output pairs can attain overall more correlation via the emulator than input and {\it simulated} output pairs. 
Note that the correlation of the {\it emulated} pairs should depend on the precision of training.
Nonetheless, the relative magnitudes of MI among six parameter-observable pairs remain the same in terms of its order.
For instance, the MI of SMF-parameter pairs is still higher than that of SFRD-parameter pairs except for $\asntwo$ and $\aagntwo$.

In addition to the gap between simulations and the emulator, the MI among the parameters shows considerable differences in magnitude.
$\asnone$ is found to be the most relevant parameter to both SFRD and SMFs, whereas the AGN feedback parameters ($\aagnone$ and $\aagntwo$) have MI that is one or two orders of magnitude lower than other parameters.
That is, $\asnone$ can precisely be inferred or estimated in terms of inference and machine learning.
On the other hand, not only the AGN parameters can be hardly constrained but also the AGN physics themselves has a negligible impact on the SFRD and the SMFs in the TNG suite of the CAMELS simulations that have low resolution and limited volume (refer to Section \ref{sec:agn_weak_correlation} for the SMFs and Figure 9 of \citet[]{camels2021ApJ...915...71V}) for the SFRD.
Such impact of relative magnitudes among parameter-SFRD pairs can also be found in \citet[Figure 9]{camels2021ApJ...915...71V} which demonstrates that the SFRD is only sensitive to $\om$, $\sig$, $\asnone$, and $\asntwo$, but not $\aagnone$ and $\aagntwo$.

The values of the MI are also in line with the performance of inference with respect to the parameters. 
For instance, the MI between $\asntwo$ and the SFRD, which is an exceptional case, is larger than that between $\asntwo$ the SMFs, which is in line with the relative error of $\asntwo$ in the SFRD ({\it fifth} rows) being also smaller than that of the SMFs ({\it sixth} row).
The ratio among the MIs of each parameter is inversely proportional to that of the relative error and/or the variance for all inferences presented in this work. 
In addition, the convergence depends on the degree of the correlation of the pairs, implying that more training data---i.e. simulations---are required to converge in the SFRD (compare Figures \ref{fig:emulated_sfrd_stability} and \ref{fig:emulated_smf_stability}).
Lastly, this is also in agreement with previous work.
The CAMELS introduction paper \citep{camels2021ApJ...915...71V} builds a fully-connected neural network to predict cosmological and astrophysical parameters taking the SFRD as input. 
With the neural network, $\Omega_{m}$, $\sigma_8$, $A_\mathrm{SN1}$, and $A_\mathrm{SN2}$ are predicted relatively well compared to $A_\mathrm{AGN1}$ and $A_\mathrm{AGN2}$ \citep[Figure 11]{camels2021ApJ...915...71V}.

\begin{figure}[t!]
    \centering
    \includegraphics[width=0.5\textwidth]{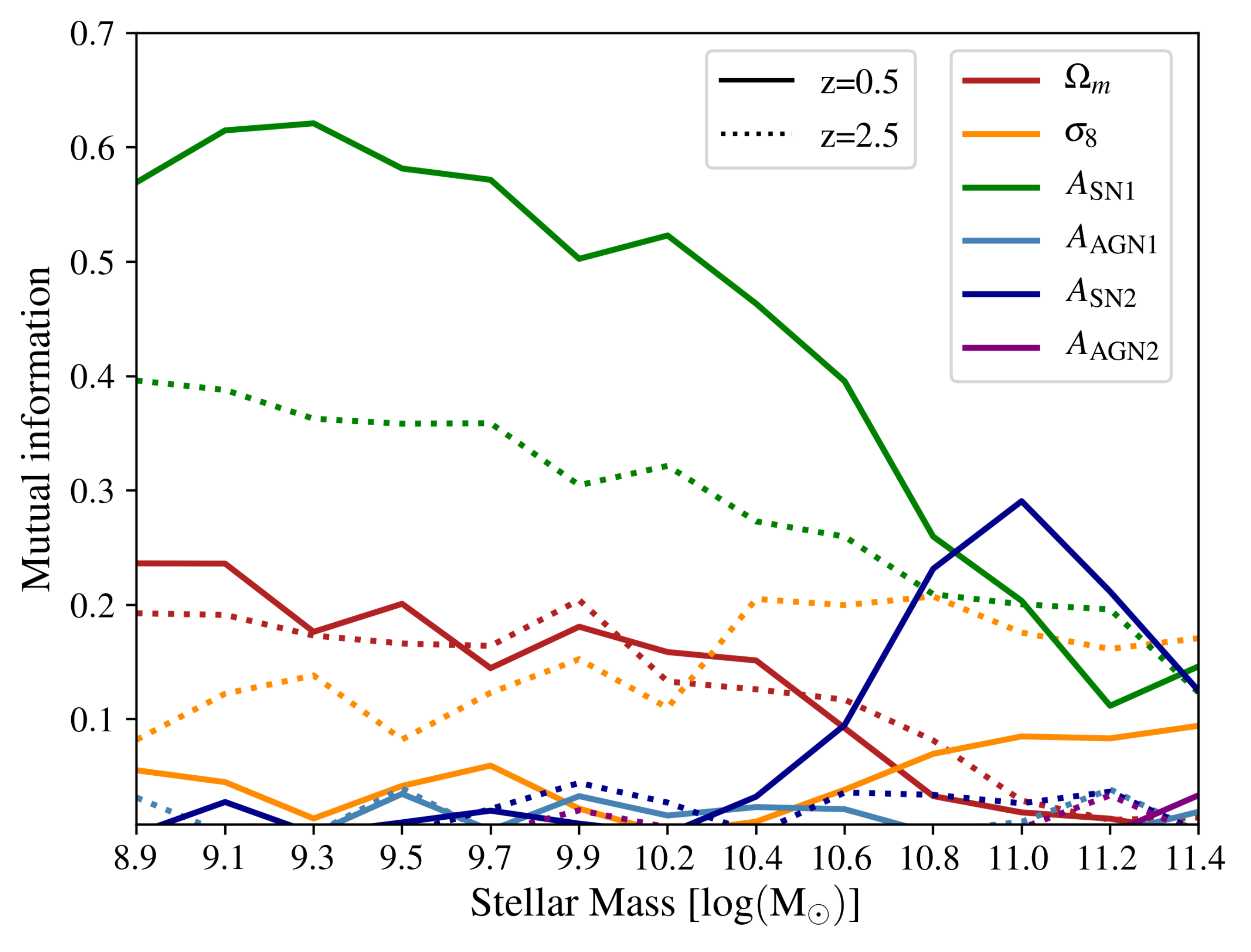}
    \caption{Mutual information of the stellar mass function (SMF) at $z = 0.5$ ({\it solid}) and $2.5$ ({\it dotted}) and each parameter as a function of stellar mass.
    Each colored line indicates mutual information between one parameter and the SMF.}
    \label{fig:mi_smf}
\end{figure}

Figure \ref{fig:mi_smf} shows the MI between each parameter and the SMF at $z=0.5$ ({\it solid}) and $2.5$ ({\it dotted}) as a function of stellar mass of each bin.
At $z=0.5$, $\asnone$ is dominant over other parameters especially for $M_{\star}\lesssim10^{10.8}\msun$, while $\om$ also has appreciable MIs. 
In the higher mass end ($M_{\star}\gtrsim10^{10.8}\msun$), $\asntwo$ is most effective, followed by $\sig$ and $\asnone$, whereas the MIs of $\aagnone$ and $\aagntwo$ are negligible across the entire stellar-mass range.
In the case of the higher redshift ($z=2.5$), the order of the relative magnitudes of the MIs among parameters has changed.
$\asnone$ is still the most dominant parameter in the low mass end, but the gap with $\om$ is smaller. 
The MI of $\sig$ is drastically larger overall and dominates the high mass end, whereas $\asntwo$ becomes negligible.  
Compared to $z=0.5$ ({\it solid} lines), $\om$ ({\it red dotted}) remains approximately the same and $\sig$ ({\it yellow dotted}) is significantly increased, while the $\asnone$ ({\it green dotted}) and $\asntwo$ ({\it blue dotted}) are appreciably decreased, which, overall, leads to the increase in the ratio of MI of the cosmological parameters to the astrophysical parameters.
This can be indicative that cosmology is more influential at early times than late times compared to astrophysical phenomena.

The MI can be an indirect but crucial measure for the degree of the relative impact of parameters on observables. 
We can extract physical insights from Figure \ref{fig:mi_smf} as follows: 
(1) The impact of cosmology ($\om$ and $\sig$) diminishes at lower redshift in terms of the SMF.
(2) The energy budget of supernova-driven winds ($\asnone$) is of paramount importance in the galaxies of stellar mass $\lesssim 10^{10.8}\msun$.
(3) For massive galaxies ($M_{\star} \gtrsim 10^{10.8}\msun$), the stellar wind velocity ($\asntwo$) has more effect than its energy budget ($\asnone$).
(4) The portion of dark matter ($\om$) in the universe has more impact on lower stellar-mass galaxies than higher-mass galaxies, 
(5) whereas the density fluctuation ($\sig$) is more in massive galaxies.
(6) The kinetic feedback of the black holes ($\aagnone$ and $\aagntwo$) are weakly related with the SMF regardless of both redshift and stellar mass of the galaxies.
In the following section, the impact of the AGN parameters on the SMFs is discussed further.

\begin{figure*}[t!]
    \centering
    \includegraphics[width=\textwidth]{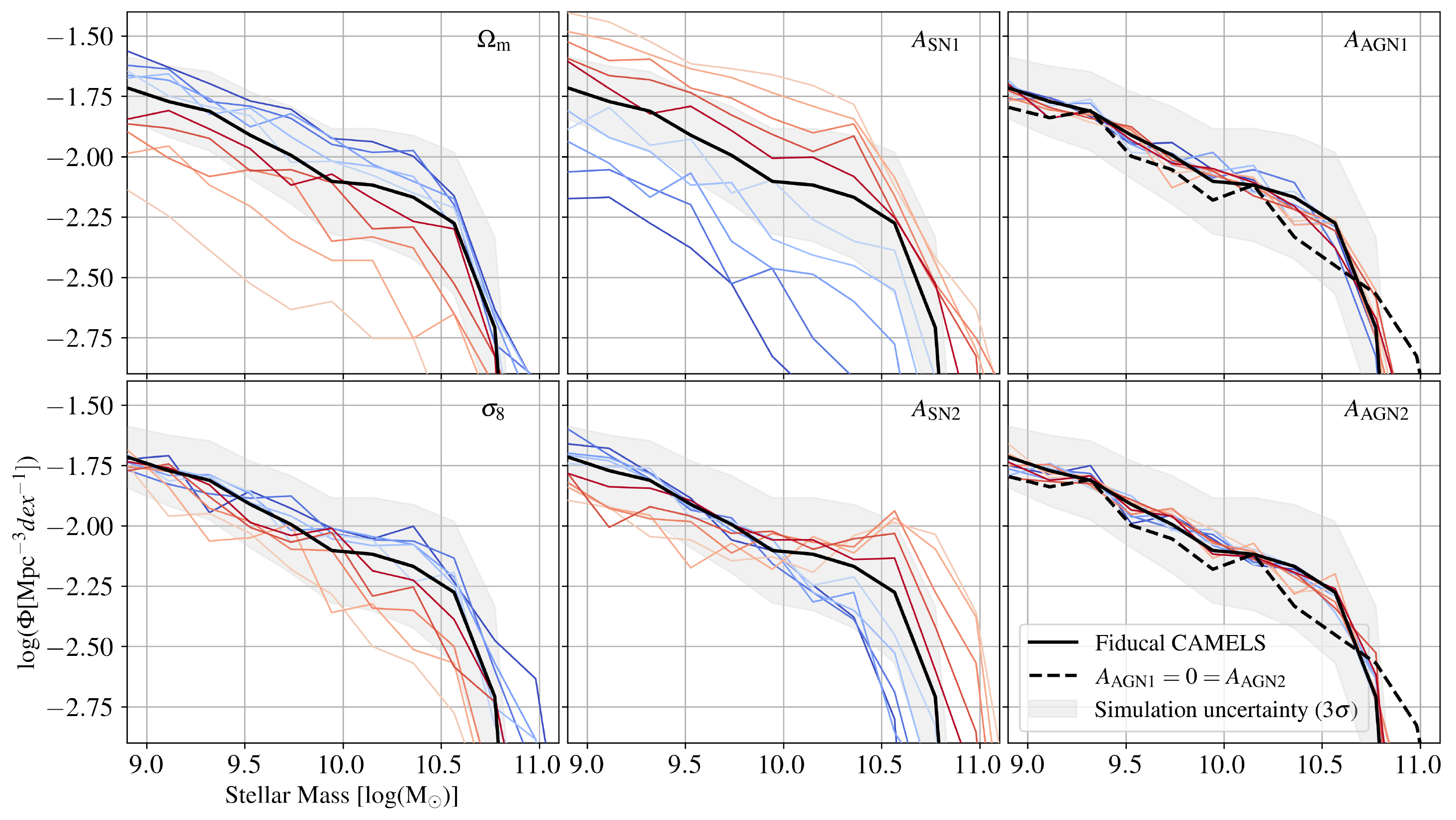}
    \caption{
    Simulated stellar mass functions at $z=0.5$ from the 1P set (higher parameter values with bluer curves and lower values with redder curves), from the fiducial parameter combination ({\it black solid}), and without the AGN feedback ({\it black dashed}).
    The {\it light grey region} indicates the $3\sigma$ region of the simulation uncertainty.
    The variations of $\om$, $\sig$, $\asnone$, and $\asntwo$ ({\it left and middle panels}) lead to considerable changes in the SMFs, whereas the variations of $\aagnone$ and $\aagntwo$ ({\it right panels}) have a negligible impact on the SMFs. 
    However, the inclusion of the kinetic AGN feedback obviously has an impact on the SMF by comparison between the weakest AGN feedback ({\it reddest}) and no kinetic AGN feedback ({\it black dashed}) (see Section \ref{sec:agn_weak_correlation} for details).
    }
    \label{fig:1p_smf}
\end{figure*}

\subsubsection{Negligible Impact of AGN Kinetic Feedback on Galactic Stellar Mass}
\label{sec:agn_weak_correlation}
We have seen the weak correlation between the observables and AGN feedback parameters from both the variances of the posteriors and the mutual information.
In this section, we directly investigate the impact of AGN parameters on the SMFs, compared with that of other parameters.
The AGN parameters control the energy budget of the low-accretion mode (kinetic feedback) of black holes ($\aagnone$) and its burstiness ($\aagntwo$) (refer to Section \ref{sec:method_camels}).
The kinetic wind from the black holes can play a more crucial role in quenching massive galaxies than thermal feedback can, since the kinetically-injected energy is less vulnerable to gas (over-)cooling, especially for dense gas where the cooling time is short 
\citep[]{bower2006MNRAS.370..645B,croton2006MNRAS.365...11C,fabian2012ARA&A..50..455F,dubois2013MNRAS.433.3297D,roasa-guevara2015MNRAS.454.1038R,daniel2017MNRAS.464.2840A,zinger2020MNRAS.499..768Z,piotrowska2022MNRAS.512.1052P,wellons2022arXiv220306201W}.
Furthermore, \citet{terrazas2020MNRAS.493.1888T} finds that in the TNG model, the galaxy can be quenched whenever the cumulative kinetic feedback energy of the central black hole exceeds the gravitational binding energy of the gas within the galaxy. 
In this regard, the lack of correlation between the AGN parameters and the SMFs in \rv{the CAMELS-TNG}\footnote{\rv{Note that the CAMELS-SIMBA shows more variations with respect to the AGN parameters.}}, even at the higher mass end, seems puzzling.

Shown in Figure \ref{fig:1p_smf} are the SMFs at $z=0.5$ from the simulations in the 1P set and the SMFs from the simulations with no kinetic feedback ($\aagnone=0=\aagntwo$, {\it black dashed}) and the fiducial parameter combination ({\it black solid}).
Note that the 1P set consists of 61 simulations varying only one parameter at a time \citep[see Section 3.3.2]{camels2021ApJ...915...71V} and all the simulations presented in this section are performed with the same initial conditions. 
In each panel, the bluer the SMF, the higher the parameter value, and the redder the SMF, the lower the parameter value.
In comparison to the simulation uncertainty ({\it light grey region}), the four parameters ($\om$, $\sig$, $\asnone$, and $\asntwo$) except the AGN parameters show significant changes of the SMFs in response to variations of their values, which is in line with the previous section.
One interesting feature is that the impact of 
$\asntwo$ is applied conversely to the low mass end and the high mass end of the SMFs. 
For instance, the increase in the supernova-driven wind speed ($\asntwo$) effectively suppresses the massive population while keeping the lower-mass population similar or increasing it.

In contrast, the changes in the AGN parameters ({\it right panels}) have a negligible impact on the SMFs. 
However, this does not imply that AGN kinetic feedback lacks any effect at all, since even the weakest AGN feedback in the 1P set ({\it the reddest line in the right panels}) suppresses the massive population compared to our new simulation that lacks kinetic feedback altogether ($\aagnone=0=\aagntwo$)\footnote{The new simulation does include thermal AGN feedback.}, as expected in many papers \citep[]{page2012Natur.485..213P, dubois2013MNRAS.433.3297D,terrazas2020MNRAS.493.1888T,su2021MNRAS.507..175S}.
In addition, Figure 8 of \citet[]{pillepichconvergence2018MNRAS.473.4077P} shows the clear impact of black hole kinetic feedback on the stellar masses of massive galaxies. 
The negligible effects of the AGN parameters in the CAMELS simulations can be explained by a combination of the following: 
(1) Due to the small box size $\sim (25 \,\mathrm{Mpc}/h)^3$, the CAMELS simulations usually lack massive galaxies subject to the AGN kinetic feedback.
(2) The impact of weakest AGN feedback parameters of the CAMELS simulations is already so effective that further increases of these parameters cannot lead to significant changes in the SMFs.

\begin{figure*}[ht!]
    \centering
    \includegraphics[width=\textwidth]{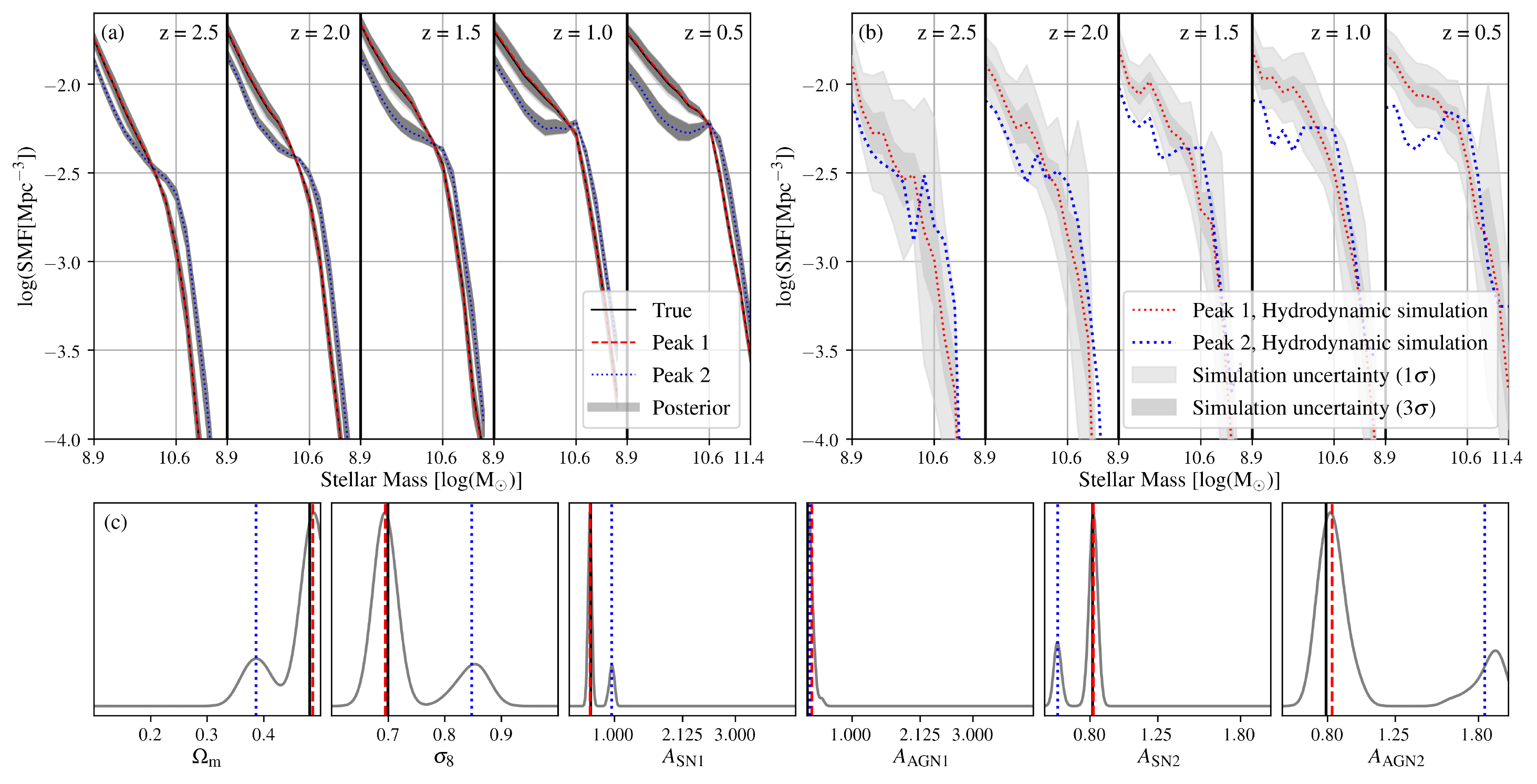}
    \caption{
    (a) Stellar mass functions from the emulator; 
    (b) Stellar mass functions from the cosmological simulation;
    (c) One-dimensional projections of the inferred posterior based on Figure \ref{fig:emulated_sfrd_bimodality}.
    The two peaks that produce the nearly identical SFRDs, which are degenerate, in Figure \ref{fig:emulated_sfrd_bimodality} are drawn with the {\it red dashed} and {\it blue dotted} lines consistently. For details, refer to Section \ref{sec:bimodality_sfrd}.
    Note that there exists an offset between the emulated SMFs and simulated SMFs. 
    This is because the emulated SMFs are re-scaled, whereas the simulated SMFs come right from the cosmological simulations.
    Given that both simulated and emulated SMFs from the two peaks lie outside the $3\sigma$ region of the simulation uncertainty, the degeneracy can be broken with the SMFs (see Section \ref{sec:degeneracy_breaking}).
    }
    \label{fig:smf_degeneracy_breaking}
\end{figure*}

\subsection{Degeneracy Broken with Stellar Mass Functions}
\label{sec:degeneracy_breaking}
We have seen the bimodality of the inferred posterior density and degeneracy in SFRD in Section \ref{sec:bimodality_sfrd} and Figures \ref{fig:emulated_sfrd_bimodality} and \ref{fig:sfrd_degeneracy}.
In this section, we discuss how degeneracy in the SFRD can be broken with both the {emulated} SMF and the {simulated} SMF.
Figure \ref{fig:smf_degeneracy_breaking} illustrates the marginal distributions ({\it bottom}) that are exactly the same marginals as in Figure \ref{fig:sfrd_degeneracy}, derived by inference on the SFRD, and the corresponding {\it emulated} SMFs ({\it top left}) and the {\it simulated} SMFs ({\it top right}).
In contrast to the case of the SFRD where the two {\it emulated} SFRDs from the two peaks of the posteriors ({\it red dashed} and {\it blue dotted}) are nearly on top of each other, the corresponding {\it emulated} SMFs for the two separate set of parameters are clearly different, as shown in panel (a). 
The standard deviation of the SMFs ({\it grey}) in panel (a), which is 0.081, is an order of magnitude higher than that of the SMFs from the convergent posterior in Figure \ref{fig:emulated_smf_no_uncertainty}, which is only 0.007.
In addition, the relative error of the SMFs from peak 2 ({\it blue dotted}) with respect to the SMFs from peak 1 ({\it red dashed}) is 59\%, which is far larger than 0.79\% from the SFRDs in Figure \ref{fig:sfrd_degeneracy}.
Therefore, we can conclude that the SMFs can break the degeneracy of the SFRDs in terms of the emulators.

Also, we perform new simulations with the parameter combinations of the peak 1 ({\it red dashed}) and peak 2 ({\it blue dotted}) to study whether the SMFs can break the degeneracy even under the simulation uncertainty.
In panel (b) of Figure \ref{fig:smf_degeneracy_breaking}, the {\it red dashed} and {\it blue dotted} lines indicate the simulated SMFs from the two peaks.
Similarly to the emulated SMFs, the two simulated SMFs do not coincide with each other.
However, since the simulation is affected by the simulation uncertainty, such as cosmic variance and butterfly effect (refer to Section \ref{sec:cosmic_variance_buttefly_effect}), we include the confidence regions ({\it grey} in panel (b)) of the simulation uncertainty.
We use $\pm 1\sigma$ ({\it grey}) and $\pm 3\sigma$ ({\it light grey}) regions that correspond to 68.1\% and 99.7\% confidence levels for the Gaussian distribution, respectively.
The standard deviations of the SMF are directly calculated from the simulations in the CV set as a function of stellar mass.
The confidence regions are drawn with respect to the {\it red dashed} curves.
Panel (b) ({\it top right}) demonstrates that the low mass ends of the {\it blue dotted} SMFs noticeably fall outside the $3\sigma$ regions at $z\geq1.5$.
By the definition of $3\sigma$ regions that corresponds to 99.7\% confidence level, the {\it blue dotted} SMFs have {\it only 0.3\%} chance that they share the same origin as the {\it red dashed} SMFs.

Both emulated and simulated results support two main conclusions: (1) There exists degeneracies in the SFRD (refer to Section \ref{sec:bimodality_sfrd}); (2) The SMFs can break the degeneracy in the SFRD (Figure \ref{fig:smf_degeneracy_breaking}).
Taken together with the correlation analysis in Section \ref{sec:correlation_between_obs_and_params}, these conclusions further support the notion that the higher the correlation between observable and parameters, the stronger the observable constrains the parameters; here, the SMF is shown to be a stronger constraint than the SFRD.

\begin{figure*}[t!]
    \centering
    \includegraphics[width=\textwidth]{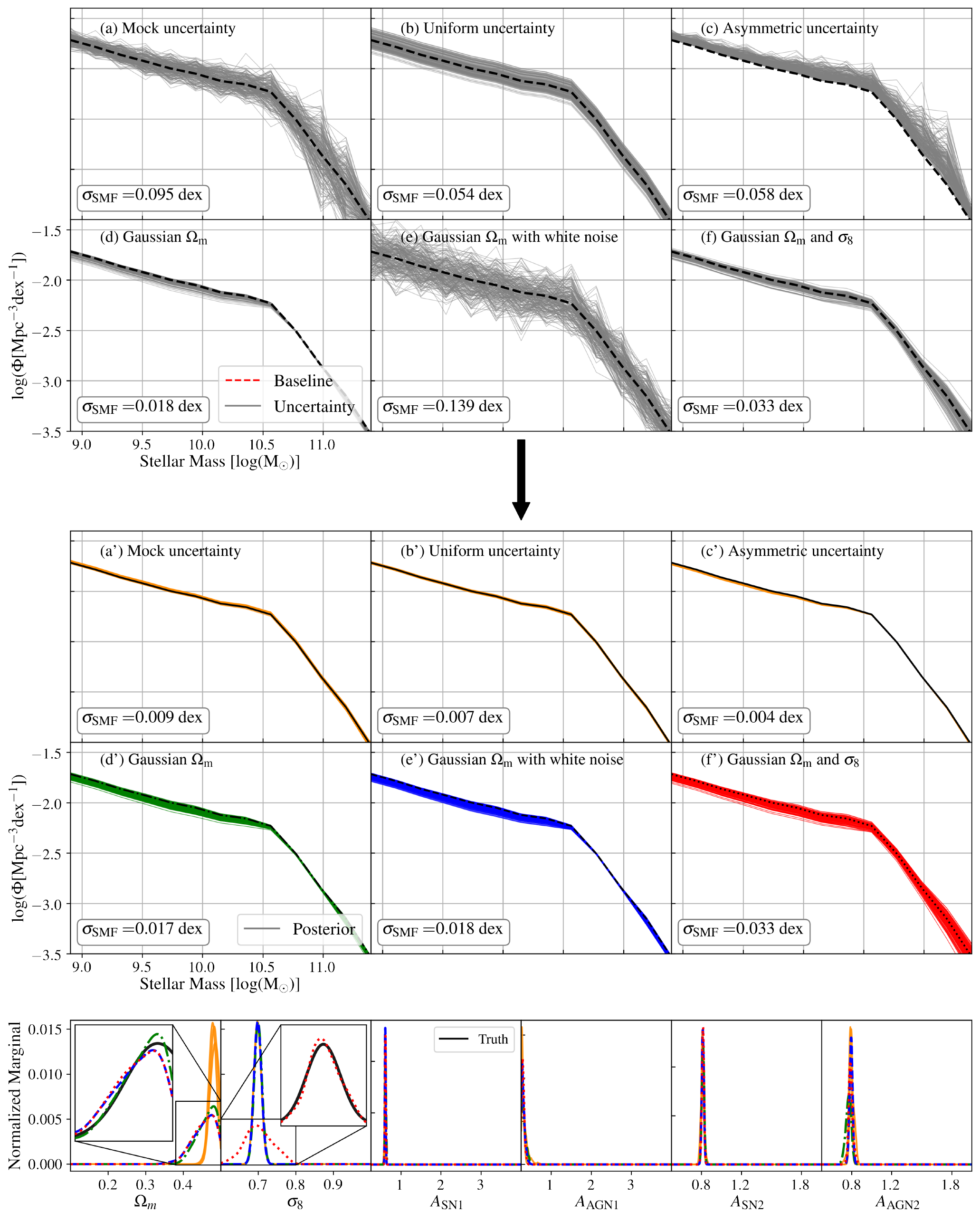}
    \caption{
    The {\it upper} six panels show the stellar mass functions (SMF) at $z=0.5$ with six different types of uncertainties.
    The {\it middle} six panels show the inferred SMFs at $z=0.5$ with six different types of uncertainties.
    The {\it lower} six panels show six marginal distributions of the inferred posteriors from the SMFs with six different types of uncertainties.
    The marginals are color-coded as follows: Mock uncertainty, Uniform, Asymmetric White Noise (all three shown in {\it yellow}), Gaussian $\om$ ({\it green}),  Gaussian $\om$ with White Noise ({\it blue}), Gaussian $\om$ and $\sig$({\it red}).
    In the zoom-in panels, the Gaussian distributions implemented in the uncertainties (d), (e), and (f) are shown in {\it black solid} lines.
    }
    \label{fig:various_uncertainty}
\end{figure*}

\subsection{Uncertainty in Implicit Likelihood Inference}
\label{sec:various_types_of_uncertainties}
We have briefly discussed that the mock uncertainty applied to the SMFs has negligible impact on the variance of posteriors in Section \ref{sec:smf_uncertainty} unlike the case of the SFRD in Section \ref{sec:uncertainty_sfrd}. 
To elucidate the origin and implications of these results, and what type of uncertainty should be adopted for ILI, here we perform tests of ILI using various types of uncertainties.
We adopt six different uncertainties as follows: 
(a) {\it Mock uncertainty} (that which we have used throughout this paper) is modelled to mimic the simulation uncertainty from the CV set in Appendix \ref{apx:cosmicvariance_butterflyeffect}; 
(b) {\it Uniform uncertainty} is made of the univariate Gaussian distribution. Random variables are drawn from the Gaussian and added to the SMF uniformly with respect to stellar mass, leading to overall shifts in normalization;
(c) {\it Asymmetric Uncertainty} is simply modelled as the modulus of the mock uncertainty such that the uncertainty only goes in the positive direction.
(d) For {\it Gaussian $\om$ uncertainty}, given a set of parameters $(\om^0, \sig^0, \asnone^0, \aagnone^0, \asntwo^0, \aagntwo^0)$, we generate the SMFs out of $(\om',$ $\sig^0,$ $\asnone^0,$ $\aagnone^0,$ $\asntwo^0,$ $\aagntwo^0)$ where $\om'$ $\sim$ $\mathcal{N}(\om^{0},$ $\sigma_{\om})$. Here, $\sigma_{\om}$ is set to 0.04;
(e) We model {\it Gaussian $\om$ uncertainty with white noise} by adding Gaussian white noise directly to the SMFs with the {\it Gaussian $\om$} uncertainty;
(f) {\it Gaussian $\om$ and $\sig$ uncertainty} is modelled similarly to {\it Gaussian $\om$} except that we additionally vary $\sig$ as $\sig'$ $\sim$ $\mathcal{N}(\sig^0,\sigma_{\sig})$ with $\sigma_{\sig}$ of 0.04.
The visual description of above six uncertainties is shown in the {\it top} panels of Figure \ref{fig:various_uncertainty}.

\begin{table}[t]
\centering
\caption{The standard deviations of the inferred marginal distributions in Figure \ref{fig:various_uncertainty}.}
\begin{tabular}{?c?c|c|c|c|c|c?}
\thickhline
& $\sigma_{\om}$ & $\sigma_{\sig}$ & $\sigma_{\asnone}$ & $\sigma_{\aagnone}$ & $\sigma_{\asntwo}$ & $\sigma_{\aagntwo}$\\ 
\thickhline
{a'}& 0.005 &0.001 & 0.006 & 0.084 & 0.006 & 0.010\\ 
\cline{1-7}
{b'}& 0.007 & 0.002 & 0.012& 0.056& 0.008 & 0.023\\ 
\cline{1-7}
{c'}& 0.002 & 0.001 & 0.006 & 0.027& 0.004 & 0.011\\ 
\cline{1-7}
{d'}& 0.025 & 0.001 & 0.001& 0.055& 0.006 & 0.027\\ 
\cline{1-7}
{e'}& 0.030& 0.001 & 0.005& 0.042& 0.004 & 0.012\\ 
\cline{1-7}
{f'}& 0.028& 0.036 & 0.007& 0.030& 0.003 & 0.019\\ 
\thickhline
\end{tabular}
\label{table:various_uncertainty}
\end{table}

We perform ILI from the emulated SMF at $z=0.5$ that is generated with the same parameters that are used in Section \ref{sec:performance_smf}, together with the six uncertainties, respectively. 
Figure \ref{fig:various_uncertainty} shows the marginals of the inferred posteriors ({\it bottom}) and the corresponding SMFs ({\it middle}).
The corresponding posterior ({\it middle}) inferred from each implemented uncertainty ({\it top}) (a, b, c, d, e, and f) is shown in the panels (a', b', c', d', e', and f'), respectively. 
Also, the inferred SMFs ({\it middle}) and marginal densities ({\it bottom}) are consistently color-coded as follow: {\it orange dotted}---(a', b', and c'), {\it green solid}---(d'), {\it blue dotdash}---(e'), and {\it red dashed}---(f'). 
We color (a', b', and c') with the same color since the implemented uncertainties (a, b, and c) share the same property that the uncertainty injected to the SMFs is {\it uncorrelated} with the parameters.

In the case of a, b, and c, 
the standard deviations of the inferred SMFs (a', b', and c') ({\it first row} of th {\it middle} panel of Figure \ref{fig:various_uncertainty}) is an order of magnitude smaller than that of the implemented uncertainty ({\it fisrt row} of the {\it top} panel).
The marginal distributions (a', b', and c') ({\it yellow}) are nearly on top of each other (see Table \ref{table:various_uncertainty}).
Most importantly, the standard deviations of both the SMFs and the marginals are close to that of the inference from the SMFs {\it without uncertainty} in Figure \ref{fig:emulated_smf_no_uncertainty} of Section \ref{sec:performance_smf}, which are $\bar{\bs{\sigma}}_{\rm smf} = 0.007$ dex and $\bs{\sigma}_{(\om, \sig, \asnone,\aagnone, \asntwo, \aagntwo)}$=(0.006, 0.002, 0.024, 0.076, 0.006, 0.018).
This leads us to the proposition that {\it uncorrelated uncertainty} has essentially no impact on inference.

In the case of the uncertainties directly injected to $\om$ and $\sig$, the inferred SMFs (d', f') and the implemented uncertainty (d, f) have approximately the same standard deviations (see Figure \ref{fig:various_uncertainty}).
On the other hand, the standard deviation of the inferred SMFs (e') are completely different from that of the implemented uncertainty (e), since the white noise that is uncorrelated with SMFs in (e) is cancelled out during the inference.
Shown in the zoom-in panels are the analytic Gaussian distributions ({\it black solid}) that are used to generate the uncertainties (d, e, f).
In $\om$, the three inferred marginals---{\it green dotdashed, blue dashed, red dotted}---are in a relatively good agreement with the analytic lines ({\it black solid}).
Note that the marginal (e') has not been affected by the white noise.
Due to the range limit of $\om$, the inferred marginals are skewed such that the probability density near $\om=0.5$ drops sharply compared to the analytic line, whereas in the $\sig$ panel, the marginal (f) ({\it red dotted}) precisely matches the analytic line ({\it black solid}) (see Table \ref{table:various_uncertainty}).

The implemented uncertainties (d, e, f) are {\it correlated} since the uncertainty is directly injected to the parameter such that there exists $\btheta'$ satisfying $f(\btheta') = f(\btheta + \bs{Z}(\eta)) $ for $\btheta' \in \mathbb{R}$. 
Meanwhile, the uncertainty (e) is the {\it partially correlated uncertainty} since it includes the {\it uncorrelated} part as well.
In contrast to {\it uncorrelated} uncertainty (a, b, and c), the variance of {\it correlated} uncertainty (d and f) can successfully be captured in the posteriors (d' and f').
Moreover, in the case of {\it partially correlated} uncertainty (e), the uncorrelated part is cancelled out, leaving only variance of the {\it correlated uncertainty} in the posterior distribution.
This is strong evidence that {\it uncorrelated uncertainty} has negligible impact on ILI and highlights the importance of well-establish mock uncertainty that can reproduce the correlation between the simulation uncertainty and the parameters.
From another point of view, if the simulation uncertainty is not correlated with the parameters, performing ILI using cosmological simulations can naturally eliminate the effects of the simulation uncertainty.
Note that it is not yet proven that the simulation uncertainty is correlated with the parameters.

\begin{figure*}[t!]
    \centering
    \includegraphics[width=\textwidth]{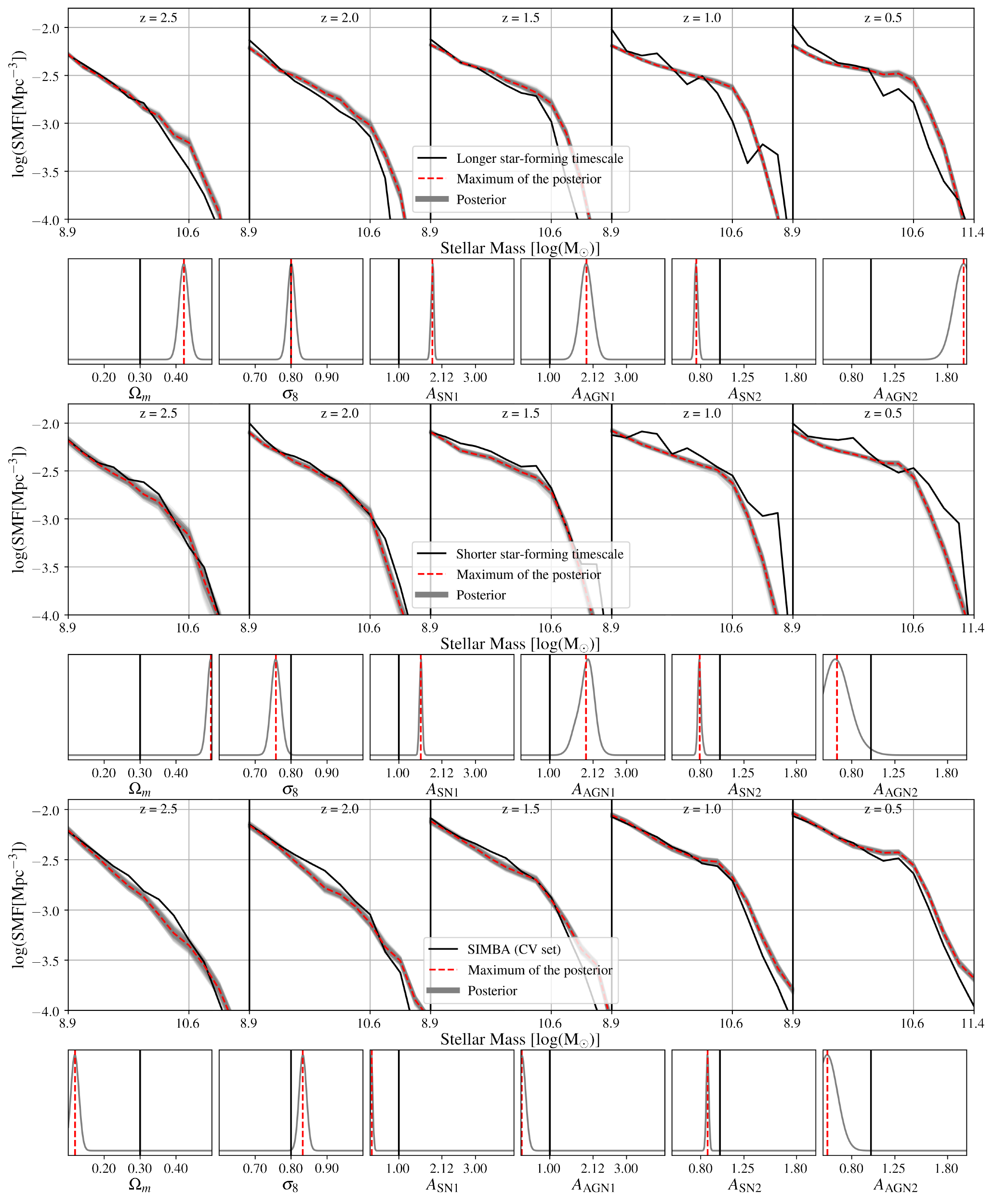}
    \caption{
    {\it Top}: Stellar Mass functions and the marginal posterior densities that are inferred from the SMFs with the six fiducial parameters and a longer star-formation timescale. 
    {\it Middle}: Same, but with a shorter star-formation timescale. 
    {\it Bottom}: Stellar Mass functions and the marginal posterior densities that are inferred from the average SMFs out of the CV set of the SIMBA, rather than IllustrisTNG, suite. 
    }
    \label{fig:add_params}
\end{figure*}

\subsection{Physical Limitations of Simulations}
\label{sec:discussion_physical_limits}
\subsubsection{Mismatch between Inferred SMFs and Observed SMFs}
\label{sec:mismatch_SMFs}
We have seen the significant mismatch between the inferred SMFs and the observed SMFs in Section \ref{sec:observation_smf}.
There are four primary issues in the SMFs inferred from the observed SMFs shown in Figure \ref{fig:obs_smf}: 
(1) The population of massive galaxies ($M_{\star}\gtrsim{10^{10}}\msun$) in the inferred SMFs are located far below the observed SMFs across all redshifts;
(2) The inferred SMFs have a `shoulder' at $M_{\star} = [10^{10.5}, 10^{11}]\msun$, which does not appear in the observations.
(3) The evolution of emulated SMFs with respect to redshift $d\phi(M,z)/dz$ in the low-mass end is smaller than that of the observed SMFs;
(4) the SMF-stellar mass slope changes in time are appreciably different between simulations and observations.

The mismatch in the high mass end has been a crucial problem in the inference from the five concatenated observed SMFs in Section \ref{sec:observation_concatenated_smf}.
The difference between the inferred SMFs and the observed SMFs is approximately 0.3 dex on average \rv{(or the relative error of 41\%)}.
Concerning physical models in cosmological simulations, there can be two physical mechanisms that can control the populations of massive galaxies: (1) large-scale structure formation and evolution and (2) astrophysical feedback.
In the context of structure formation, a sufficient amount of matter ($\om$) and large density fluctuations ($\sig$) can lead to a high abundance of massive halos and galaxies.
On the other hand, it is well known that AGN feedback is a dominant factor in quenching star formation in massive galaxies (for a seminal review, see \citet{fabian2012ARA&A..50..455F,somerville2015ARA&A..53...51S}), whereas stellar feedback is more effective in low-mass galaxies.
However, since AGN feedback is ineffective as we have seen in Section \ref{sec:agn_weak_correlation}, the dependence of the massive-galaxy population on cosmology should considerably increase.
The compensation of cosmology can be found in Section \ref{sec:observation_individual_smf}:
In short, the inferences from the individual SMFs in Figure \ref{fig:obs_smf_seperated} have recovered the population of massive galaxies by having significant larger $\sig$ compared to the posterior inferred from the concatenated SMFs in Figure \ref{fig:obs_smf}.
That being said, the inferences cannot find a set of parameters that can reproduce the evolution of SMFs as seen in the observations.

In addition, Figure \ref{fig:obs_smf_seperated} exhibits several notably different properties for observations and emulators.
The emulated SMFs tend to have a `shoulder' at $M_{\star} = [10^{10.5}, 10^{11}]\msun$, which does not appear in the observed SMFs at all.
Although we can find this property in the actual hydrodynamic simulation, the re-scaling might enhance such properties (refer to Section \ref{sec:method_rescaling} and Appendix \ref{appendix:rescaling}).
Figure \ref{fig:rescaling} demonstrates that the re-scaled SMFs ({\it red}) have more notable shoulders than the SMFs without re-scaling ({\it red dashed}). 
Thus, such shoulders can be attributed to the re-scaling and/or resolution effects.

The rate of evolution of the SMFs in cosmological simulations is relatively low compared to the rate of evolution in observations.
In addition, the slope of the SMFs with respect to stellar mass at the low-mass end hardly changes across redshift in the observations, whereas the slope in simulations tends to decrease as time goes by.
Such discrepancies between simulations and observations can be evidence for the limits of physical models in hydrodynamic simulations and/or the limited dimensions of parameter space.
We anticipate that this problem can be at least partially alleviated by introducing additional dimensions in the parameter space.

\subsubsection{An Extra Parameter: Star Formation Timescale}
\label{sec:extra_params}
In this section, we investigate whether the parameters can be successfully inferred from a simulated observable that comes from a higher-dimensional parameter space that has an {\it extra} dimension.
Thus far, we have performed ILI on six parameters from observables---SFRD or SMFs---that are obtained either by a specific set of six parameters, or from actual observations.
Here, we perform ILI on the same six parameters from SMFs that are simulated from the six fiducial parameters while varying one additional subgrid model parameter, the star formation timescale.
The two simulations are performed with longer (twice of the fiducial) and shorter (half of the fiducial) star formation timescale than the fiducial run (hereafter, LT and ST denote the longer and shorter star formation timescale simulations, respectively).
In LT and ST, the same random seed as in the fiducial run is adopted to minimize the effect of simulation uncertainty through cosmic variance.
Lastly, to approximate emulator counterparts of SMFs for LT (ST), we calculate the ratios of the SMFs bin by bin from LT (ST) to the simulated fiducial SMFs, and then apply them to the emulated SMFs of the fiducial parameters to obtain emulated SMFs for LT (ST).
Here, we assume that the changes of star formation timescale do not lead to any significant changes in other physical properties and observables except SMFs.

Figure \ref{fig:add_params} shows the marginal posterior densities and SMFs inferred from the emulated SMFs from LT and ST ({\it top} and {\it middle} panels, respectively).
In the case of LT, the peaks of the marginal densities of the inferred posteriors ({\it red dashed}) considerably deviates from the fiducial parameters ({\it black solid}), except for $\sig$.
The low-mass end of the inferred SMFs ({\it red dashed}) at $z=2.5, 2.0,$ and $1.5$ is in a relatively good agreement with the SMFs from LT ({\it black solid}), whereas at $z=1.0$ and $0.5$, the inferred SMFs and the SMFs of LT show a complete mismatch.
In addition, the LT has a high evolutionary rate of the SMFs and tends to have steeper slopes as redshift decreases, which the inferred SMFs fail to match as in the inference from observations (see Figure \ref{fig:obs_smf}).
Similar to LT, the inferred SMFs are in good agreement with the SMFs of ST at $z=2.5, 2.0$ and $1.5$, whereas the high-mass end of the inferred SMFs at low redshifts is clearly lacking compared to the SMFs of ST, which we have also seen in the inference from the observation in Figure \ref{fig:obs_smf}.

We have found that the SMFs from the seven-dimensional parameter space including an extra parameter---star formation timescale---are not necessarily reproduced by points in the conventional six-dimensional parameter space that we have used throughout this work.
LT and ST have intrinsic properties that cannot be reproduced from points in the six-dimensional parameter space (e.g., higher rates of evolution in the SMFs, slope evolution in the SMFs, and large populations of massive galaxies).
The results indicate that the introduction of extra dimensions or parameters has the potential to resolve the problem of the significant mismatch between the five concatenated observed SMFs and observations.

\subsubsection{Inference from SIMBA}
\label{sec:simba}
In addition to the extra parameter, we also perform ILI from the SMFs that are obtained by the fiducial simulation from the CV set of the SIMBA suite (refer to Section \ref{sec:method_camels} or \citet[]{camels2021ApJ...915...71V}). 
Likewise, we re-scale the SIMBA SMFs to obtain the emulated SIMBA SMFs in the same way as we have in Section \ref{sec:extra_params}, assuming that the same re-scaling relation is applicable.
Note that although we make inference from a SIMBA-derived observable, we use the same emulators that we have used so far, which are trained on the TNG suite of the CAMELS simulations.

The {\it bottom} row of Figure \ref{fig:add_params} illustrates the inferred SMFs ({\it red dashed} and {\it grey}) and the emulated SIMBA SMFs ({\it black solid}).
The inferred SMFs from the maximum of the posteriors ({\it red dashed}) seem to match the SIMBA SMFs relatively well even though the relative error is $\sim 21\%$.
Unlike LT and ST in Section \ref{sec:extra_params}, the inferred SMFs reasonably follow the trends of the SMFs, such as slopes and rates of evolution.
Still, the difference between inferred parameters and fiducial parameters has not been narrowed.
Such deviations between the inferred and fiducial parameters can be attributed to differences of physical model between TNG and SIMBA.

\subsection{Caveats \& Physical Interpretation of Inference}
\label{sec:physical_interpretation}
This section discusses how the inferred parameters can be interpreted in an astrophysical sense.
We defer the physical interpretation of the inference to the last section because the inferred parameters contain very little meaningful physics at this time.
Prior to the physical interpretation, it is imperative to understand the key factors that influence inferences on parameters: e.g. emulator, simulation uncertainty, resolution convergence, and limited parameter space.

First, we have employed the emulators for computational efficiency, paying the price of discrepancies between simulations and emulators.
The discrepancies inevitably propagate to the inferred posterior and lead to deviations from the posterior that would have been inferred if actual cosmological simulations were employed in the inference.
Assuming that we can replace the emulators with actual cosmological simulations, the next questions shall be: ``Will the posterior inferred from the actual cosmological simulations contain robust information on physics?''
Or, more specifically, ``Will the cosmological parameters inferred from the cosmological simulations, not the emulators, be physical and comparable to the pre-existing estimations?''
To answer these questions, resolution convergence and the limited dimension of parameter space, along with simulation uncertainty, should be taken into consideration.

\subsubsection{Resolution Effect}
\label{sec:discussion_resolution_effect}
The resolution effects in hydrodynamic simulations have been discussed in many papers \citep[and also see Sections \ref{sec:method_rescaling} and Appendix \ref{appendix:rescaling} in this paper]{lia2000A&A...360...76L,lia2002MNRAS.330..821L,ceverino2008IAUS..245...33C,nagamine2010AdAst2010E..16N,hubber2013MNRAS.432..711H,daniel2014ApJ...782...84A,regan2014MNRAS.439.1160R,snaith2018MNRAS.477..983S,pillepichconvergence2018MNRAS.473.4077P,pillepichrescaling2018MNRAS.475..648P}.
In general, it is believed that the hydrodynamic simulations are sensitive to (spatial and mass) resolution.
On the other hand, pure hydrodynamic simulations themselves without subgrid models, such as radiative cooling, are highly likely to be convergent at some level (refer to \citet[]{hubber2013MNRAS.432..711H} for resolution convergence of both SPH and AMR codes).
This implicitly indicates that the resolution effect is likely to be attributed to response of subgrid models to the different resolutions.
In addition, cosmological hydrodynamic simulations with subgrid models can become convergent once it reaches a certain resolution and beyond \citep[]{lia2000A&A...360...76L, lia2002MNRAS.330..821L,ceverino2008IAUS..245...33C,hopkins2018MNRAS.480..800H}.  
The above arguments suggest that resolution effects can generally be incorporated into the subgrid models---astrophysical parameters.
This is often called weak convergence that requires re-calibration for each resolution of simulations, whereas strong convergence requires that the simulation results not change once the subgrid models are fixed \citep[]{schaye2015MNRAS.446..521S}.

We turn our attention to the resolution effect in this work.
As discussed in Section \ref{sec:method_rescaling}, we re-scale the obserables---SFRD and SMF---to minimize the the resolution effect according to \citet[]{pillepichrescaling2018MNRAS.475..648P}.
However, aside from the inaccuracy of re-scaling, there is a problem with re-scaling.
We find that the resolution effect has a significant dependence on the position in parameter space, which is in line with the above arguments. 
In principle, the re-scaling must be a function of parameters $\btheta$, whereas we construct the re-scaling function only based on the fiducial parameters in this work.
Due to computational cost, it is not realistically possible to obtain the full re-scaling relation as a function of parameters by running high-resolution simulations over the entire parameter space \citep[]{ho2022MNRAS.509.2551H}.
Hence, the inference results near the fiducial parameters can be physically comparable to TNG100-1, which is the target resolution of the re-scaling, but other regions of the parameter space may be difficult to grasp.

\subsubsection{Limited Parameter Space}
In this work, we have seen the limits of the six-dimensional model parameter space:
e.g., weak correlation between AGN parameters and observables in Section \ref{sec:agn_weak_correlation}, failure of inference from the observed SMF in Section \ref{sec:mismatch_SMFs}, and potential of the extended parameter space in Section \ref{sec:extra_params}.
The limits of parameter space can affect not only the performance or accuracy of inference but also the physical interpretation of the inferred posterior.
In addition, the limitations of the subgrid model and the limited dimensions of the parameter space can lead to overfitting or over-fine-tuning of the parameters to the target observable, even if the inferred physics is not realistic.
For instance, since our emulators (or simulations) are insensitive to the AGN feedback, the cosmological and stellar feedback parameters should be adjusted to control the population of the high-mass end in Section \ref{sec:observation_individual_smf}.
The parameter space of the TNG universe or the SIMBA universe is likely to be insufficient to describe the real universe.
Thus, we will ultimately move towards extending the parameter space beyond the current TNG universe so that we can resolve the problems of inference failure and secure the physical meaning of the posteriors.

\subsubsection{\rv{Physics of Inferred Parameters}}
\rv{
The posterior distributions inferred from the observations have shown appreciable discrepancy with what researchers usually believe, especially in the cosmological parameters. 
The discrepancies mostly stem from the physical limitations of the simulations upon which the emulators are built, but not from the inference procedure itself. 
Considering that the inference is a procedure that finds the best combinations of the parameters that can match a target observable, we can find some problems in (the diversity of) the ``combinations'', but not ``finding''. 
For instance, in Section \ref{sec:extra_params}, we have directly observed how the limits of parameter space or an additional dimension can have impact on the inference. 
Also, the lack of physics in one part can lead to a (unwanted) compensatory reaction to another part, in Section \ref{sec:mismatch_SMFs}. 
Hence, the inferred parameters in this work are largely affected by intrinsic features of imperfection of simulations such as the limited parameter space volume, resolution effects, subgrid models, and simulation box size.
}

\section{Summary}
\label{sec:summary}
In order to calibrate cosmological simulations against observations, we have employed implicit likelihood inference (ILI) that enables rigorous Bayesian inference in a computationally efficient way by adopting neural density estimators (NDE) that evaluate the likelihood instead of an explicit analytic likelihood in conventional Bayesian inference (Section \ref{sec:method_ILI}).
In addition, for computational efficiency, we have adopted emulators that are trained on $\sim$ 1000 cosmological simulations from the CAMELS project (specifically, those based on the IllustrisTNG framework) to predict simulated observables, taking as input the cosmological and astrophysical parameters, and used these emulators as surrogates to the cosmological simulations (Section \ref{sec:emulator_as_surrogate}).
Using the emulators, we have conducted ILI on the cosmological and astrophysical parameters ($\om$, $\sig$, stellar wind feedback, and kinetic black hole feedback) from the cosmic star formation rate density (SFRD) and stellar mass functions at different redshifts (SMFs) and retrieved 6-dimensional posterior distributions of the parameters.
\\
\\

We summarize our results as follows:
\begin{itemize}
\item The posteriors inferred from emulated SFRD and SMFs converge to their true values with relative errors of less than 1\% in either SFRD or SMFs (Sections \ref{sec:performance_sfrd} and \ref{sec:performance_smf}, respectively). 
However, the SFRD requires an order of magnitude more training data to converge than the SMFs do, having two convergent stages---an unstably convergent stage and a stably convergent stage (Figure \ref{fig:emulated_sfrd_stability}).
    
    \item In the unstably convergent stage, the posterior distribution is bimodal with two {\it degenerate} peaks (\ref{sec:bimodality_sfrd}).
However, the {\it degeneracy} in the SFRD, which is also confirmed with new cosmological simulations (i.e., it is not an artefact of the emulator), is broken with the SMFs in both the emulator and cosmological simulations (Section \ref{sec:degeneracy_breaking}).
This indicates that the SMFs provide stronger constraints for the parameters.
    
    \item In inferences with the mock uncertainty that we add to the emulators to mimic the simulation uncertainty (Appendix \ref{apx:cosmicvariance_butterflyeffect}), the posterior inferred from the emulated SFRD has successfully captured the variance of the mock uncertainty in the parameters, whereas the SMFs cannot capture any variances in the posterior.
In Section \ref{sec:various_types_of_uncertainties}, we show that uncorrelated uncertainties have a negligible impact on ILI.
More work will be required to build a robust model that can precisely reproduce the simulation uncertainty (note that the actual simulation uncertainty might be also uncorrelated with the parameters). 
In addition, an emulator or theoretical model that can be used to marginalize the simulation uncertainty will be another direction to future work (refer to Appendix \ref{apx:marginalization}).
    
\item \rv{Employing ILI on the observationally-driven SFRD and SMFs of \citet{lejasmf2020ApJ...893..111L,msLeja2021arXiv211004314L}, we find the inferred SFRD that matches the target observation with a relative error of only 4.1\% (Section \ref{sec:observation_sfrd}). 
However, the inferred parameters show notable discrepancy with the fiducial values. 
Moreover, the posterior distribution for the cosmological parameters barely includes the values of the standard cosmology. 
Meanwhile, the similarly inferred SMFs show significant discrepancies with the target observed SMFs (Section \ref{sec:observation_smf}).}


\item \rv{The inconsistency between the inferred parameters and standard values could potentially originate from several causes: the limited simulation box size, the resolution effect, the limited parameter space, and the intrinsic physical limits of cosmological simulations with the TNG framework (Section \ref{sec:discussion_physical_limits} and \ref{sec:physical_interpretation}).}

    \item Using mutual information, we measure the correlation between parameter-observable pairs and find that the performance of inference for each observable largely depends on the correlation between the parameters and observable (Section \ref{sec:correlation_between_obs_and_params}).
Also, we confirm that the amount of information in the SMFs is relatively higher than that of the SFRD, which is in line with our inference results.
    
\item In both correlations and inferences, we find that the AGN parameters (black hole kinetic feedback) are most weakly correlated with both SFRD and SMFs.
This can be attributed to the relatively insignificant impact of the black hole kinetic feedback parameters, within the range varied in CAMELS, on both the formation of galaxies with high stellar mass and star formation in massive galaxies compared to cosmological parameters and stellar wind feedback (Section \ref{sec:agn_weak_correlation}).
    
    \item In this work, we have refrained from conducting physical interpretations of the inferred parameters because the inference result is sensitive to emulator accuracy, resolution effects, simulation uncertainty, and inaccuracy from limited parameter space (Sections \ref{sec:physical_interpretation}).

    \item This work is only a cornerstone of calibrating cosmological simulations against observations and provides considerable insights into future directions.
In future work, we will focus on resolution convergence, simulation uncertainty, the extension of the parameter space, as well as the number of target observables, and inference (only) with cosmological simulations without emulator bias.
\end{itemize}

\acknowledgments
We thank Joel Leja for providing preliminary analysis results.
The Flatiron Institute is supported by the Simons Foundation.
J.K. acknowledges support by Samsung Science and Technology Foundation under Project Number SSTF-BA1802-04, and by the POSCO Science Fellowship of POSCO TJ Park Foundation. His work was also supported by the National Institute of Supercomputing and Network/Korea Institute of Science and Technology Information with supercomputing resources including technical support, grants KSC-2020-CRE-0219 and KSC-2021-CRE-0442. \rv{DAA acknowledges support by NSF grants AST-2009687 and AST-2108944, CXO grant TM2-23006X, and Simons Foundation award CCA-1018464. DN acknowledges funding from the Deutsche Forschungsgemeinschaft (DFG) through an Emmy Noether Research Group (grant number NE 2441/1-1).}

\appendix

\begin{figure}[t!]
    \centering
    \includegraphics[width=\textwidth]{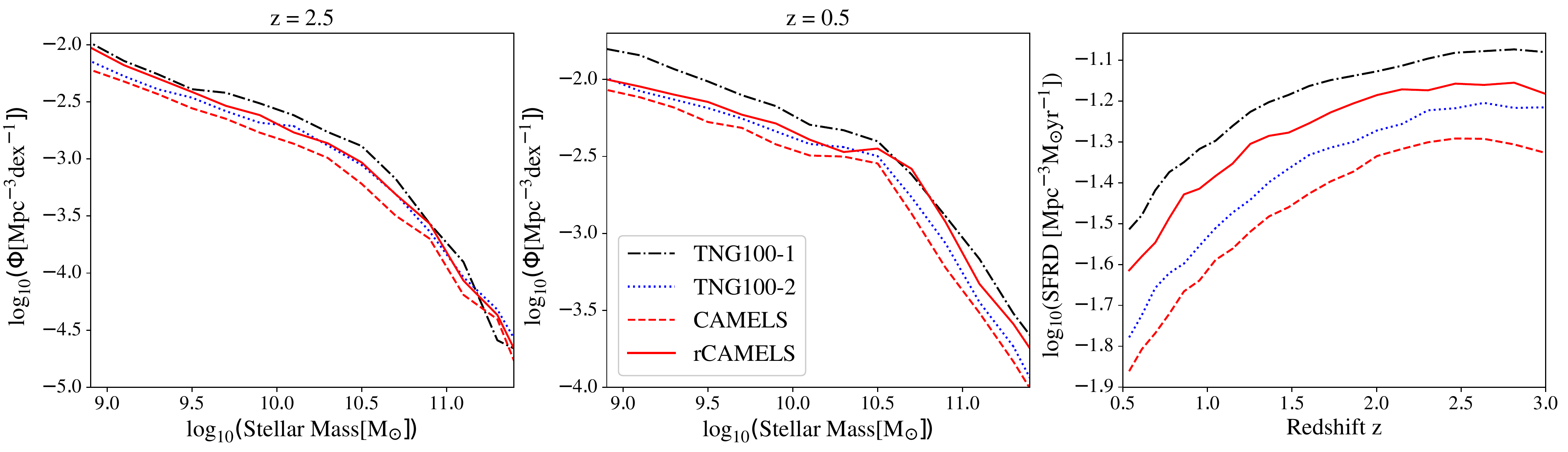}
    \caption{
    {\it Left} and {\it Middle}: Stellar mass functions at $z=2.5$ and $0.5$. 
    {\it Right}: Cosmic star formation rate densities. 
    The difference in TNG100-1 ({\it black dash-dotted}) and TNG100-2 ({\it blue dotted}) illustrates the resolution effect of the TNG simulations.  
    The fiducial CAMELS simulation ({\it red dashed}) is re-scaled with the ratio between TNG100-1 and TNG100-2.
    The re-sacled CAMELS simulation is shown in {\it red solid} lines.
    }
    \label{fig:rescaling}
\end{figure}

\section{Rescaling}
\label{appendix:rescaling}
In Section \ref{sec:method_rescaling}, we have discussed the re-scaling of observables to account for the incomplete resolution convergence of our simulations.
We adopt the re-scaling method introduced in \citet[Appendix A]{pillepichrescaling2018MNRAS.475..648P}.
There, the stellar mass functions (SMF) of TNG300-1 is re-scaled using the ratio between TNG100-2 and TNG100-1.
The re-scaling is based on the resolutions of TNG100-2 and TNG300-1 being essentially the same, and the SMFs of the two simulations being in a good agreement despite the difference in the simulation volumes.
As a result, the re-scaled SMF of TNG300-1 (rTNG300-1) coincides with that of TNG100-1 with high accuracy.

Figure \ref{fig:rescaling} shows resolution effects in each observable in comparison between TNG100-1 ({\it black dash-dotted}) and TNG100-2 ({\it blue dotted}). 
The fiducial CAMELS ({\it red dahsed}), computed as the average of the CV set, is overall slightly lower than TNG100-2 in both SMF and SFRD.
This is because the resolution of the CAMELS simulations is lower than TNG100-2. 
For dark matter particles, the mass and spatial resolution in CAMELS are $\sim 9.67\times10^7\msun/h$ and $\sim 2 \kpc$ comoving, whereas TNG100-2 has a mass resolution of $\sim 5.97\times10^7\msun$ and a spatial resolution of $\sim 1.48 \kpc$ comoving.
In this work, we ignore the discrepancies between the CAMELS simulations and TNG100-2.
We re-scale the SMF by estimating stellar mass ($M_{\star}$) as a function of each bin of halo mass ($M_{\rm halo}$) as follows:
\begin{equation}
M_{\rm \star, rCAMELS}(M_{\rm halo}) = M_{\star, \rm CAMELS}(M_{\rm halo}) \times\frac{\left<M_{\star, \rm TNG100-1}(M_{\rm halo})\right>}{\left<M_{\star, \rm TNG100-2}(M_{\rm halo})\right>},
\end{equation}
where $\left<\cdot\right>$ stands for average over all halos in each bin to which $M_{\rm halo}$ belongs.
We multiply the stellar mass of each CAMELS halo by the corresponding re-scaling factor (the last fractional term) for that halo mass.
The re-scaled SMF of the CAMELS simulations ({\it red solid} in Figure \ref{fig:rescaling}) at $z=2.5$ ({\it left}) is in a good agreement with TNG100-1 ({\it black dash-dotted}), whereas the re-scaled SMF at $z=0.5$ ({\it center}) deviates appreciably from TNG100-1, especially in the low-mass end.
The discrepancy is attributed to the resolution limit that leads to the lower bounds for both dark matter mass ($\sim 10^8\msun$) and stellar mass ($\sim 10^7\msun$) of halos.
Since the mass-resolution limits of halos cause lack of galaxy population in the vicinity of the limits, the construction of the re-scaling factor becomes unfeasible in the halo-mass range of $[10^8, 10^9 ]\msun$, which largely affects the low-mass end of SMFs.
This results in the significant discrepancy between the SMFs of rescaled CAMELS and TNG100-1 in low-mass end at $z=0.5$. 
In this work, we have not employed any post-processing for the zero-stellar mass galaxies whose stellar mass is not resolved due to the mass-resolution limit.

Similarly, we re-scale the SFRD by multiplying the SFR of the CAMELS simulations by the re-scaling factor as a function of bin of halo mass as follows:
\begin{equation}
\mathrm{SFR}_{\rm rCAMELS}(M_{\rm halo}) =
\mathrm{SFR}_{\rm CAMELS}(M_{\rm halo}) 
\times\frac{\left<\mathrm{SFR}_{\rm TNG100-1}(M_{\rm halo})\right>}
{\left<\mathrm{SFR}_{\rm TNG100-2}(M_{\rm halo})\right>},
\end{equation}
where ${\dd M_{\star}}/{\dd t}$ is a star formation rate of galaxy.
The re-scaled SFRD in Figure \ref{fig:rescaling} ({\it right, red solid}) has a consistent offset from TNG100-1 ({\it black dash-dotted}), which is simply attributed to the difference between TNG100-2 ({\it blue dotted}) and the CAMELS ({\it red dashed}).

\begin{figure*}[t!]
    \centering
    \includegraphics[width=\textwidth]{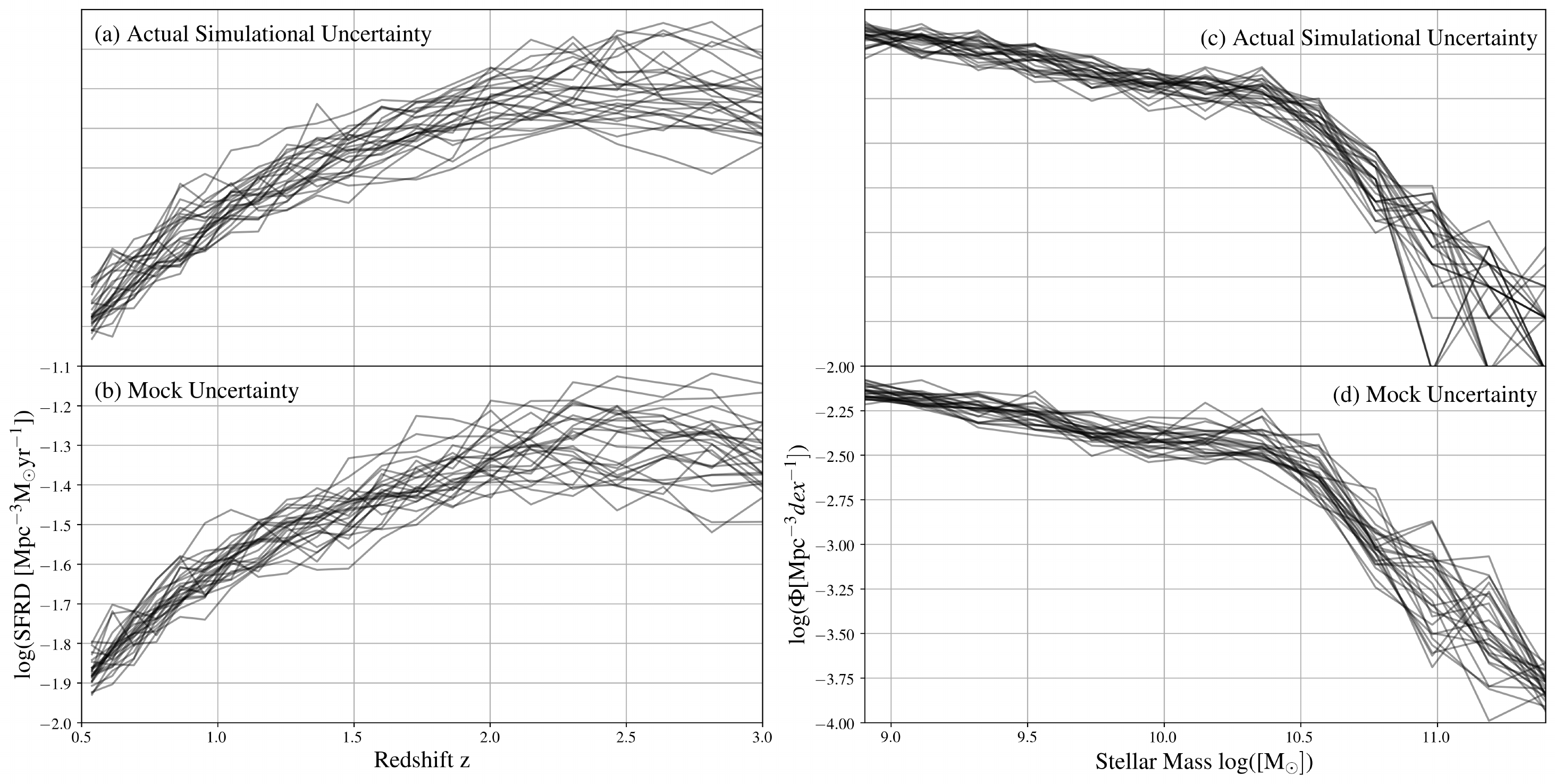}
    \caption{(a) Cosmic star formation rate densities from 27 simulations of the CV set;
    (b) Mean of the 27 cosmic star formation rate densities from the CV set + Mock uncertainty;
    (c) Stellar mass functions at $z=0.5$ from 27 simulations of the CV set; 
    (d) Mean of the 27 stellar mass functions at $z=0.5$ from the CV set + Mock uncertainty.
    Each panel consists of 27 different curves.}
    \label{fig:simulation_uncertainty}
\end{figure*}

\section{Simulation Uncertainty and Mock Uncertainty}

\label{apx:cosmicvariance_butterflyeffect}
In Sections \ref{sec:cosmic_variance_buttefly_effect} and \ref{sec:uncertainty_sfrd}, we have discussed the simulation uncertainty that originates from various sources of randomness of cosmological simulations. 
In this section, we study the simulation uncertainty quantitatively using the CV set and model the mock uncertainty.
Figure \ref{fig:simulation_uncertainty} illustrates 27 SFRDs ({\it top left}, (a)) and 27 SMF at $z=0.5$ ({\it top right}, (c)) from 27 cosmological simulations in the CV set, which represents the simulation uncertainty (refer to Section \ref{sec:method_camels} for details of the CV set).
We measure the standard deviations of 27 SFRDs and 27 SMFs as $\sigma_{\rm sim, sfr} = 0.057$ dex and $\sigma_{\rm sim, smf} = 0.111$ dex, respectively.

We model the mock uncertainty of the SFRD and SMF using a modified six-dimensional Gaussian noise in the form
\begin{equation*}
   \bold{Z}(\bx) = \bold{C}^{\rm T}
   \frac{ \exp(-\frac{1}{2}(\bx-\boldsymbol{\mu})^{\rm T}\boldsymbol{\Sigma}^{-1}(\bx-\boldsymbol{\mu}))}
   {\sqrt{(2\pi)^6}|\boldsymbol{\Sigma}|}
\end{equation*}
with $\bold{C} = \boldsymbol{I}$ for the SFRD and $[0.3,\,0.34,\,0.38,\,0.43,\,0.47,\,0.51,\,0.56,\,0.6,\,0.7,\,0.9,\,1.5,\,1.7,\,0.5]$ for the SMF, $\boldsymbol{\mu} = \boldsymbol{0}$, and $\boldsymbol{\Sigma} = \sigma e^{\gamma\boldsymbol{\Gamma}}$ where
\begin{equation*}
    \boldsymbol{\Gamma} = 
    \begin{bmatrix}
    0&1&2&3&4&5\\
    1&0&1&2&3&4\\
    2&1&0&1&2&3\\
    3&2&1&0&1&2\\
    4&3&2&1&0&1\\
    5&4&3&2&1&0\\
    \end{bmatrix}.
\end{equation*}
Here, $(\sigma, \gamma)$ are $(0.02, 0.2)$ for the SFRD and $(0.02, 2)$ for the SMF. Note that the parameters are empirical.
The parameters of the mock uncertainty are tuned such that the mock uncertainty can have a visually similar form and similar standard deviations to the simulation uncertainty.
In Figure \ref{fig:simulation_uncertainty}, panels (b) and (d) exhibit the 27 SFRDs and 27 SMFs with mock uncertainty.
Here, 27 SFRDs and 27 SMFs are generated by adding the mock uncertainty to the mean of 27 SFRDs and 27 SMFs from the CV set, respectively.
The mock uncertainty-implemented SFRDs and SMFs are visually in a good agreement with SFRDs and SMFs from the simulation uncertainty.
The standard deviations of the mock uncertainty of SFRD and SMF are $\sigma_{\rm mock, sfr}=0.061$ dex and $\sigma_{\rm mock, smf}=0.096$ dex, which approximates to that of the simulation uncertainty.
Note that in principle, the mock uncertainty depends on the cosmological and astrophysical parameters.
Nevertheless, we model the mock uncertainty as if the simulation uncertainty is consistent over the entire parameter space since it is computationally impossible to perform simulations across parameter space to obtain the simulation uncertainty as a function of parameters.

\section{Architecture of emulators}
\label{appendix:arch}
In this section, we describe the details of the neural network architectures used for emulators.
The structures and hyper-parameters of the emulators are automatically optimized using {\tt Optuna}. 
\begin{table}[H]
\centering
\begin{tabular}{c|c|c|c}
\thickhline
Layer & Number of neurons & Dropout & Activation fuctions \\ 
\thickhline
Input& 6 &  & \\ 
 \hline
Fully-connected& 871 & 0.25 & Leaky ReLU\\ 
 \hline
Fully-connected& 100 & 0.26 & Leaky ReLU\\ 
 \hline
Output& 21 & & \\ 
\thickhline
\end{tabular}
\caption{Star formation rate density. Learning rate and weight decay are 0.0028 and 1.25e-06.}
\end{table}

\begin{table}[H]
\centering
\begin{tabular}{c|c|c|c}
\thickhline
Layer & Number of neurons & Dropout & Activation fuctions \\ 
\thickhline
Input& 6 &  & \\ 
 \hline
Fully-connected& 451 & 0.31 & Leaky ReLU\\ 
 \hline
Fully-connected& 727 & 0.76 & Leaky ReLU\\ 
 \hline
Fully-connected& 851 & 0.69 & Leaky ReLU\\ 
 \hline
Fully-connected& 890 & 0.31 & Leaky ReLU\\ 
 \hline
Fully-connected& 825 & 0.61 & Leaky ReLU\\ 
 \hline
Output& 21 & & \\ 
\thickhline
\end{tabular}
\caption{Stellar mass functions at $z=2.5$. Learning rate and weight decay are 0.00015 and 0.00013.}
\end{table}

\begin{table}[H]
\centering
\begin{tabular}{c|c|c|c}
\thickhline
Layer & Number of neurons & Dropout & Activation fuctions \\ 
\thickhline
Input& 6 &  & \\ 
 \hline
Fully-connected& 283 & 0.28 & Leaky ReLU\\ 
 \hline
Fully-connected& 738 & 0.32 & Leaky ReLU\\ 
\hline
Output& 21 & & \\ 
\thickhline
\end{tabular}
\caption{Stellar mass functions at $z=2.0$. Learning rate and weight decay are 0.00051 and 0.00031.}
\end{table}

\begin{table}[H]
\centering
\begin{tabular}{c|c|c|c}
\thickhline
Layer & Number of neurons & Dropout & Activation fuctions \\ 
\thickhline
Input& 6 &  & \\ 
 \hline
Fully-connected& 828 & 0.26 & Leaky ReLU\\ 
 \hline
Fully-connected& 845 & 0.25 & Leaky ReLU\\ 
\hline
Fully-connected& 567 & 0.27 & Leaky ReLU\\ 
\hline
Output& 21 & & \\ 
\thickhline
\end{tabular}
\caption{Stellar mass functions at $z=1.5$. Learning rate and weight decay are 2.4e-05 and 5.7e-05.}
\end{table}

\begin{table}[H]
\centering
\begin{tabular}{c|c|c|c}
\thickhline
Layer & Number of neurons & Dropout & Activation fuctions \\ 
\thickhline
Input& 6 &  & \\ 
 \hline
Fully-connected& 244 & 0.78 & Leaky ReLU\\ 
 \hline
Fully-connected& 626 & 0.22 & Leaky ReLU\\ 
\hline
Output& 21 & & \\ 
\thickhline
\end{tabular}
\caption{Stellar mass functions at $z=1.0$. Learning rate and weight decay are 0.00040 and 0.00013.}
\end{table}

\begin{table}[H]
\centering
\begin{tabular}{c|c|c|c}
\thickhline
Layer & Number of neurons & Dropout & Activation fuctions \\ 
\thickhline
Input& 6 &  & \\ 
 \hline
Fully-connected& 890 & 0.20 & Leaky ReLU\\ 
 \hline
Fully-connected& 298 & 0.23 & Leaky ReLU\\ 
\hline
Fully-connected& 836 & 0.31 & Leaky ReLU\\ 
\hline
Fully-connected& 600 & 0.62 & Leaky ReLU\\ 
\hline
Output& 21 & & \\ 
\thickhline
\end{tabular}
\caption{Stellar mass functions at $z=0.5$. Learning rate and weight decay are 4.3e-05 and 3.9e-07.}
\end{table}

\section{Marginalization of simulation uncertainty in Emulator}
\label{apx:marginalization}
This section discusses how emulators marginalize the simulation uncertainty in the LH set that the emulators are trained on.
The LH set, consisting of 1000 cosmological simulations, suffers from the simulation uncertainty since the initial conditions of the 1000 simulations are all different, namely it is dominated by cosmic variance, since the scale of our box size is $25\,\mathrm{Mpc}/h$.
The simulation uncertainty plays a role as an intrinsic noise in the simulated data in training the emulators, which potentially degrades the accuracy of emulators. 
Thus, we study whether the emulators can marginalize the noise during training or how much the emulator suffers from the simulation uncertainty.
If the emulators can marginalize the simulation uncertainty completely, then the emulator prediction should be equivalent to the mean of the results of a hypothetical suite with a large number of emulations for each point in parameter space---hereafter the uncertainty-marginalized ideal simulation.

We first describe the simulated observable $g$ (SFRD or SMF) as follows: 
\begin{equation}
    g= g(\btheta, \delta(\lambda, \btheta)),
\end{equation}
where $\btheta$ is a set of parameters and $\delta(\lambda, \btheta)$ describes the initial conditions with a random seed $\lambda$ for cosmological simulations.
Here, the marginalization of simulation uncertainty---i.e., the emulator prediction is equivalent to the uncertainty-marginalized ideal simulation---can be written as $\bar{g}(\btheta) = \left<g(\btheta, \delta_{}(\lambda, \btheta))\right>_{\lambda}$ by averaging $g$ over the sources of randomness $\lambda$.
The physical analogy of the mean of the simulation uncertainty is the mean of cosmic variance by performing the infinite-volume simulation (only if we ignore the effects of the long-wave limit in power spectrum in cosmic variance).

The emulator prediction $f$ can be written with respect to the ideally-marginalized prediction---the uncertainty-marginalized ideal simulation---in the form
\begin{equation}
    f= f(\btheta) = \bar{g}(\btheta) + \epsilon(\btheta)+m(\btheta).
\end{equation}
Here, we separate the inaccuracy of the emulator into the training error $\epsilon(\btheta)$ and the marginalization error $m(\btheta)$.
Then, the bias of the emulator with respect to simulations in the LH set, $b_{\rm LH}$, can be written as 
\begin{equation}
    b_{\rm LH}= 
    \left<g(\btheta,\delta(\lambda_{\btheta}, \btheta)) - f(\btheta)\right>_{\btheta_{\rm LH}}=
    \left<g(\btheta,\delta(\lambda_{\btheta}, \btheta))\right>_{\btheta_{\rm LH}}-\left<\bar{g}(\btheta)\right>_{\btheta_{\rm LH}} - \left<\epsilon(\btheta)\right>_{\btheta_{\rm LH}}-\left<m(\btheta)\right>_{\btheta_{\rm LH}},
\end{equation}
where $\btheta_{\rm LH}$ denotes $\btheta \in \Theta_{\rm LH}$ and $\Theta_{\rm LH}$ is a set of 1000 parameters in the LH set. Here, we write $\lambda$ as $\lambda_{\theta}$ because in the LH set, $
\lambda$ are already determined depending on $\btheta$. 
Assuming that the average of simulations over the LH set can average not only the simulations from different parameters but also marginalize their simulation uncertainties such that it can be approximately equal to the average of the uncertainty-marginalized ideal simulations over the LH set---i.e., $\left<g(\btheta,\delta(\lambda_{\btheta},\btheta))\right>_{\btheta_{\rm LH}} \sim \left<\left<g(\btheta, \delta(\lambda, \btheta))\right>_{\lambda}\right>_{\btheta_{\rm LH}}\equiv\left<\bar{g}(\btheta)\right>_{\btheta_{\rm LH}}$, we obtain
\begin{equation}
    b_{\rm LH}\simeq
    -\left<\epsilon(\btheta)\right>_{\btheta_{\rm LH}}-\left<m(\btheta)\right>_{\btheta_{\rm LH}}.
\end{equation}
The empirical biases $b_{\rm LH}$ for the SFRD and SMFs are 0.0026 dex and -0.0014 dex, respectively. 
Compared to the standard deviations of the simulation uncertainty for the SFRD and SMFs ($\sim 0.06$ dex and $\sim 0.1$ dex), the emulators have relatively small bias, which indicates that the mean of emulators and the mean of simulations are in a good agreement.

Secondly, we estimate the variance of emulator prediction with respect to the uncertainty-marginalized ideal simulation in terms of the LH set as follows:
\begin{equation}
\begin{split}
    \hat{\sigma}_{\rm LH}^2&= 
\left<\left(\bar{g}(\btheta)- f(\btheta)\right)^2\right>_{\btheta_{\rm LH}}\\
&=
\left<\left(\bar{g} (\btheta)-g(\btheta,\delta(\lambda,\btheta))+g(\btheta,\delta(\lambda,\btheta))- f(\btheta)\right)^2\right>_{\btheta_{\rm LH}}.
\end{split}
\end{equation}
With the definitions $A\equiv \bar{g}(\btheta) - g(\btheta,\delta(\lambda_{\btheta},\btheta))$, $B\equiv g(\btheta,\delta(\lambda_{\btheta},\btheta))-f(\btheta)$ and $C\equiv\bar{g}(\btheta)-f(\btheta)$, we obtain
\begin{equation}
    \begin{split}
        \tilde{\sigma}_{\rm LH}^2=
\left<
C^2
\right>_{\btheta}&_{\rm LH}=
\left<
A^2
\right>_{\btheta_{\rm LH}}
+
\left<
B^2
\right>_{\btheta_{\rm LH}}
+2
\left<
AB
\right>_{\btheta_{\rm LH}}
\\&=   
\left<
A^2
\right>_{\btheta_{\rm LH}}
+
\left<
B^2
\right>_{\btheta_{\rm LH}}
+2
\left<
A(C-A)
\right>_{\btheta_{\rm LH}}
\\&=   
-\left<
A^2
\right>_{\btheta_{\rm LH}}
+
\left<
B^2
\right>_{\btheta_{\rm LH}}
+2
\left<
AC\right>_{\btheta_{\rm LH}}
\\&\simeq
-\left<
A^2
\right>_{\btheta_{\rm LH}}
+
\left<
B^2
\right>_{\btheta_{\rm LH}}.
\end{split}
\end{equation}
In the third line, $\left<AC\right>_{\btheta_\mathrm{LH}}$ is the covariance of $A$ and $C$ where $A$ and $C$ are the simulation uncertainty and the deviations of the emulator from the ideally-marginalized prediction, respectively.
Since the error of emulator and simulation uncertainty are, in principle, fully independent, we assume that $\left<AC\right>_{\btheta_\mathrm{LH}}$ approximately vanishes.
$\left<A^2\right>_{\btheta_{\rm LH}}$ is the variance of simulation uncertainty and $\left<B^2\right>_{\btheta_{\rm LH}}$ is the variance of emulator with respect to simulation in the LH set, both of which are measurable quantities. 
$\left<A^2\right>_{\btheta_{\rm LH}}$ and $\left<B^2\right>_{\btheta_{\rm LH}}$ for the SFRD are 0.036 dex$^2$ and 0.004 dex$^2$, and $\left<A^2\right>_{\btheta_{\rm LH}}$ and $\left<B^2\right>_{\btheta_{\rm LH}}$ for the SMFs are 0.369 dex$^2$ and 0.035 dex$^2$.
Thus, $\hat{\sigma}^2_{\rm LH}$  for the SFRD and SMFs are 0.032 dex$^2$ and 0.334 dex$^2$, respectively.
As a result, the emulators have negligibly small biases but are deviated from the mean of the simulation uncertainty at a similar level to the variance of the simulation uncertainty. 
Therefore, we can conclude that the emulators cannot properly marginalize the simulation uncertainty.

\section{Definition of degeneracy}
\label{apx:definition_degeneracy}
We have encountered degeneracy in Section \ref{sec:bimodality_sfrd} that discusses the bimodality of the SFRD.
In this section, we probabilistically discuss the relation between uncertainty and degeneracy and mathematically define degeneracy for our purposes.
In terms of parameter-observable pair, the degeneracy originates from {\it indistinguishability} among these pairs.
Given an arbitrary observation $\bx_0$, if there exists a set of parameters $\boldsymbol{\Theta}_{\rm degen}$ such that $\boldf(\btheta)|_{\btheta \in \boldsymbol{\Theta}_{\rm degen}} = \bx_0$ where $\boldf(\btheta)$ is e.g.~a theoretical model, a simulation, and a fast approximation method that predict observable $\bx$ as a function of parameters $\btheta$, then one {\it cannot pinpoint} the parameters $\btheta$ from which the given observation $\bx_0$ comes.
In this case, the pairs $\{(\btheta, \bx_0)|\btheta \in \boldsymbol{\Theta}_{\rm degen}\}$ are said to be degenerate with respect to observation $\bx_0$.
This can be usually seen in quantum systems such as the spin triplet state under no magnetic field.
In consideration of an arbitrary uncertainty $\boldsymbol{Z}(\eta)$ in observable\footnote{
For example, the cosmological simulations $\boldf$ reproduce different output $\bx$ depending on the initial conditions $\eta$ even with the same set of physical and free parameters $\btheta$. i.e.~$\boldf(\btheta) = \bx + \bs{Z}(\eta)$ where $Z({\eta})$ stands for cosmic variance.}
, the pairs can be written in either $(\btheta, \bx + \boldsymbol{Z}(\eta))$ or $(\btheta+\boldsymbol{\epsilon}(\eta), \bx)$
\footnote{
Given a model $\boldf(\btheta)$ that predicts observable $\bx$ taking parameters $\btheta$ as input, the uncertainty in observable $\boldsymbol{Z}(\eta)$ can be propagated onto parameters $\btheta$ as follows:
$\bx + \boldsymbol{Z}(\eta)\xrightleftharpoons[\bx=\boldf(\btheta)]{\btheta=\bs{g}(\bx)}\btheta+\boldsymbol{\epsilon}(\eta)$
where $\eta$ is a random seed.
Here, $\bs{g}(\bx)$ can be a set of locally defined functions that satisfy $\bx=f(\btheta)$.
\label{footnote:propagation_uncertainty}}
where $\eta$ is a random seed (Section \ref{sec:uncertainty_sfrd} and Appendix \ref{apx:cosmicvariance_butterflyeffect}).
One can notice that $\boldsymbol{Z}$ and $\boldsymbol{\epsilon}$ are basically random variables that require a probabilistic treatment.
In the following section, degeneracy will be discussed in a probabilistic manner.

The ideal probabilistic inference naturally traces the propagation of uncertainty in observation onto each parameter\footref{footnote:propagation_uncertainty}.
Nevertheless, the consistent, robust confinement for the posterior density is necessary to define a finite region of the degenerate parameter space since an arbitrary inferred posterior density is generally well-defined over the entire parameter space.

We define degeneracy as follows.
\\
Given an arbitrary probability density $p(\bs{\theta})$ and $\zeta_{\rm thres}$, 
there exist $p_{\rm thres}$ and $\bs{\Theta}_{\rm degen}$ satisfying $\bs{\Theta}_{\rm degen} = \{{\bs{\theta} \,|\, p(\bs{\theta}) \geq p_{\rm thres}}\}$ 
such that 
\begin{equation}
\label{equation:degeneracy_continuous}
    \hspace{3mm}({\rm continuous})\hspace{7mm}
\frac{\int_{\bs{\theta} \in \bs{\Theta}_{\rm degen}}p(\bs{\theta})d\bs{\theta}}
    {\int_{\bs{\theta} \in \mathcal{V}} p(\bs{\theta})d\bs{\theta}}=\zeta_{\rm thres} 
\end{equation}
or 
\begin{equation}
\label{equation:degeneracy_discrete}
\centering
    \hspace{5mm}({\rm discrete})\hspace{9mm}
    \frac{\sum_{\bs{\theta} \in \bs{\Theta}_{\rm degen}}p(\bs{\theta})\Delta\bs{\theta}}
    {\sum_{\bs{\theta}\in\mathcal{V}}p(\bs{\theta})\Delta\bs{\theta}} = \zeta_{\rm thres} 
\end{equation}
where $0 < \zeta_{\rm thres} < 1$ and $\mathcal{V}$ stands for the entire parameter space.
Here, the parameters $\bs{\theta}$ in $\bs{\Theta}_{\rm degen}$ are said to be degenerate. 
Note that the threshold value $\zeta_{\rm thres}$ is a free parameter. 

In case of a high dimensional problem with an intractable probability distribution, it is practically impossible to numerically integrate the probability distribution $p(\bs{\theta})$ over $\mathcal{V}$ or $\bs{\Theta}_{\rm degen}$ even if $\bs{\Theta}_{\rm degen}$ is known. 
In low dimensional problems, simple quadrature methods can work. 
For example, we can evaluate a function on a fixed grid of points, then apply the trapezoid rule. 
However, in high dimensions, the number of grid points grows exponentially.
\footnote{In our case, six dimensional parameter space requires $(10^2)^6=10^{12}$ grid points where we space each axis with 100 grid points, which means that $10^{12}$ arithmetic calculations at least are needed.}  
Hence, MCMC methods are widely adopted to integrate high dimensional functions.

We utilize the MCMC sampling to estimate the degenerate parameter space. 
The MCMC provides millions of parameter sets such that the ratio of the number of parameter sets in each bin, $N(\Delta\btheta_{\rm bin})$, to the total number of the samplings, $N_{\rm total}$, represents the approximate probability of that bin.
In other words, $p(\btheta)\Delta\btheta_{\rm bin} \simeq N(\Delta\btheta_{\rm bin})/N_{total}$ where $\Delta\btheta_{\rm bin}$ is the size of each bin. 
We can rewrite the left-hand side of Eq. \ref{equation:degeneracy_discrete} in the form 
\begin{equation}
    \frac{
    \sum_{\bs{\theta} \in \bs{\Theta}_{\rm degen}}p(\bs{\theta})\Delta\bs{\theta}
    }
    {
    \sum_{\bs{\theta}\in\mathcal{V}}p(\bs{\theta})\Delta\bs{\theta}
    } =  
    \frac{
    \sum_{\bs{\theta} \in \bs{\Theta}_{\rm degen}} N(\Delta\btheta_{\rm bin})
    }
    {
    N_{\rm total}
    }.
\end{equation}
Then,
\begin{equation}
    \sum_{\bs{\theta} \in \bs{\Theta}_{\rm degen}} N(\Delta\btheta_{\rm bin})
    = \zeta_{\rm thres}N_{\rm total}.
\end{equation}
Together with $\bs{\Theta}_{\rm degen} = \{{\bs{\theta} \,|\, p(\bs{\theta}) \geq p_{\rm thres}}\}$, 
we can write 
$\bs{\Theta}_{\rm degen} = \argmax_{D \in \mathcal{D}}p_{\rm thres}; \mathcal{D}=\{D|\tilde{p}(\btheta)>p_{\rm thres} \forall\btheta; \btheta \in D ,\,n(D)=\zeta_{\rm thres} N_{\rm total},\, D \subset \bs{\Theta}_{\rm MCMC} \}$ 
where $\bs{\Theta}_{\rm MCMC}$, $\tilde{p}$ and $n(\cdot)$ are the set of sampled parameters from MCMC, an arbitrary surrogate posterior function, such as the NDE or kernel density estimation, and the cardinality of a set.
The set of degenerate points in parameter space, $\bs{\Theta}_{\rm degen}$, is a subset of $\bs{\Theta}_{\rm MCMC}$ such that the sum of probability of the parameters in the subset is equal to $\zeta_{\rm thres}$ (refer to Eq. \ref{equation:degeneracy_discrete}).
In practice, we identify $\bs{\Theta}_{\rm degen}$ among $\bs{\Theta}_{\rm MCMC}$ as follows:
(1) Estimate $\tilde{p}(\btheta)$ for all $\btheta \in \bs{\Theta}_{\rm MCMC}$;
(2) Pair $\btheta$ and $\tilde{p}(\btheta)$ into $(\btheta, \tilde{p}(\btheta))$;
(3) Sort $(\btheta, \tilde{p}(\btheta))$ in a descending order along $\tilde{p}(\btheta)$;
(4) Identify the first $\zeta_{\rm thres}N_{\rm total}$ number of $\btheta$ of the sorted pairs as $\bs{\Theta}_{\rm degen}$.

\bibliography{main}{}
\bibliographystyle{aasjournal}

\end{document}